\documentclass[prb,aps,amssymb,twocolumn,showpacs,floatfix]{revtex4}
\usepackage[dvips]{graphicx}
\renewcommand{\theequation}{\arabic{section}.\arabic{equation}}
\def\be{\begin{equation}}
\def\ee{\end{equation}}
\def\bea{\begin{eqnarray}}
\def\eea{\end{eqnarray}}

\begin{document}  
 
\title{Electron transport in disordered Luttinger liquid}
\author{I.V.~Gornyi$^{1,*}$}
\author{A.D.~Mirlin$^{1,2,\dagger}$}
\author{D.G.~Polyakov$^{1,*}$}
\affiliation{$^{1}$Institut f\"ur Nanotechnologie,
Forschungszentrum Karlsruhe, 76021 Karlsruhe, Germany \\
$^{2}$Institut f\"ur Theorie der kondensierten Materie, Universit\"at
Karlsruhe, 76128 Karlsruhe, Germany}

\date{\today}

\begin{abstract} We study the transport properties of interacting electrons in
a disordered quantum wire within the framework of the Luttinger liquid
model. We demonstrate that the notion of weak localization is applicable to
the strongly correlated one-dimensional electron system. Two alternative
approaches to the problem are developed, both combining fermionic and bosonic
treatment of the underlying physics.  We calculate the relevant dephasing
rate, which for spinless electrons is governed by the interplay of
electron-electron interaction and disorder, thus vanishing in the clean
limit. Our approach provides a framework for a systematic study of mesoscopic
effects in strongly correlated electron systems. \end{abstract}

\pacs{71.10.Pm, 73.21.-b, 73.63.-b, 73.20.Jc}

\maketitle

\section{Introduction}
\label{Intro}

Mesoscopics of strongly correlated electron systems has emerged as an area of
great interest to both experimental and theoretical communities working in the
field of nanoscale physics. Recently, progress in manufacturing of nanodevices
has paved the way for systematic transport measurements on narrow quantum
wires with a few or single conducting channels.  Most prominent examples of
these are: single-wall carbon nanotubes,\cite{Carbon}
cleaved-edge,\cite{auslaender02} $V$-groove,\cite{palevski05} and crystallized
in a matrix\cite{zaitsev-zotov00} semiconductor quantum wires, quantum Hall
edges running in opposite directions and interconnected by means of
tunneling,\cite{kang00,grayson05} polymer nanofibers,\cite{aleshin04} and
metallic nanowires.\cite{slot04,venkataraman06}

Much experimental attention has been focused on the electrical properties of
single-wall nanotubes
(Refs.\onlinecite{bockrath99,yao99,yao00,shea00,nygard00,bockrath01,
postma01,krupke02,mason04,man05,wei05,leek05,man06,kasumov99,morpurgo99,
jarilloherrero06} and references therein).  In particular, evidence has
emerged pointing towards the existence of Luttinger liquid in metallic
single-wall carbon nanotubes, as expected for strongly interacting electrons
in one dimension (1d). The Luttinger liquid behavior was observed via the
power-law temperature and bias-voltage dependence of the current through
tunneling contacts attached to the nanotubes. In the past few years,
technological advances have made possible the fabrication of contacts between
nanotubes and metallic leads with a very low contact resistance, which allows
one to observe mesoscopic effects.\cite{man05,wei05,leek05,man06} Further,
techniques to grow very long (of the mm scale) nanotubes have been developed
and the corresponding transport measurements performed.\cite{li04}

On the theoretical side, the challenge is to expand the ideas that have been
developed for mesoscopic disordered systems on one side and for strongly
correlated clean systems on the other.  The situation is particularly
interesting in 1d, where both disorder and electron-electron (e-e)
interaction, even if they are weak, modify dramatically the large-scale,
low-energy physics of the problem. In a clean wire, e-e interaction leads to
the formation of a Luttinger-liquid ground state. The properties of the
Luttinger liquid without impurities and in the presence of a single barrier
are known in great detail, see
Refs.~\onlinecite{solyom79,voit94,schulz95,gogolin98,schulz00,giamarchi04} for
review. Much less is known about the Luttinger liquid in the presence of many
impurities (disordered Luttinger liquids). The influential works by Apel and
Rice\cite{apel82a} and by Giamarchi and Schulz\cite{giamarchi88} have defined
the state of the art in this area for two decades. Recent years have seen a
revival of interest in disordered Luttinger
liquids,\cite{li02,nattermann03,giamarchi04a,
glatz04,malinin04,artemenko05,chudnovskiy05,rosenow06,altland06} largely
motivated by the technological advances mentioned above. However, the very
applicability of such key notions of the mesoscopic physics as weak
localization (WL) and dephasing to a disordered 1d system, not to mention a
detailed analysis of these effects, remained an open problem. This problem is
a subject of the present paper. Throughout the paper we study spinless
(spin-polarized) electrons. Spin-related effects will be considered in
Ref.~\onlinecite{yashenkin06}.

The structure of the paper is as follows. In Sec.~\ref{II} we summarize the
earlier achievements in the mesoscopics of higher-dimensional systems on one
side and in the theory of strongly correlated 1d systems on the
other. Section~\ref{III} contains an exact formulation of the problem, as well
as a discussion of our strategy in solving it. In Sec.~\ref{IV} we present the
bosonization method and the renormalization-group technique that we use to
account for the Luttinger-liquid renormalizations. Having analyzed in
Sec.~\ref{V} the e-e inelastic scattering rate in 1d, we turn to the
calculation of a weak-localization correction to the conductivity and the
associated dephasing rate in Sec.~\ref{VI}. We also compare this dephasing
rate with the one that governs the damping of Aharonov-Bohm
oscillations. Section~\ref{VII} is devoted to the analysis of the transport
properties of a disorder Luttinger liquid by using an alternative
approach---the ``functional bosonization''---which allows one to treat the
renormalization and the inelastic scattering on an equal footing. Our results
are summarized in Sec.~\ref{VIII}. Technical details of the calculations are
presented in several Appendices. Some of the results of this paper were
published in a brief form in Letter Ref.~\onlinecite{gornyi05a}.

\section{Background}
\label{II}
\setcounter{equation}{0}


In this introductory section, we briefly summarize the known results for
higher-dimensional mesoscopic systems (Sec.~\ref{IIa}) and for strongly
correlated 1d systems (Sec.~\ref{IIb}). These will serve as a starting point
for our work.

\subsection{Mesoscopics of higher-dimensional systems}
\label{IIa}

During the last three decades, mesoscopic physics of low-dimensional diffusive
systems has been attracting a great deal of
attention.\cite{altshuler85,meso91, imry97,efetov97,aleiner99,mirlin00} It
has been recognized that the effects of the quantum interference and e-e
interaction become strongly enhanced in disordered systems of reduced
dimensionality $d$, affecting in an essential way the low-temperature
transport properties of the systems. These effects lead to a set of remarkable
phenomena, including the WL, mesoscopic conductance fluctuations, and the
Altshuler-Aronov interaction-induced quantum correction to the
conductivity. At still lower temperatures, these effects may drive the system
towards strong localization. We now briefly remind the reader the basic facts
in this field.

\subsubsection{Moderately low temperatures: Weak-localization regime}
\label{IIa1}

The quantum localization is the most prominent manifestation of the quantum
interference in disordered systems. At not too low temperatures, the
localization effects are cut off by inelastic processes, resulting in a WL
correction\cite{gorkov79} $\Delta\sigma_{\rm wl}$ to the Drude conductivity
$\sigma_{\rm D}$, 
\bea 
\label{1} 
{\Delta\sigma_{\rm wl}\over \sigma_{\rm D}} &\simeq& -
\int_{L_\phi^{-1}}^{l^{-1}} {(d{\bf q}) \over \pi\nu_0 D q^2} \nonumber \\
&\sim& -{1\over \nu_0 D} \left\{ \begin{array}{ll} \ln(\tau_\phi/\tau)~, &
\qquad {\rm 2d}~, \\[0.4cm] (D\tau_\phi)^{1/2}~, & \qquad {\rm quasi-1d}~,
\end{array} 
\right.  
\eea 
where $D$ is the diffusion constant, $\nu_0$ the density of states, $l$ the
mean free path, $\tau$ the transport mean free time, $\tau_\phi$ and 
$L_\phi=(D\tau_\phi)^{1/2}$ are the interaction-induced
dephasing time and length, respectively. We use notation $(d{\bf
q})=d^dq/(2\pi)^d.$ Throughout the paper $\hbar=1$. Another manifestation of
the quantum coherence are mesoscopic fluctuations of the conductance and of
other observables.

As far as e-e interaction in the WL regime is concerned, it is responsible for
two distinctly different main effects. First, it renormalizes elastic
scattering on impurities through the creation of virtual electron-hole
excitations (screening) with a high energy transfer $\omega$ larger than the
temperature $T$, $T\alt |\omega| \alt \epsilon_F$. Here $\epsilon_F$ is the
Fermi energy. At sufficiently high $T$ (in the ballistic regime $T\tau\gg 1$),
i.e., when the relevant spatial scales are smaller than $l$, the screening can
be described in terms of a dressing of impurities by Friedel
oscillations\cite{rudin97,zala01} and yields a $T$-dependent Drude
conductivity.\cite{zala01} At lower $T$ (in the diffusive regime $T\tau\ll
1$), effects of this type generate the Altshuler-Aronov correction to the
conductivity,\cite{altshuler85}
\bea
\label{2}
{\Delta\sigma_{\rm AA}\over\sigma_{\rm D}} &\simeq& 
-\alpha \int_{L_T^{-1}}^{l^{-1}} 
{(d{\bf q}) \over \pi\nu_0 D  q^2} \nonumber \\
&\sim& {\alpha\over \nu_0 D} 
\left\{
\begin{array}{ll}
\ln(T\tau)~, & \qquad {\rm 2d}~, \\[0.4cm]
-(D/T)^{1/2}~, & \qquad {\rm quasi-1d}~,
\end{array}
\right. 
\eea
where $L_T\sim (D/T)^{1/2}$ is the thermal length. This correction is
perturbative both in the effective interaction constant $\alpha$ and in the
disorder strength. The Coulomb interaction corresponds to $\alpha\sim 1$. The
$T$ dependence of $\Delta\sigma_{\rm AA}$ is determined by the thermal
smearing of the distribution function, which introduces an infrared cutoff on
spatial scales of the order of $L_T$. Using the renormalization-group (RG)
approach within the replicated $\sigma$-model, Finkel'stein extended this
theory to include the interaction nonperturbatively.\cite{finkelstein83}

Effects related to the interaction-induced renormalizations and screening are
also responsible for a zero-bias anomaly\cite{altshuler85,finkelstein83} in
the tunneling density of states as a function of electron energy,
$\nu(\epsilon)$.  It is worth noting that, in the case of Coulomb interaction,
$\nu(\epsilon)$ may become strongly
suppressed,\cite{finkelstein83,levitov97,kamenev99}
\bea 
{\nu(\epsilon)\over \nu_0} &\sim& \exp\left\{-{1\over 8\pi^2 \nu_0
  D}\left[\,\ln^2\left(D\kappa^2/|\epsilon|\right)\right.\right.\nonumber \\
&-&\left.\left. 
\ln^2(D\kappa^2\tau)\,\right]\right\}\alt 1~,\qquad {\rm 2d}~,  
\label{3}
\eea
where $\kappa^{-1}$ is the screening radius, while the conductivity can
still remain close to its Drude value ($\Delta\sigma_{\rm AA}/\sigma_{\rm
D}\ll 1$).

The second effect of interaction is inelastic e-e scattering which breaks the
phase coherence.\cite{altshuler85} A key concept in the localization theory of
a disordered Fermi liquid is that of the dephasing rate $\tau_\phi^{-1}$. The
phase-breaking processes are of crucial importance in the problem of transport
since without dephasing the conductivity of low-dimensional systems would be
zero at any $T$, see Sec.~\ref{IIa2}. Two qualitatively different sources of
dephasing are possible: (i) scattering of electrons by external excitations
(in practice, phonons) and (ii) e-e scattering (at low $T$ the dephasing is
mostly due to e-e interactions). In either case, at sufficiently high $T$, the
dephasing rate $\tau_\phi^{-1}$ is high, so that the localization effects are
reduced to the WL correction to the Drude conductivity.

A characteristic energy transfer in the real inelastic processes that
determine the phase relaxation is restricted by temperature: $|\omega|\alt T$,
so that the dephasing rate vanishes at zero temperature. The infrared
divergency characteristic of a diffusive system makes it necessary to
introduce a self-consistent cutoff in the equation for the dephasing
rate:\cite{altshuler85}
\bea
\label{4}
\tau_\phi^{-1} &\sim & 
{\displaystyle  
\alpha^2 T \int _{L_\phi^{-1}}^{L_T^{-1}} 
{(d{\bf q}) \over \pi\nu_0 D q^2}
}
\nonumber \\
&\sim& 
\left\{
\begin{array}{ll}
{\displaystyle {\alpha^2 T\over \nu_0 D}
\ln\left({\nu_0 D\over \alpha^2}\right)}~, 
&\quad {\rm 2d}~, 
\\[0.4cm] 
{\displaystyle\left({\alpha^2 T \over \nu_0 D}\right)^{2/3}D^{1/3}}~, &\quad
{\rm quasi-1d}~. 
\end{array}
\right. 
\eea
A rigorous calculation necessitates employing the path-integral
technique.\cite{altshuler82}

The different energy scales relevant to the disorder renormalization due to
virtual processes with $T<|\omega|<\epsilon_F$ and to the dephasing due to
real inelastic scattering with $|\omega|\alt T$ allow for a straightforward
separation of these two effects of e-e interaction. At sufficiently high $T$,
the conductivity
\be
\label{5}
\sigma(T)\simeq\sigma_{\rm D}+\Delta\sigma_{\rm wl}+\Delta\sigma_{\rm AA}
\ee
is close to the Drude value $\sigma_{\rm D}$. It is worth mentioning that for
weak interaction ($\alpha\ll 1$) the WL correction [Eq.~(\ref{1}) with
$\tau_\phi$ given by Eq.~(\ref{4})] is much stronger than the Altshuler-Aronov
correction [Eq.~(\ref{2})]. The WL correction grows with lowering $T$ and
eventually becomes strong ($|\Delta\sigma_{\rm wl}|/\sigma_{\rm D}\sim 1$)
when $L_\phi$ reaches $\xi$ or, equivalently, when $\tau_\phi^{-1}$ becomes of
order $\Delta_\xi$. Here $\xi$ is the localization length and $\Delta_\xi$ is
the corresponding energy level spacing. At this temperature,
$|\Delta\sigma_{\rm AA}|/\sigma_{\rm D}\ll 1$ for $\alpha\ll 1$. This means
that in a weakly interacting system the strong localization occurs due to the
growth of the interference-induced WL correction. In the next
subsection, we briefly overview the strong localization regime which emerges
at low temperatures, when $\tau_\phi^{-1} \alt \Delta_\xi.$


\subsubsection{Low temperatures: Fate of dephasing in the strongly localized
regime}
\label{IIa2}

In a pathbreaking paper\cite{anderson58} Anderson demonstrated that a quantum
particle may become localized by a random potential. In particular, in
noninteracting systems of one- or two-dimensional geometry even weak disorder
localizes all electronic states,\cite{abrahams79,gorkov79} thus leading to the
exactly zero conductivity, $\sigma(T)=0$, whatever temperature $T$. A nonzero
$\sigma(T)$ in such systems may only occur due to inelastic scattering
processes leading to the dephasing of electrons.

As discussed in Sec.~\ref{IIa1}, the WL correction (\ref{1}) to the Drude
conductivity of a diffusive system behaves as $\tau_\phi^{(2-d)/2}$
($\ln\tau_\phi$ for $d=2$) and diverges with lowering temperature for $d\leq
2$. This singular behavior of the perturbative corrections is a manifestation
of strong Anderson localization. It prompts a question as to how the
disordered interacting low-dimensional systems conduct at low temperatures,
when the localization effects are no longer weak.

It is commonly believed that transport in the localized regime is of the
hopping nature. Mott\cite{mott74} has developed the picture of
variable-range-hopping (VRH) as the transport mechanism in the insulating
state. The VRH mechanism of Ref.~\onlinecite{mott74} is well justified in the
presence of delocalized phonons which provide inelastic scattering between
single-particle states with different energies (phonon-assisted hopping). The
finite-temperature conductivity is then proportional to the strength of the
electron-phonon coupling and governed by Mott's VRH:\cite{mott74}
$\sigma(T)\propto \exp[-(T_0/T)^{\mu}]$ with $\mu = 1/(d+1)$ depending on the
spatial dimensionality $d$. In the presence of a long-range Coulomb
interaction, the Coulomb gap in the tunneling density of states
modifies\cite{shklovskii84} the VRH exponent, $\mu={1\over 2}$.

But what if the coupling to phonons is negligibly weak? It was proposed in
Ref.~\onlinecite{fleishman78} that the e-e interaction by itself is sufficient
to induce VRH at low $T$. Although the idea\cite{fleishman78} of a phononless
VRH was widely exploited in the community, it is, however, in conflict with
the argument by Fleishman and Anderson\cite{fleishman80} that elementary hops
in the low-$T$ limit are forbidden for a short-range interaction (including
the case of $1/r$ Coulomb interaction in $d<3$), because energy conservation
cannot be respected when an electron attempts a real transition by exciting an
electron-hole pair. Since there is a mismatch in the positions of the energy
levels of individual single-particle localized states, inelastic scattering is
necessary to broaden these levels to allow for hopping between them. It was
argued in Ref.~\onlinecite{fleishman80} that at low $T$ a short-range e-e
interaction alone is not sufficient to broaden the localized states and to
induce a nonzero conductivity. The situation is particularly interesting in
1d and 2d, where no mobility edge exists, activation to which otherwise might
give $\sigma (T)\neq 0$.

Very recently, the problem was reconsidered in
Refs.~\onlinecite{gornyi05b,basko06}. It was found there that the Anderson
localization in Fock space crucially affects the conductivity at low $T$. As a
result, the phononless VRH should not exist; instead, the system undergoes a
transition at certain temperature $T_c$ below which the conductivity vanishes,
in agreement with Ref.~\onlinecite{fleishman80}.

The mechanism of low-$T$ transport in low-dimensional disordered systems is
thus completely determined by inelastic processes which govern the lifetime of
localized states. This lifetime may be viewed as a low-temperature
reincarnation of the dephasing time that controls the cutoff of
quantum-interference effects at higher $T$. Therefore, an analysis of the WL
effects turns out to be important also for understanding of the establishment
of the strong localization with lowering $T$.

\subsection{Strongly correlated 1d electron systems}
\label{IIb}

This paper is concerned with transport in a single-channel quantum wire, where
both disorder and interaction more strongly modify the transport properties as
compared to higher dimensions. In the absence of disorder, the e-e
correlations drive the system into the non-Fermi liquid state known as
Luttinger liquid. Yet another peculiarity of the single-channel 1d system is
that the ballistic motion on short scales crosses over in the absence of
interaction directly to the localization regime, with no diffusive dynamics on
intermediate scales. We are now going to expand on these known features of 1d
electronic systems.

\subsubsection{Noninteracting case: Anderson Localization}
\label{IIb1}

We first briefly recall the peculiarities of 1d noninteracting disordered
systems. In the absence of interaction, transport in a disordered
single-channel 1d gas has been studied in great
detail.\cite{gertsenshtein59,mott61,berezinskii73,gogolin75,thouless77,
abrikosov78,berezinskii79,gogolin82,lifshitz88} It is known that arbitrarily
weak disorder leads to the localization of all electronic states in this
system, with the localization length $\xi$ being the mean free path. The dc
conductivity is then zero, $\sigma(T)\equiv 0$ for any $T$, since the
temperature merely determines the distribution function over the localized
states.

The characteristic spatial and temporal scales of the problem are probed in
the absence of interactions by the ac response at external frequency
$\Omega$. The appropriate technique for the calculation of the response, which
exploits the separation of fast and slow degrees of freedom in the case of
weak disorder, $\epsilon_F\tau\gg 1$, was put forward in a pioneering paper by
Berezinskii;\cite{berezinskii73} later, alternative approaches were developed
by Abrikosov and Ryzhkin\cite{abrikosov78} and by Berezinskii and
Gor'kov.\cite{berezinskii79} In the limit of low frequency, $\Omega\tau\ll 1$,
the leading contribution to the ac conductivity is given by the
Berezinskii-Mott formula:\cite{berezinskii73,gogolin75} 
\be 
\label{6} 
{\rm Re}\,\sigma(\Omega)/\sigma_{\rm D} =
4(\Omega\tau)^2\ln^2(\Omega\tau)~.  
\ee 
In the opposite (ballistic) limit of high frequency, $\Omega\tau\gg 1$, the
conductivity $\sigma(\Omega)$ is approximately equal to the Drude
formula, $\sigma_{\rm D}(\Omega) = \sigma_{\rm D}/(1-i\Omega\tau)$, with a
quantum interference (WL) correction arising at fourth order in the expansion
in $(\Omega\tau)^{-1}\ll 1$. At $\Omega\tau\sim 1$, the ballistic regime
crosses over directly to the strongly localized regime, Eq.~(\ref{6}), so that
the diffusive regime is absent in the single-channel quantum wire without e-e
interaction.

\subsubsection{Clean quantum wires with interaction: Luttinger liquid}
\label{IIb2}
 
The interest in the role of e-e interaction in quantum wires is largely
inspired by the fact that in 1d the interaction breaks down the Landau's Fermi
liquid approach. Owing to the particular geometry of the Fermi surface,
systems of dimensionality one are unique in that the e-e correlations change
the noninteracting picture completely and lead to the formation of a strongly
correlated state. A remarkable example of the correlated 1d electron phase is
the Luttinger-liquid model (for a review see, e.g.,
Refs.~\onlinecite{solyom79,voit94,schulz95,gogolin98,schulz00,giamarchi04}).
The Luttinger-liquid correlations show up in a power-law singularity of the
tunneling density of states at low energy $\epsilon$ counted from the Fermi
surface,
\be
\nu(\epsilon)\sim \nu_0 \left(|\epsilon|/\Lambda\right)^\beta~.
\label{7}
\ee
Here $\nu_0$ is the density of states in the absence of interaction, $\Lambda$
is the ultraviolet cutoff (determined by the band width, the Fermi energy, or
the range of interaction). The exponent $\beta$ is determined by the
interaction strength and depends on the system geometry (tunneling into the
bulk of a wire differs from tunneling into the end of a wire). The suppression
of the tunneling density of states in Luttinger liquid is similar to the
interaction-induced zero-bias anomaly in higher dimensions, Eq.~(\ref{3}).

According to a common wisdom, electronic excitations do not exist in Luttinger
liquid and a proper language to describe the strongly correlated electrons is
the language of bosonic excitations. This notion is based on the fact that the
bosonic density-fields in a clean Luttinger liquid do not decay in the
linearized-spectrum approximation, when a finite curvature of the electron
spectrum is neglected.

\subsubsection{Renormalization of an impurity in Luttinger liquid}
\label{IIb3}

Interactions between oppositely moving electrons generate charge- and
spin-density wave correlations that lead to striking transport properties of
Luttinger liquid in the presence of impurities. The density-wave correlations
yield a strong renormalization of an
impurity:\cite{gorkov73,mattis74,luther74,kane92,furusaki93,matveev93} if the
bare strength of the impurity is small, the renormalized backscattering
amplitude $U(2k_F,\epsilon)$ seen by an electron with energy $\epsilon$ scales
as
\be 
U(2k_F,\epsilon)\propto \left(\Lambda/|\epsilon|\right)^a~,
\label{8}
\ee 
where $a$ is determined by the interaction strength. As a result, a single
impurity in the infinite Luttinger liquid with a repulsive interaction
effectively decouples the wire into two independent pieces, thus leading to
a vanishing linear conductance at zero temperature. This conclusion remains
valid no matter how weak the interaction is.

\subsubsection{Many impurities: Disordered Luttinger liquid}
\label{IIb4}

The conductivity of a disordered Luttinger liquid was discussed for the first
time more than 30 years ago in Refs.~\onlinecite{gorkov73,mattis74,luther74}
in terms of the renormalization of individual impurities (see also
Refs.~\onlinecite{chui77,suzumura83} for early attempts to construct the phase
diagram of a disordered Luttinger liquid). In the beginning of the 80's, Apel
and Rice\cite{apel82a} combined the scaling arguments developed for
noninteracting systems\cite{thouless77,abrahams79,anderson80} with the results
of the theory of a clean Luttinger liquid. Later, the problem of transport in
a disordered Luttinger liquid was studied by perturbative in disorder methods
based on bosonic field theories in
Refs.~\onlinecite{giamarchi88,furusaki93a,maslov95}. A small correction to the
quantized conductance of a short quantum wire due to a Gaussian disorder was
derived in Ref.~\onlinecite{maslov95} (for a multichannel wire see
Ref.~\onlinecite{sandler97}). A detailed exposition of these issues can be
found in the recent lecture course Ref.~\onlinecite{maslov05}.

The RG equations for a disordered Luttinger liquid were derived by Giamarchi
and Schulz.\cite{giamarchi88} Their approach bears a similarity with the
Finkel'stein RG\cite{finkelstein83} developed earlier for higher-dimensional
diffusive systems (both RG schemes treat disorder perturbatively and e-e
interaction exactly). The flow equations of Ref.~\onlinecite{giamarchi88}
describe, in addition to the renormalization of impurities by interaction
(Sec.~\ref{IIb3}), also a renormalization of the interaction by disorder.  A
sufficiently strong attractive interaction was shown to yield a
metal-insulator transition\cite{giamarchi88} at zero $T$ with changing
interaction strength.  A repulsive interaction was predicted to enhance the
disorder-induced localization. In the latter case, the RG equations show that
at a certain temperature (such that $L_T\sim \xi$) the renormalized disorder
becomes strong. This means that the disordered interacting 1d system enters a
``strong-coupling regime", where the Drude-like approach breaks down. At
higher temperatures, these equations describe the temperature dependence of an
effective scattering time renormalized by interaction (which can be viewed as
screening of impurities by Friedel oscillations), thus yielding a
temperature-dependent Drude conductivity,
\be
\sigma_{\rm D}(T)\propto \left(T/\Lambda\right)^{2a}.
\label{9}
\ee 
Recently, a crossover between Eqs.~(\ref{9}) and (\ref{2}) with
increasing number of channels was studied in Ref.~\onlinecite{altland06}. 

It is important to emphasize that the RG approach formalizes the
renormalization of the Drude conductivity due to the $T$-dependent screening
but does not capture the essential physics of the Anderson localization. In
particular, the RG equations of Ref.~\onlinecite{giamarchi88} miss the
interference effects (coherent scattering on several impurities) that lead to
the WL correction to the conductivity. As we demonstrate below, it is the WL
correction that governs the temperature dependence of the conductivity at
moderately high temperatures for the case of weak interaction and eventually
drives the quantum wire into the strong-localization regime with decreasing
temperature. Another related deficiency of the RG scheme of
Ref.~\onlinecite{giamarchi88} is the treatment of the scalings with length and
$u/T$ as interchangeable.  While this approach is justified for the ``elastic
renormalization" of disorder (i.e., the screening effects), it does not
properly account for the dephasing (since the dephasing length $L_\phi$ is in
general different from $L_T$, see Sec.~\ref{IIIb1} and Sec.~\ref{IVc}).

\subsubsection{Dephasing and localization in disordered Luttinger liquid}
\label{IIb5}

As already mentioned in Sec.~\ref{IIb4}, a key ingredient of transport theory
as regards the WL and the interaction-induced dephasing in a strongly
correlated 1d system is missing. Some papers (e.g.,
Refs.~\onlinecite{giamarchi88,furusaki93}) have suggested that the dephasing
length that controls localization effects in a disordered Luttinger liquid is
$L_T=u/T$, where $u$ is the plasmon velocity. According to this approach, the
interference effects get strong with lowering $T$ at $L_T\sim\xi$. An
alternative approach\cite{apel82a} is predicated on the assumption that the
dephasing rate is determined by the single-particle properties of a clean
Luttinger liquid.  On top of that, one might think that since in the case of
linear dispersion the clean interacting electron system can be equivalently
represented in terms of {\it noninteracting} bosons, the interaction should
not induce any dephasing at all. The conductivity would then be exactly zero
at any $T$. As we demonstrate below, none of the approaches captures the
essential physics of dephasing in the conductivity of a disordered 1d system.

As for the strongly localized regime, recently Ref.~\onlinecite{nattermann03}
used bosonization to study the problem and concluded that transport is of 
variable-range-hopping character. Subsequently, Ref.~\onlinecite{malinin04}
argued that this result requires a nonzero coupling to a thermal bath, in
agreement with Ref.~\onlinecite{fleishman80}, see Sec.~\ref{IIa2}.

\subsection{Summary of known results: 1d vs 2d}
\label{IIc}

We summarize the basic facts discussed in Secs.~\ref{IIa} and \ref{IIb} in
Table~\ref{tab1}: for simplicity, we compare here 2d and single-channel 1d
systems. It turns out that the difference between the disordered Luttinger
liquid and the higher-dimensional (quasi-1d and 2d) systems is not so strong
as one might think judging by the proverbial non-Fermi-liquid character of the
Luttinger liquid.

\begin{table}[tbp]
\caption{The 2d-1d vocabulary. \label{tab1}}
\begin{ruledtabular}
\begin{tabular}{cc}
2d  & 1d  \\[0.3cm]
 \colrule
disorder, no interaction:              & disorder, no interaction: \\
$\sigma(T)=0$                          & $\sigma(T)=0$ 
\\[0.5cm]
interaction, common tool:              & interaction, common tool:\\
fermionic diagrams                     & bosonization
\\[0.5cm]
field theory: $\sigma$-model           & field theory:\\
$Q$-matrices                           & bosonic density fields
\\[0.5cm]
zero-bias anomaly in                   & Luttinger (``non-Fermi")
liquid,      \\ 
tunneling DoS,\ Eq.~(\ref{3})        & power-law DoS,\ Eq.~(\ref{7})
\\[0.5cm]
T-dependent screening                  & renormalization of impurity \\
(Friedel oscillations)                 & (Friedel oscillations),
Eq.~(\ref{8})     
 \\[0.5cm]
Altshuler-Aronov correction            & renormalization of disorder\\
to the conductivity, Eq.~(\ref{2})    & $\ \longrightarrow\
\sigma_{\rm D}(T)$, Eq.~(\ref{9}) 
\\[0.5cm]
Finkel'stein RG:                       & Giamarchi-Schulz RG:\\
disorder (perturbative)                &disorder (perturbative)\\
\& interaction (exact)                 & \& interaction (exact)
\\[0.5cm]
$L_T\sim \xi:\ $ strong                & $L_T\sim \xi:\ $ strong coupling\\
Altshuler-Aronov corrections           & in Giamarchi-Schulz RG \\
$\neq$ Anderson localization           & $\neq$ Anderson localization
\\[0.5cm]  
$L_\phi\sim \xi, \quad \Delta\sigma_{\rm wl}\sim \sigma_{\rm D}:$
& 
$L_\phi\sim \xi, \quad \Delta\sigma_{\rm wl}\sim \sigma_{\rm D}:$\\
strong Anderson localization           & strong Anderson localization 
\\[0.5cm]
weak localization correction           & this work,  \\
to the conductivity,\ Eq.~(\ref{1})   & Secs.~\ref{VI},\ref{VII}
\\[0.5cm]
dephasing time,\ Eq.~(\ref{4})        & this work, Secs.~\ref{VI},\ref{VII}

\end{tabular}
\end{ruledtabular}
\end{table}

\section{Disordered Luttinger liquid}
\label{III}
\setcounter{equation}{0}

\subsection{Formulation of the problem}
\label{IIIa}

Let us now introduce the model we are going to study. We consider a
single-channel infinite wire at sufficiently low temperatures, much lower than
the bandwidth and the Fermi energy. The effective low-energy theory can then
be described by the Luttinger
model,\cite{solyom79,voit94,schulz95,gogolin98,schulz00,giamarchi04} with a
linear dispersion of the electronic spectrum $\epsilon_k=kv_F$, where $v_F$ is
the Fermi velocity. We use the notion of right- and left-movers corresponding
to two Fermi points $k\sim \pm k_F$. The fermionic operators $\psi_\mu$ are
then labeled by $\mu=\pm$ for the right-/left-movers.

We consider a finite-range (screened) pairwise e-e interaction potential
$V(x-x')$. We need only its Fourier transforms at $q=0$ and $q=\pm 2k_F$:
$V_{\rm f}$ and $V_{\rm b},$ respectively, which give the amplitudes of
forward and backward scattering. In terms of the
$g$-ology,\cite{solyom79,voit94,giamarchi04} the interaction in a homogeneous
system is characterized by three coupling constants $g_1=V_{\rm b}$ (e-e
backscattering), $g_2=V_{\rm f}$ (forward scattering between right- and
left-movers), and $g_4=V_{\rm f}$ (forward scattering of right-/left-movers on
right-/left-movers). In several places below, it will be convenient to treat
$g_2$ and $g_4$ as independent parameters.

Next, we introduce weak ($\epsilon_F\tau_0\gg 1$) white-noise disorder,
described by the correlation function
\be
\langle U(x)U(x') \rangle=\delta(x-x')/2\pi\nu_0\tau_0~,
\label{10}
\ee
where $\tau_0$ is the elastic scattering time and $\nu_0=1/\pi v_F$ is the
density of states per spin in the absence of interaction. 

The Hamiltonian for spinful electrons takes the form:
\begin{eqnarray}
H&=&
H_{\rm kin}+H_{\rm ee}+H_{\rm dis}~,\label{11}
\\[0.3cm]
H_{\rm kin}&=&
\sum_{k,\mu=\pm,\sigma}v_F(\mu k - k_F)\psi^\dagger_{\mu\sigma}(k) 
\psi_{\mu\sigma}(k)~,
\label{12}
\\
H_{\rm ee}&=&{1\over 2}\sum_{\mu,\sigma,\sigma'}\int dx
\left\{n_{\mu,\sigma} \, g_2 \, n_{-\mu,\sigma'}
+
n_{\mu,\sigma}\, g_4 \, n_{\mu,\sigma'} \right.
\nonumber \\
&+&\left. \psi^\dagger_{\mu,\sigma} \psi_{-\mu,\sigma} \, g_1 \,
\psi^\dagger_{-\mu,\sigma'} \psi_{\mu,\sigma'}
\right\}~,
\label{13} 
\\
H_{\rm dis}&=&\sum_{\sigma}\ \int dx
\left\{ {\cal U}_b^* \ \psi^\dagger_{+,\sigma} \psi_{-,\sigma} +
 {\cal U}_b\ \psi^\dagger_{-,\sigma}\psi_{+,\sigma}\right\}
\nonumber 
\\
 &+& \sum_{\sigma}\ \int dx
\left\{ {\cal U}_f^* \ \psi^\dagger_{+,\sigma} \psi_{+,\sigma} +
 {\cal U}_f\ \psi^\dagger_{-,\sigma}\psi_{-,\sigma}\right\}~. \nonumber \\
\label{14}
\end{eqnarray}
Here
\be
n_{\mu,\sigma}(x)= \psi^\dagger_{\mu,\sigma}(x) \psi_{\mu,\sigma}(x)
\label{15}
\ee
is the fermionic density, ${\cal U}_b$ and ${\cal U}_f$ are backscattering and
forward-scattering amplitudes due to disorder. It can be
shown\cite{giamarchi88,giamarchi04} that the forward-scattering amplitude can
be gauged out in the calculation of the conductivity. Therefore, from now on
we set ${\cal U}_f\equiv 0$. The correlation functions for ${\cal U}_b(x)$ 
read
\bea
\langle{\cal U}_b(x){\cal U}_b(x')\rangle&=& 0~,
\label{16}
\\
\langle{\cal U}_b(x){\cal U}^*_b(x')\rangle&=&\langle U(x)U(x') \rangle~.
\label{17}
\eea

Two points are worth mentioning here. First, all the parameters of the above
Hamiltonian include, in general, high-energy renormalization effects (similar
to the Fermi-liquid renormalization in higher dimensionalities) and should be
considered as effective couplings of the low-energy theory. Second, the
low-energy theory is only then well-defined when supplemented by an
ultraviolet energy cutoff $\Lambda$. The latter is determined by the Fermi
energy, the bandwidth, or the spatial range of interaction in the original
microscopic theory, whichever gives the smallest cutoff.

Let us now concentrate on the spinless model. In this case, two essential
simplifications occur. First, the interaction-induced backward scattering and
forward scattering relate to each other as direct and exchange processes.
Therefore, the backscattering $V_{\rm b}$ only appears in the combination
$V_{\rm f}-V_{\rm b}$ and thus merely redefines the parameters of the Luttinger
model, formulated\cite{solyom79,voit94} in terms of forward-scattering
amplitudes only. In other words, the term with $g_1$ in Eq.~(\ref{13}) has
the same structure as the $g_2$-term, so that we can absorb $g_1$ into $g_2$,
$${\tilde g}_2\equiv V_{\rm f}-V_{\rm b}=g_2-g_1,$$ and start our
consideration with the following form of the Hamiltonian (\ref{11}):
\begin{eqnarray}
H_{\rm kin}&=&
\sum_{k,\mu=\pm}v_F(\mu k - k_F)\psi^\dagger_{\mu}(k) \psi_{\mu}(k)~,
\label{18}
\\
H_{\rm ee}&=&{1\over 2}\sum_{\mu=\pm}\int dx
\left\{n_{\mu} \, {\tilde g}_2 \,
n_{-\mu}
+ n_{\mu} \, g_4 \, n_{\mu}
\right\}~,
\label{19} 
\\
H_{\rm dis}&=& \int dx
\left\{ {\cal U}^*_b(x) \  \psi^\dagger_{+} \psi_{-} + 
{\cal U}_b(x)\ \psi^\dagger_{-}\psi_{+}\right\}~.
\label{20}
\end{eqnarray}
In what follows we omit the tilde over the shifted coupling $g_2$. The second
simplification is related to the $g_4$-term in Eq.~(\ref{19}).  Naively, one
could think it is simply zero, since it describes a local interaction between
identical fermions. In fact, the situation is slightly more tricky. While
yielding no genuine interaction, this term generates a shift of the velocity
due to a quantum anomaly, $v_F\to v^*_F$, where
\be
v^*_F=v_F + {g_4\over 2\pi}~.
\label{21}
\ee
This is also straightforwardly seen in the bosonization approach, see
Sec.~\ref{IV}.

Therefore, the complete set of parameters defining the spinless problem
includes $v_F^*$, $g_2$, $\tau_0$, and $\Lambda$, with the interaction
characterized by a single coupling $g_2$. It is customary to define the
corresponding dimensionless parameter in the form
\be
K=\left( {1- g_2/2\pi v^*_F \over 1 + g_2/2\pi v^*_F} \right)^{1/2}~.
\label{22}
\ee
Note that in a 1d system with $g_2=g_4$ Eq.~(\ref{22}) is identical to another
conventional representation of $K$ in terms of the bare velocity $v_F$,
\be
K={1\over (1+g_2/\pi v_F)^{1/2}}~.
\label{23}
\ee
For the case of a weak interaction (considered in the present paper), $K
\simeq 1- \alpha$, where
\be
\alpha = {g_2\over 2\pi v_F^*} \simeq {g_2\over 2\pi v_F} \ll 1
\label{24}
\ee
is the dimensionless strength of the interaction.

\subsection{Qualitative discussion: Luttinger-liquid renormalization vs 
Fermi-liquid dephasing}
\label{IIIb}

Having formulated the model in Sec.~\ref{IIIa}, we turn now to a discussion
of our strategy in its analysis. In principle, one could proceed in a number
of alternative ways:

(i) First of all, one can bosonize the action and try to study the transport
properties of interacting electrons in the bosonic language,
Sec.~\ref{IV}. While this approach is very powerful for the problem of 1d
interacting fermions, it turns out to be poorly suited for the analysis of
transport problems in the presence of disorder. For this reason, we develop an
alternative ``two-step procedure".

(ii) The two-step procedure combines the language of bosonization for high
energies with a fermionic treatment of processes with low-energy transfers.
This method is discussed on the qualitative level in the remainder of this
subsection and is implemented in Secs.~\ref{IV},\ref{VI} below.

(iii) Finally, one more alternative is the ``functional bosonization"
approach.\cite{fogedby76,lee88,kopietz97,yurkevich02,lerner05} This method
also combines the languages of fermionic (electrons) and bosonic (plasmons)
excitations, but in contrast to the two-step procedure this is done in a
uniform fashion, so that the processes with high and low energy transfers are
treated on the same footing. This method yields results identical to the
two-step approach and is presented in Sec.~\ref{VII}.

Let us describe now the ideas underlying our two-step approach. The method is
based on the separation of two types of processes: (i) ``elastic
renormalization'' of disorder, which is associated with virtual transitions
with energy transfers $|\omega|\agt T$ and (ii) inelastic processes with
$|\omega|\alt T$ which lead to dephasing in quantum-interference effects and
thus to a destruction of the Anderson localization. This separation of
energy scales is very much similar to that in physics of mesoscopic phenomena
in higher-dimensional case, see Sec.~\ref{IIa}.

\subsubsection{Step 1: Luttinger-liquid renormalization}
\label{IIIb1}

We begin by considering the Drude conductivity under the condition that the
dephasing time $\tau_\phi$ is much shorter than the transport time of elastic
scattering off disorder $\tau$. The source of the strong dephasing may be
external (say, interaction with phonons, or Coulomb interaction with
``environment'') or intrinsic, that is the inelastic e-e interaction within
the quantum wire. In the latter case, the limit $\tau_\phi\ll\tau$ is achieved
at sufficiently high $T$, as shown below in Sec.~\ref{VI}. For
$\tau_\phi\ll\tau$, the Anderson-localization effects are strongly suppressed
by dephasing.

To leading order in $\tau_\phi/\tau\ll 1$, the conductivity is given by the
Drude formula
\be 
\sigma_{\rm D}=e^2\rho v_F^2\tau~,
\label{25}
\ee
where $\rho$ is the compressibility. The Drude conductivity
(\ref{25}) depends on $T$ through a $T$-dependent renormalization
of the static disorder:\cite{mattis74,luther74,giamarchi88,kane92,matveev93}
\begin{equation}
\tau(T)=\tau_0(T/\Lambda)^{2a}~.
\label{26} 
\end{equation} 
For spinless electrons, the exponent in Eq.~(\ref{26}) reads
\be
a = 1- K \simeq \alpha~.
\label{27}
\ee
In principle, the interplay of disorder and interaction leads to the
renormalization of both disorder, Eq.~(\ref{26}), and
interaction.\cite{giamarchi88} However, the value of $\alpha$ in
Eq.~(\ref{27}) is given by the bare interaction constant (the one in a clean
system), see Sec.~\ref{IV} for details.

The Luttinger-liquid renormalization (\ref{26}) is similar to the
Altshuler-Aronov corrections\cite{altshuler85} [Eq.~(\ref{2})] in higher
dimensionalities. The underlying physics of the elastic renormalization of
disorder can be described in terms of the $T$-dependent screening of
individual impurities and scattering by slowly decaying in real space Friedel
oscillations, similarly to higher-dimensional systems.\cite{rudin97,zala01} At
this level, the only peculiarity of Luttinger liquid as compared to higher
dimensionalities is that the renormalization of $\tau$ is more singular and
necessitates going beyond the Hartree-Fock (HF) approach even for the weak
interaction.\cite{polyakov03}
 
Technically, Eq.~(\ref{26}) can be derived by a variety of methods: either
fermionic\cite{solyom79,matveev93} or bosonic.\cite{giamarchi88,kane92} In
particular, a conceptually important framework is based on the bosonic RG
procedure developed by Giamarchi and Schulz.\cite{giamarchi88} This approach
allows one to integrate out all energy scales between $\Lambda$ and $T$, which
are responsible for the elastic Luttinger-liquid renormalization due to
virtual e-e scattering processes. We will discuss the features of the bosonic
RG approach\cite{giamarchi88} relevant to our analysis in more detail in
Sec.~\ref{IV}.

The renormalization of $\tau$ stops with decreasing $T$ at $T\tau(T)\sim 1$.
Physically, this is due to the fact that the long-range Friedel oscillations
are cut off even at zero $T$ on the spatial scale of the order of the
disorder-induced mean free path. The condition 
\be 
T\tau(T)\sim 1 \label{28}
\ee 
gives the zero-$T$ localization length 
\be 
\xi\propto\tau_0^{1/(1+2a)}~.
\label{29} 
\ee 
It is important to stress, however, that Eq.~(\ref{28}) does not correctly
predict the onset of localization, in contrast to the argument made in
Refs.~\onlinecite{giamarchi88,furusaki93}. This can be seen, in particular, by
noting that the temperature $T\sim \tau^{-1}$ does not depend on the strength
of interaction for small $\alpha$, whereas it is evident that for
noninteracting electrons there is no interaction-induced dephasing (the
dephasing length is infinite, $L_\phi=\infty$) and hence $\sigma(T)=0$ for any
$T$ due to the Anderson localization, see Sec.~\ref{IIb1}. The onset of strong
localization occurs at $L_\phi\sim\xi$, as in the case of higher
dimensionality, see the discussion in Sec.~\ref{IIa}. The temperature $T_1$
below which the Anderson-localization effects become strong is determined in
1d by the condition
\be
\tau(T_1)\sim \tau_\phi(T_1)~.
\label{30}
\ee
At this temperature, the quantum-interference effects leading to the WL
correction $\Delta\sigma_{\rm wl}$ to the conductivity are no longer weak,
\be
|\Delta\sigma_{\rm wl}(T_1)|\sim \sigma_{\rm D}(T_1)~.
\label{31}
\ee

\subsubsection{Step 2: Fermi-liquid dephasing}
\label{IIIb2}

It is worth stressing that the Luttinger-liquid physics in the present problem
is fully accounted for by Eq.~(\ref{26}), i.e., all the Luttinger-liquid
power-law singularities are now incorporated in the renormalization of
$\tau(T)$.  At this point, we are left with the system of interacting
electrons with a new effective ``Fermi energy'' $T$. These electrons are
scattered by disorder with renormalized $\tau_0\to \tau(T),$ which is still
weak for $T\tau(T)\gg 1$.

What remains after the renormalization is the e-e scattering with energy
transfers $|\omega| \alt T$. The interaction is thus still important and leads
to real inelastic processes. The physics of these processes is in essence
equivalent to the Fermi-liquid physics of dephasing in higher
dimensionalities. Having integrated out the virtual transition, we have to
choose a controllable method of dealing with the dephasing. The bosonization
approach (whose advantage is that it can be straightforwardly formulated for
the arbitrary strength of interaction) is not particularly beneficial
here. Although, in principle, it contains all effects of the Anderson
localization in 1d, the machinery of bosonization is poorly suited to deal
with such ingredients of the localization theory as the interference and
dephasing. Even for noninteracting electrons, it is by far not straightforward
to derive the Anderson localization [for instance, to obtain the
Berezinskii-Mott formula (\ref{6}) for the case of weak disorder] in the
language of bosonization. For this reason, we refermionize the theory at this
step, Secs.~\ref{V},\ref{VI}.

\section{Bosonization approach}
\label{IV}
\setcounter{equation}{0}

\subsection{Bosonized action}
\label{IVa}

As discussed in Sec.~\ref{IIIb}, the first part of our program, namely the
renormalization of the model due to high-energy processes, is done most
efficiently in the framework of the bosonization technique. In this section,
we present a calculation of $\tau(T)$, using the RG scheme developed by
Giamarchi and Schulz.\cite{giamarchi88} After a brief description of the
derivation of the RG equations, we will discuss an important subtlety of the
scheme,\cite{giamarchi88} namely a mixing of disorder and interaction.

For simplicity, we consider here the spinless case (in notation of
Ref.~\onlinecite{giamarchi88}, we set $g_{1\perp}=g_{2\perp}=0$ and hence
$u_\rho=u_\sigma,\ K_\rho=K_\sigma$). As discussed in Sec.~\ref{IIIa}, for
spinless electrons the backscattering amplitude $g_1$ can be absorbed into
$g_2$, and $g_4$ into $v_F^*$, so that only one interaction coupling constant
remains ($g_2$), yielding the dimensionless Luttinger-liquid parameter $K$,
Eq.~(\ref{22}).

Using the boson representation of the fermionic operators,\cite{giamarchi04}
the Hamiltonian can be expressed in terms of the boson operators $\phi$ and
$\Pi$ that satisfy the commutation relation
$[\Pi(x),\phi(x')]=-i\delta(x-x').$ In a given realization of disorder the
Hamiltonian reads
\bea
\label{32}
H&=&{1\over 2\pi}\int dx 
\left\{ u K (\pi\Pi)^2 +{u\over K}(\partial_x\phi)^2
\right\}\nonumber \\
&+& {1\over 2\pi \lambda } \int dx 
\left\{ {\cal U}_b^*(x)\exp(2i\phi)+ {\rm H.c.} \right\} 
\eea
with 
\be
\label{33}
u=\left[(v^*_F)^2-g_2^2/4\pi^2\right]^{1/2}~.
\ee
The ultraviolet length scale $\lambda$ is related to the energy cutoff
$\Lambda$ as
\be
\lambda=u/\pi \Lambda~.
\label{34}
\ee
We will concentrate on the case of e-e repulsion, implying that $K<1$. The
disorder-induced forward scattering amplitude ${\cal U}_f(x)$ is
absorbed\cite{giamarchi88} in the quadratic part of $H$ as a shift of
$\phi(x)$.
 
In order to perform the disorder averaging, one introduces replicas, $\phi_n$.
In the resulting action the interaction (Luttinger-liquid) term $S_{LL}$ is
quadratic in fields $\phi_n,$ while disorder generates a term $S_D$ of
$\cos2\phi$-type which breaks the Gaussian character of the theory:
\begin{eqnarray}
S&=&\sum_n S_{LL}[\phi_n]+\sum_{n,m}S_D[\phi_n,\phi_m]~,
\label{35}\\
S_{LL}&=&{1\over u K} \int {dx d\tau\over 2\pi} 
\left\{  [\partial_\tau \phi_n(x,\tau)]^2 +
u^2[\partial_x \phi_n(x,\tau)]^2 \right\}~,
\nonumber 
\\
\label{36} \\
S_D&=&-D_b\int\!{dx \,d\tau\, d\tau' \over (2\pi \lambda)^2}
\cos\{2[\phi_n(x,\tau)-\phi_m(x,\tau')]\}~,
\nonumber
\\
\label{37}
\end{eqnarray}
where $D_b=v^*_F/2\tau_0$. The integration over the imaginary times is
performed within the interval $0\leq \tau \leq 1/T.$

The conductivity is expressed in terms of the current-current correlation
function at zero momentum via the Kubo formula:
\be
\sigma(\Omega)= -{e^2\over i\Omega}
\left[{v_F\over\pi}-\int_0^{1/T} \!\!\!d\tau \,e^{i\Omega_n\tau} 
{\cal K}(\tau) \right]_{i\Omega_n\to \Omega+i 0}
\label{38}
\ee
with
\bea
{\cal K}(\tau)=
\int dx \langle T_\tau j(x,\tau) j(0,0) \rangle~,
\label{39}
\eea
where in the bosonic language the currents are expressed via the field
$\phi(x,\tau)$ as
\be
j(x,\tau) = {i\over \pi }\partial_\tau \phi(x,\tau)~.
\label{40}
\ee
The first (diamagnetic) term in Eq.~(\ref{38}) is canceled by a contribution
of the singularity at $\tau=0$ in the second term. It is worth noting that,
strictly speaking, the correlation function $\langle T_\tau \partial_\tau
\phi(x,\tau) \partial_\tau \phi(0,0) \rangle$ is a {\it reducible} (with
respect to the interaction) correlator, when it is calculated with the use of
$S_{LL}$ as given by Eq.~(\ref{36}). On the contrary, the conductivity is
given by the {\it irreducible} current-current correlator. However, at zero
external momentum $Q$, the two correlators are equal to each other, since the
polarization operator vanishes at $Q=0$ for arbitrary nonzero frequency. This
should be contrasted with a calculation of the conductance of a finite
interacting wire: the latter is related to the response at finite $Q$ and the
Kubo formula should be supplemented with an analysis of the effect of
renormalization of the electric field by e-e
interaction.\cite{kawabata96,oreg96}

Note that the authors of Ref.~\onlinecite{giamarchi88} did not calculate the
conductivity using the Kubo formula (\ref{38}), (\ref{39}),
(\ref{40}). Instead, they found the renormalized value of $1/\tau$ and
substituted it into the Drude formula, Eq.~(\ref{25}) (which can be
represented as a certain approximation in the framework of the memory-function
formalism, see, e.g., Refs.~\onlinecite{luther74,giamarchi91}). In view of
technical difficulties, in this paper we do not attempt to evaluate the
conductivity, including the localization effects, by using the Kubo formula in
the bosonic language either. Having used the bosonization for the analysis of
the high-energy renormalizations, we refermionize the theory and calculate the
conductivity in the fermionic language, Sec.~\ref{VI}.

\subsection{Giamarchi-Schulz RG}
\label{IVb}

Giamarchi and Schulz\cite{giamarchi88} performed the renormalization of the
coupling constants $K,u,$ and $D_b$ in the action (\ref{35}) upon
rescaling the ultraviolet cutoff $\lambda \to L=\lambda\exp(\ell).$ Their
analysis is done to first order in the dimensionless strength of disorder
\be
{\cal D}=2 D_b \lambda/\pi u^2~.
\label{41}
\ee 
To this order, one can omit replica indices in Eq.~(\ref{35}).  Before
presenting the RG equations, we discuss a subtle point of the approach
developed in Ref.~\onlinecite{giamarchi88}. Specifically, the coupling
constant $\tilde{K}$ in terms of which the RG equations are obtained in a
natural way, is not simply $K$ but rather contains an admixture of
disorder. We now briefly reproduce the corresponding argument from
Ref.~\onlinecite{giamarchi88}. In the fermionic representation, the part of
the action related to the disorder-induced backscattering acquires, after the
disorder averaging, the form
\bea
S_D&=&-D_b \int dx\, d\tau\, d\tau^\prime \ \psi^\dagger_{+}(x,\tau) 
\psi_{-} (x,\tau) \nonumber \\
&\times& \psi^\dagger_{-}(x,\tau^\prime) \psi_{+}(x,\tau^\prime)~.
\label{42}
\eea
In order to derive the RG equations, Giamarchi and Schulz singled out the
ultraviolet contribution of close times $u|\tau-\tau^\prime|\leq \lambda,$
splitting the time integration into two parts:
\bea
S_D&=&-D_b \int_{u|\tau-\tau^\prime|>\lambda } dx\, d\tau\, d\tau^\prime  
\nonumber \\
&\times&
\psi^\dagger_{+}(x,\tau) \psi_{-} (x,\tau) 
 \psi^\dagger_{-}(x,\tau^\prime) \psi_{+}(x,\tau^\prime)
\label{43}
\\
&-&{2 D_b \lambda \over u} \int dx\, d\tau \nonumber \\
&\times&\psi^\dagger_{+}(x,\tau) 
\psi_{-} (x,\tau) \psi^\dagger_{-}(x,\tau) \psi_{+}(x,\tau)~.\nonumber
\\
\label{44}
\eea
The second term in Eq.~(\ref{44}) is local in time and thus equivalent to an
interaction term generated by $g_1$-processes. For this reason, the authors of
Ref.~\onlinecite{giamarchi88} redefined the interaction constant,
$\tilde{g_1}\equiv g_1-2 D_b \lambda/u$. For the case of spinless electrons
considered here, this is equivalent to a modification of the constant $g_2$:
\be
\tilde{g_2} = g_2+2 D_b \lambda/ u~.
\label{45}
\ee

In terms of the modified interaction constants the RG
equations\cite{giamarchi88} (in the spinless case only three equations out of
original six remain) are:
\bea
d \tilde{K}/d\ell&=&-
\tilde{K}^2 
{\cal D }/2~,
\label{46}
\\[0.2cm]
d \tilde{u}/d\ell&=&- \tilde{u}   \tilde{K}
{\cal D }/2~,
\label{47}
\\[0.2cm]
d {\cal D }/d\ell&=&(3-2 \tilde{K}) {\cal D }~,
\label{48}
\eea
where $\tilde{u}$ and $\tilde{K}$ are related to the modified
interaction constant
\be
\tilde{g_2}(\ell)= g_2(\ell) + \pi u {\cal D}(\ell)
\label{49}
\ee
according to Eqs.~(\ref{33}) and (\ref{22}), respectively.

The modification of the plasmon velocity $u$ by disorder, as well as its
renormalization, can be neglected in Eq.~(\ref{49}), since this generates
higher powers of ${\cal D}$. Then Eq.~(\ref{47}), which describes the
renormalization of $u$, turns out to have no influence on the renormalization
of disorder and interaction, described by the remaining two equations. [In
fact, Eqs.~(\ref{46}),(\ref{47}) show that $\tilde{K}/\tilde{u}$ is the
invariant of the flow.] Therefore, we will concentrate on the coupled set of
the flow equations (\ref{46}), (\ref{48}). It follows from Eq.~(\ref{48}) that
for $\tilde{K}<3/2$ the strength of disorder ${\cal D}(\ell)$ increases upon
renormalization. At the same time, the value of $\tilde{K}$ decreases. 
The system flows then to the strong-coupling regime where
${\cal D}$ becomes of order one and the first-order treatment of disorder is
no longer valid.

The decrease of $\tilde{K}(\ell)$ upon renormalization
suggests that the initially weak interaction ($K$ close to 1) becomes strong
upon approaching the strong coupling regime. However, this is not so, as we
demonstrate below: the decrease of $\tilde{K}$ occurs solely due to the
admixture of increasing disorder in this quantity.

It is also worth noticing that Eq.~(\ref{46}) seems to generate finite
interaction due to the scattering off disorder in the noninteracting
case. However, as verified in Ref.~\onlinecite{giamarchi88}, this is not the
case. The point is that the condition $\tilde{K}=1$ does not mean that the
system is noninteracting, because of the mixing of interaction and disorder in
Eq.~(\ref{49}). Therefore, it is desirable to rewrite Eqs.~(\ref{46}),
(\ref{48}) in terms of the true interaction constant $K$ instead of the one
modified by the admixture of disorder, $\tilde{K}$ (see Fig.~9.6 in
Ref.~\onlinecite{giamarchi04}).

Let us first consider the case of weak interaction, $1-K\simeq
\alpha\ll 1$, retaining only the terms of first order in $\alpha$ in the RG
equations. From Eqs.~(\ref{22}),(\ref{33}),(\ref{41}), and (\ref{45}) we get,
to first order in $\alpha$,
\be
\tilde{K}=1-\alpha-{\cal D}/2+\alpha {\cal D}/2~.
\label{51}
\ee
Substituting this formula in Eq.~(\ref{48}), we express the flow equation
for the disorder strength in terms of the renormalized true interaction
constant $\alpha(\ell)$:
\be
d {\cal D } / d\ell={\cal D }+
2\alpha{\cal D } + {\cal D}^2~.
\label{52}
\ee
In this equation, the first term in the right-hand side describes Ohm's law,
while the second term ($\propto \alpha$) yields the Luttinger-liquid
renormalization of $\tau$, Eq.~(\ref{26}).

The last (quadratic-in-disorder) term in Eq.~(\ref{52}) is beyond the accuracy
of the derivation of the RG equations and should be neglected. We note,
however, that this spurious term is hidden in the RG equations of
Ref.~\onlinecite{giamarchi88}. Accidentally, this spurious quadratic
term qualitatively mimics the effect typical for the WL. Clearly, this purely
accidental similarity does not hold on the quantitative level: the true
WL---when included into consideration---would only appear in the r.h.s. of
Eq.~(\ref{52}) at ${\cal D}^3$ order.

\begin{figure}[ht]
\centerline{
\includegraphics[width=\columnwidth]{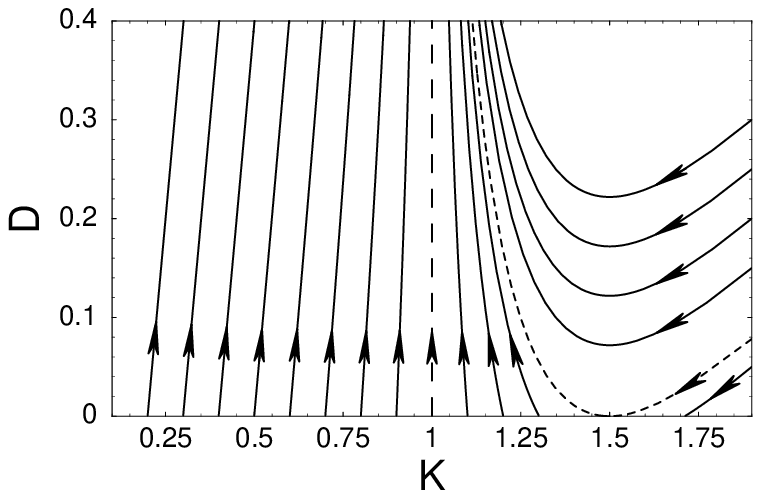}
} 
\centerline{
\includegraphics[width=\columnwidth]{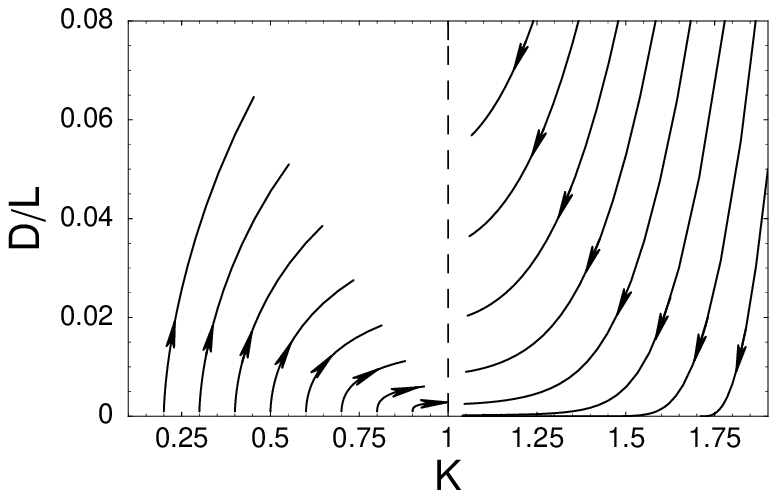}
} 
\caption{
Upper panel: the RG flows given by Eqs.~(\ref{57}),(\ref{58}) in the plane
$(K,{\cal D})$, where $K$ is the {\it true} interaction constant and ${\cal
D}$ is the strength of disorder (the dimensionless resistance). Dashed line:
the separatrix touches the quantum critical point at $K=3/2,\ {\cal
D}=0$. Long-dashed line: the line of $K=1$ separates the repulsion ($K<1$) and
attraction ($K>1$) regions. Lower panel: the RG flows from
Eqs.~(\ref{57}),(\ref{58}) in the coordinates $K$ and ${\cal D}/L$, the latter
corresponds to the resistivity. The renormalization for each curve is stopped
at ${\cal D}=1$ (the strong coupling regime). The dashed line separates the
repulsion and attraction regions: an attractive (repulsive) interaction cannot
become repulsive (attractive) in the course of the renormalization. Initial
couplings in the upper panel from left to right: ${\cal D}_0 = 0.001$, $K_0 =
0.2,\ 0.3,\ 0.4,\ 0.5,\ 0.6,\ 0.7,\ 0.8,\ 0.9,\ 1.0,\ 1.1,\ 1.2,\ 1.3$ and $K_0
= 1.9$, ${\cal D}_0 = 0.3,\ 0.25,\ 0.2,\ 0.15,\ 0.05$.  Separatrix: $K_0 =
1.9$; ${\cal D}_0 \simeq 0.079$.  Initial couplings in the lower panel: $K<1$
(from left to right): ${\cal D}_0 = 0.001$, $K_0 = 0.2,\ 0.3,\ 0.4,\ 0.5,\
0.6,\ 0.7,\ 0.8,\ 0.9$; $K>1$ (from top to bottom): $K_0 = 1.9$, ${\cal D}_0 =
0.4,\ 0.35,\ 0.3,\ 0.25,\ 0.2,\ 0.15, \ 0.10,\ 0.05$.
}
\label{f1} 
\end{figure}

Now let us turn to the flow equation (\ref{46}) for the interaction
strength. We rewrite it using Eqs.~(\ref{51}) and (\ref{52}) in the
form
\be
d\alpha /d\ell=-3\alpha{\cal D }/2- {\cal D}^2~.
\label{53}
\ee
Again, the last (quadratic-in-disorder) term here is beyond the accuracy of
the calculation and has to be neglected (this unphysical term generates
interaction from disorder). It is important that the linear-in-disorder term
in Eq.~(\ref{53}) vanishes for $\alpha=0$. In other words, disorder does
not generate interaction, which was not trivial to see from
Eq.~(\ref{46}).

To summarize, for a weak interaction, $\alpha\ll 1$, the set of two scaling
equations for $\alpha$ and ${\cal D}$ reads 
\bea d\alpha/d\ell&=&-3\alpha{\cal D}/2~, \label{54} \\ d {\cal D} /
d\ell&=&(1+2\alpha){\cal D}~.  
\label{55} 
\eea
\begin{figure}[ht]
\centerline{ \includegraphics[width=\columnwidth]{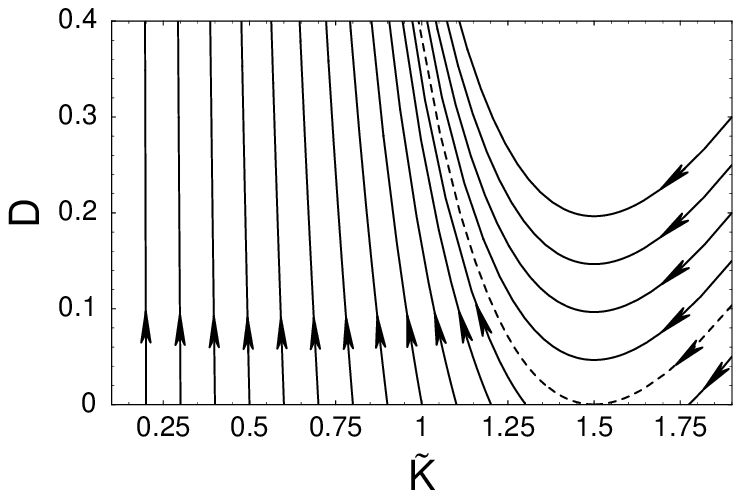} } 
\centerline{
\includegraphics[width=\columnwidth]{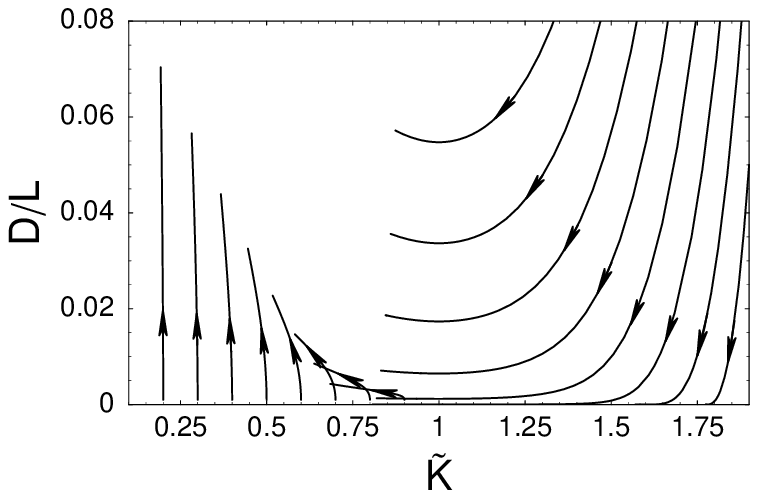} } 
\caption{ 
Upper panel: the RG flows given by Eqs.~(\ref{46}),(\ref{48}) in the plane of
the {\it modified} interaction constant ${\tilde K}$, introduced in
Ref.~\onlinecite{giamarchi88}, and the strength of disorder (the dimensionless
resistance) ${\cal D}$. Dashed line: the separatrix touches the quantum
critical point at ${\tilde K}=3/2,\ {\cal D}=0$. Lower panel: the RG flows
from Eqs.~(\ref{46}),(\ref{48}) in the coordinates ${\tilde K}$ and ${\cal
D}/L$ (corresponding to the resistivity). The renormalization for each curve
is stopped at ${\cal D}=1$ (the strong coupling regime).  Initial couplings in
the upper panel from left to right: ${\cal D}_0 = 0.001$, ${\tilde K}_0 =
0.2,\ 0.3,\ 0.4,\ 0.5,\ 0.6,\ 0.7,\ 0.8,\ 0.9,\ 1.0,\ 1.1,\ 1.2,\ 1.3$ and
${\tilde K}_0 = 1.9$, ${\cal D}_0 = 0.3, 0.25, 0.2, 0.15, 0.05$. Separatrix:
${\tilde K}_0 = 1.9$; ${\cal D}_0 \simeq 0.104$. Initial couplings in the lower
panel: ${\cal D}_0 = 0.001$, ${\tilde K}_0 = 0.2,\ 0.3,\ 0.4,\ 0.5,\ 0.6,\
0.7,\ 0.8,\ 0.9$ and ${\tilde K}_0 = 1.9$, ${\cal D}_0 = 0.4,\ 0.35,\ 0.3,\
0.25,\ 0.2,\ 0.15, \ 0.10,\ 0.05$.  
} 
\label{f2} 
\end{figure}
One sees that Eq.~(\ref{54}) yields decreasing $\alpha(\ell)$: while the
disorder strength grows with increasing $L$, the interaction gets weaker
(``disorder kills interaction"---Ref.~\onlinecite{giamarchi04}; also
Ref.~\onlinecite{giamarchi87}), in contrast to a naive expectation based on
Eq.~(\ref{46}). This resembles the behavior of the renormalized interaction in
2d diffusive systems. Specifically, according to the Finkel'stein RG
equations,\cite{finkelstein83} the combination of the singlet and triplet
amplitudes (an analog of $\alpha$ for the spinful case) which determines the
interaction-induced correction to the conductivity $\Delta\sigma_{\rm AA}$,
decreases for weak interaction ($\alpha\ll 1$) due to the disorder-induced
renormalization of interaction.

For completeness, let us also consider the RG equations for arbitrary
strength of interaction. Expanding $\tilde{K}$ in ${\cal D}$ and using
Eqs.~(\ref{33}) and (\ref{45}), we obtain
\be
\tilde{K}\simeq K  - (1+K^2){\cal D}/4~. 
\label{56}
\ee
Substituting Eq.~(\ref{56}) in Eqs.~(\ref{46}) and (\ref{48}), we
arrive at
\bea
{d K \over d\ell}&=&-{1\over 2} 
\left[K^2-\frac{(1+K^2)(3-2 K)}{2}\right]{\cal D}~,
\label{57}
\\[0.2cm]
{d {\cal D } \over d\ell}&=&(3-2K) {\cal D }~.
\label{58}
\eea
The solution to Eqs.~(\ref{57}),(\ref{58}) is shown in Fig.~\ref{f1}; for
comparison, we show also the solution to Eqs.~(\ref{46}),(\ref{48}) in
Fig.~\ref{f2}. The flow equation (\ref{57}) for $K$ tells us that the
repulsive interaction, $K<1$, gets weaker in the course of renormalization,
$dK/d\ell>0$, whatever the interaction strength is, see
Fig.~\ref{f1}. Equations (\ref{57}),(\ref{58}) reduce to the set of
Eqs.~(\ref{46}),(\ref{48}) around $K=3/2$, where the Giamarchi-Schulz RG
predicts the metal-insulator transition. Note that for $K>3/2$ the flows in
terms of $K$ and $\tilde{K}$ are qualitatively similar to each other, as is
evident from a comparison of Figs.~\ref{f1} and \ref{f2}. 

\begin{figure}[ht]
\centerline{
\includegraphics[width=\columnwidth]{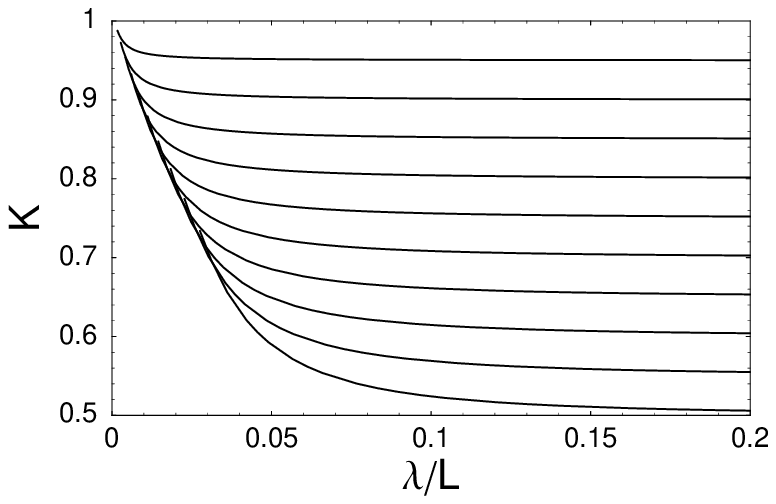}
} 
\centerline{
\includegraphics[width=\columnwidth]{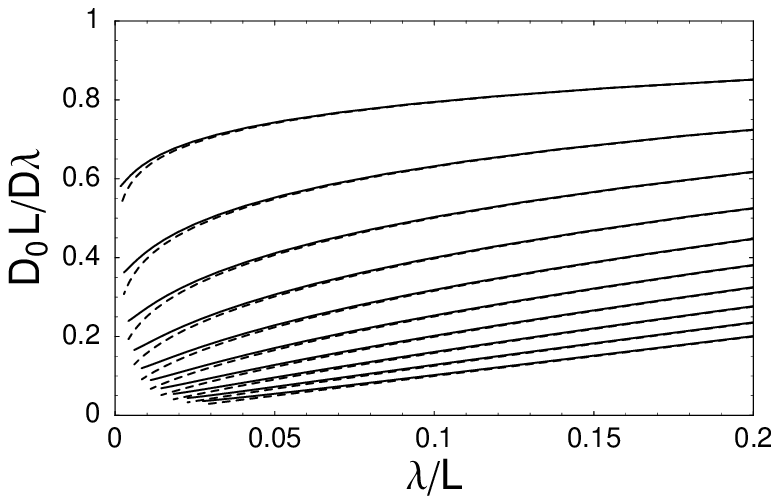}
} 
\caption{
Upper panel: the renormalization of the interaction constant $K$ due to
disorder for various initial values of $K$ in the ultraviolet limit
$\lambda/L=1$. The strength of interaction decreases upon renormalization,
$K\to 1$. Lower panel: the renormalization of the ``conductivity" $L/{\cal D}$
in units of the bare ``conductivity" $\lambda/{\cal D}_0$ (solid line). The
dashed lines describe the dependence of $L/{\cal D}$ given by Eq.~(\ref{59})
with the exponent $\alpha_0=1-K_0$, where $K_0$ is the ultraviolet value of
the Luttinger constant. Each flow is stopped at ${\cal D}=1$. Initial
couplings from top to bottom, both panels: ${\cal D}_0 = 0.001$, $K_0 = 0.95,\
0.9,\ 0.85,\ 0.8,\ 0.75,\ 0.7,\ 0.65,\ 0.6,\ 0.55,\ 0.5$.
}
\label{f3} 
\end{figure}

Let us return to the case of weak interaction and analyze the RG flow
generated by Eqs.~(\ref{54}),(\ref{55}), see Fig.~\ref{f3}. An
approximate solution of these equations in the range of weak disorder, ${\cal
D}(\ell)\ll 1$ (which is where they are valid), is easily found by an
iterative procedure. The result is:
\bea
&& {\cal D}(\ell) \simeq {\cal D}_0\left(L/\lambda\right)^{1+2\alpha_0}~,
\label{59} \\
&& \alpha (\ell)\simeq\alpha_0[\,1-3{\cal D}(\ell)/2\,]~,
\label{60}
\eea
where ${\cal D}_0$ and $\alpha_0$ are the initial (ultraviolet) values of
the coupling constants at $L\sim \lambda.$

It is worth emphasizing that whereas the correction to $\alpha_0$ is of the
order of $\alpha_0$ itself when ${\cal D}(\ell)\sim 1$, the renormalization of
disorder is governed by the initial value of the interaction constant and {\it
not} by the running coupling constant $\alpha(\ell)$. The renormalization of
the interaction constant is irrelevant in the present case, and hence the
renormalization of disorder reduces to the renormalization of an individual
impurity. As a consequence, the exponent $a$ in Eq.~(\ref{26}) for the
renormalized scattering time $\tau(T)$ is given by the bare interaction
constant (the one in a clean system), $a=a_0 \simeq \alpha_0$. Thus, in the
Drude formula (\ref{25}) the exponent of the Luttinger-liquid power-law
function $(T/\Lambda)^a$ is $T$-independent.

\subsection{Bosonization vs Localization}
\label{IVc}

As discussed in Sec.~\ref{IVb}, the solution of the RG equations flows to
the strong-coupling regime, where ${\cal D}\agt 1$ and the
perturbative-in-disorder RG equations are no longer valid. Since in the
absence of interaction the strong-coupling fixed point ${\cal D}/L\to \infty$
corresponds to the complete Anderson localization, Giamarchi and Schulz
identified the regime ${\cal D}\gg 1$ as the localized phase.

It should be stressed, however, that the notion of Anderson localization does
not follow from the RG equations of Ref.~\onlinecite{giamarchi88}. Indeed,
Eq.~(\ref{55}) does not account for the renormalization of disorder ${\cal
D}/L$ in the absence of interaction: the effects of Anderson localization
(arising due to interference terms which involve higher powers of ${\cal D}$)
are not included in this RG scheme. In the weak-coupling regime, ${\cal D}\ll
1,$ the Anderson localization would show up in an additional
$\alpha$-independent term, $f_{\rm wl}({\cal D})\sim {\cal D}^3$ on the
right-hand side of Eq.~(\ref{55}), which is beyond the accuracy of the
derivation of these equations. It is also worth noting that in the presence of
interaction all the RG trajectories for ${\cal D}/L$ tend to saturate at
finite values of ${\cal D}/L$, as can be seen from Fig.~\ref{f1}, in contrast
to what one expects for the Anderson-localization behavior. The
strong-coupling fixed point ${\cal D}\to \infty$ is therefore a manifestation
of Ohm's law, rather than of the Anderson localization.

In order to study the temperature behavior of $\sigma(T)$, the renormalization
procedure in Ref.~\onlinecite{giamarchi88} was stopped at $L=L_T\sim u/T$,
with the argument that the coherence effects disappear on length scales
greater than $L_T$ due to inelastic processes. Although this is perfectly
correct for the interaction-induced renormalization of disorder which is cut
off by the thermal smearing of the electron distribution function [similarly
to the Altshuler-Aronov correction to the conductivity in higher-$d$ systems,
see Eq.~(\ref{2})], the coherence effects relevant to the Anderson
localization are governed by the dephasing length, $L_\phi$. The condition
$L_T\sim\xi$ yielding the temperature $\sim u/\xi$ for the onset of the
Anderson localization is in general not correct.

Indeed, the additional term $f_{\rm wl}({\cal D})$ responsible for the WL
(quantum-interference) effects is cut off not by $L_T$ but rather by the
dephasing length $L_\phi$, which is parametrically different from $L_T$. In
particular, in the absence of interaction $L_\phi$ diverges, while $L_T$
remains finite, being independent of $\alpha$. For weak interaction $L_\phi\gg
L_T$ and hence, there exists an intermediate range of length scales, $L_T\alt
L \alt L_\phi$, in which only the WL term $f_{\rm wl}({\cal D})$ contributes
to the renormalization of ${\cal D}$. Thus, it turns out that the scalings
with $T$ and $L$ are not interchangeable, when $\sigma(T)$ is considered. For
$\alpha\ll 1$, the onset of strong localization is determined by the condition
\be
L_\phi\sim \xi~,
\label{61}
\ee
which in 1d is equivalent to
\be
\tau_\phi\sim\tau~.
\label{62}
\ee 

After the RG transformation, the bosonization action retains its original form
(\ref{35})--(\ref{37}), but with a renormalized ultraviolet cutoff,
$\Lambda\to T$, and a renormalized scattering rate, $\tau^{-1}_0\to
\tau^{-1}(T)$. Performing a transformation from the bosons back to fermions,
we thus return to the original fermionic theory with a corresponding
renormalization of its parameters. This theory is analyzed in the following
sections.

\section{Golden Rule in Luttinger liquid}
\label{V}
\setcounter{equation}{0}

In order to analyze the effects of dephasing, we will follow the route
suggested by earlier works on higher-dimensional (in particular, diffusive)
systems.\cite{altshuler85} Since the physics of dephasing is governed by
electronic inelastic scattering processes, a natural first step is to
calculate the e-e scattering rate $\tau_{\rm ee}^{-1}$. Indeed, for
higher-dimensional systems in the high-temperature (ballistic) regime,
$T\tau\gg 1$, the dephasing rate $\tau_\phi^{-1}$ to leading order is given by
$\tau_{\rm ee}^{-1}$, see Ref.~\onlinecite{narozhny02}. For quasi-1d and 2d
systems in the diffusive regime the scattering kernel acquires an infrared
singularity leading to a divergent $\tau_{\rm ee}^{-1}$. However, even in this
case the calculation of the e-e collision rate turns out to be instructive: a
parametrically correct result for the dephasing rate can be obtained from the
expression for $\tau_{\rm ee}^{-1}$ supplemented with an appropriate infrared
cutoff. It is thus useful to begin by analyzing $\tau_{\rm ee}^{-1}$. We will
show that this quantity is finite and meaningful in the interacting 1d system
(contrary to what one might expect based on the famous non-Fermi-liquid
character of the Luttinger liquid). On the other hand, we will demonstrate
that, while governing the dephasing of Aharonov-Bohm oscillations, $\tau_{\rm
ee}^{-1}$ does not give the dephasing rate for the WL of spin-polarized
electrons.

The Golden Rule expression for the e-e collision rate following from the
Boltzmann kinetic equation reads
\begin{eqnarray}
{1\over \tau_{\rm ee} (\epsilon)}&=&\int\! d\omega\int\!  d\epsilon' \,
{\cal K}(\omega)\nonumber \\
&\times&
 \left(f^h_{\epsilon-\omega}f_{\epsilon'}
f^h_{\epsilon'+\omega}+f_{\epsilon-\omega}
f^h_{\epsilon'}f_{\epsilon'+\omega}\right)~,
\label{63}
\end{eqnarray}
where ${\cal K}(\omega)$ is the kernel of the e-e collision integral
\begin{eqnarray}
{\rm St}\{f_\epsilon\}&=&\int\! d\omega\int\!  d\epsilon' \, 
{\cal K}(\omega) \nonumber \\
&\times& \left(-f_\epsilon
  f^h_{\epsilon-\omega}f_{\epsilon'}f^h_{\epsilon'+\omega}+ 
f^h_{\epsilon}f_{\epsilon-\omega}
f^h_{\epsilon'}f_{\epsilon'+\omega}\right)\nonumber \\
&=& -\frac{f_{\epsilon}}{\tau_{\rm ee}(\epsilon)}+\int\! d\omega\int\!
d\epsilon' \, {\cal K}(\omega)  
f_{\epsilon-\omega} f^h_{\epsilon'}f_{\epsilon'+\omega}~.\nonumber 
\label{64}
\end{eqnarray}
Here $f_\epsilon$ is the Fermi distribution function,
$f^h_{\epsilon}=1-f_{\epsilon}$.

 
To leading (second) order in interaction, the scattering kernel $K(\omega)$ in
a clean Luttinger liquid has the form
\be
K(\omega)=\eta_s[\,K^H_{++}(\omega)+K^H_{+-}(\omega)\,]+K^F(\omega)~,
\label{65}
\ee
where 
\bea
K^H_{++}&=&g_4^2 \
{1\over \pi^3\rho}\int \!{dq\over 2\pi}\, 
\left[\,{\rm Re} D_{+}(\omega,q)\,\right]^2~,
\label{66} 
\\
K^H_{+-}&=&
g_2^2 \ {1\over \pi^3\rho}\int \!{dq\over 2\pi} \,
 {\rm Re} D_{+}(\omega,q)\,{\rm Re} D_{-}(\omega,q)~,
\nonumber \\
\label{67} 
\\
K^F&=&-K^H_{++}~.
\label{68}
\eea
Here $K^H_{++}$ and $K^H_{+-}$ are related to scattering of two electrons from
the same or different chiral spectral branches, respectively, $K_F$ is the
exchange counterpart of $K^H_{++}$, $\eta_s$ is the spin degeneracy
($\eta_s=1$ for the spinless case and $\eta_s=2$ for the spinful case), and
$D_{\pm}(\omega,q)$ are the two-particle propagators, given by Eq.~(\ref{A5})
with $i\Omega_n\to \omega+i0$.

\begin{figure}[ht]
\centerline{
\includegraphics[width=0.9\columnwidth]{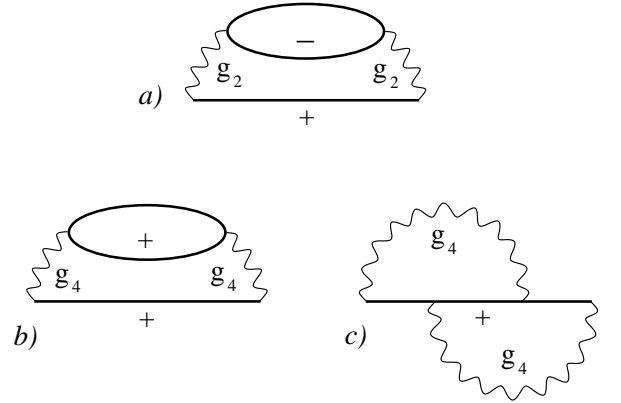}
} 
\caption{
Second-order (``Golden Rule") diagrams for the self-energy of a right-mover.
Wavy lines: interaction of the right-mover with left- ($g_2$) or right-
($g_4$) movers.  (a) $\Sigma^H_{+-}$, the Hartree contribution due to the
interaction with a left mover; (b) $\Sigma^H_{++}$, the Hartree contribution
due to the interaction with a right mover; (c) $\Sigma^F$, the Fock (exchange)
contribution.
}
\label{f4} 
\end{figure}

Substituting Eqs.~(\ref{65})--(\ref{68}) in Eq.~(\ref{63}), we obtain the
lowest-order result for the e-e scattering rate at the Fermi level
$(\epsilon=0)$ in terms of the corresponding contributions to the retarded
electronic self-energy $\Sigma^R_+$ defined by
$G^R_+(\epsilon,p)=[\,\epsilon-v_Fp-\Sigma^R_+(\epsilon,p)\,]^{-1}$, where
$G^R_+$ is the retarded Green's function for right-movers:
\bea
1/2\tau_{\rm ee}&=&-{\rm Im}\,\Sigma^R_+(0,0)~,\label{69} \\
\Sigma^R_+(0,0) &=& \Sigma^H_{++}+\Sigma^H_{+-}+\Sigma^F~.
\label{70}
\eea
The diagrams for $\Sigma^H_{+-}$, $\Sigma^H_{++}$, and $\Sigma^F$ are shown in
Figs.~\ref{f4}a, \ref{f4}b, and \ref{f4}c, respectively.  The
contribution of the diagram in Fig.~\ref{f4}b (Hartree term) is canceled
by the diagram in Fig.~\ref{f4}c (Fock term) due to the Pauli principle,
\be 
\Sigma^F=-\Sigma^H_{++}~.
\label{71}
\ee
We refer to this cancellation as the ``Hartree-Fock cancellation''.

Let us now calculate the imaginary parts of the perturbative self-energies.
Using Eq.~(\ref{A5}) to get the retarded propagators $D_\pm(\omega,q)$ by
analytically continuing $i\Omega_n\to \omega+i0$ and putting $g_2=g_4=2\pi
v_F\alpha$ we find:
\bea
&&{\rm Im}\Sigma^H_{++}=-2\pi^2\alpha^2v_F^3\int{d\omega\over 2\pi} \ 
\omega\left(\coth{\omega\over 2T}-\tanh{\omega\over 2T}\right)\nonumber \\
&&\times\,\,
\int{dq\over 2\pi} \ 
\left[\,{\rm Re} D_{+}(\omega,q)\,\right]^2\simeq
-\pi\alpha^2v_FT
\nonumber \\ 
&&\times\int_{|\omega|\alt T}\!\!\! d\omega\!\int \! dq\,
\delta(\omega-v_Fq)\delta(\omega-v_Fq)~,
\label{72}\\
&&{\rm Im}\Sigma^H_{+-}=-2\pi^2\alpha^2v_F^3\int{d\omega\over 2\pi} \ 
\omega\left(\coth{\omega\over 2T}-\tanh{\omega\over 2T}\right)\nonumber \\
&&\times\,\,
\int{dq\over 2\pi} \ 
{\rm Re} D_{+}(\omega,q) \,\,{\rm Re}D_{-} (\omega,q)\simeq
-\pi\alpha^2v_FT
\nonumber \\ 
&&\times\int_{|\omega|\alt T}\!\!\! d\omega\!\int \! dq\,
\delta(\omega-v_Fq)\delta(\omega+v_Fq)~.
\label{73}
\eea
One sees that the contribution of $\Sigma^H_{++}$ is diverging. For
spin-polarized electrons (i.e., for $\eta_s=1$) it is, however, exactly
canceled by the exchange interaction, Eq.~(\ref{71}). Indeed, as we have
discussed in Sec.~\ref{IIIa}, the $g_4$ interaction drops out of the problem
in this case, inducing only a shift of the velocity, Eq.~(\ref{21}). The
remaining term $\Sigma^H_{+-}$ is determined by $\omega,q\to 0$ and is given
by
\be
{\rm Im}\Sigma^H_{+-}=-\pi\alpha^2T/2~.
\label{74}
\ee 
Already the perturbative expansion demonstrates a qualitative difference
between two cases of spinless and spinful electrons---the spinful case will be
considered in Ref.~\onlinecite{yashenkin06}.

The characteristic frequencies in Eqs.~(\ref{72}),(\ref{73}) satisfy
$|\omega|\alt T$, because of the factor $\coth (\omega/2T)-\tanh (\omega/2T)$,
even if we would perform this calculation in the original theory with the
ultraviolet cutoff $\Lambda$. For the sake of comparison, we present also the
corresponding contribution to the real part of the self-energy:
\bea 
&&{\rm Re}\,\Sigma_+(\epsilon,p)=-2\pi^2\alpha^2v_F^2\,
\int{d\omega\over 2\pi}\int{dq\over 2\pi}
\nonumber \\ 
&&\times\,\left\{ \coth {\omega\over 2T}\,{q\over \epsilon+\omega-v_F(p+q)}
\,\delta (\omega+v_Fq)\right.  \nonumber \\
&&\left. \hspace{4mm}+\tanh {\omega+\epsilon\over 2T}\,{q\over \omega+v_Fq}\,
\delta\, [\,
\epsilon+\omega-v_F(p+q)\, ]\right\}~, \nonumber
\eea
which for $T\gg |\epsilon\pm v_Fp|$ gives
\be
{\rm Re}\,\Sigma_+(\epsilon,p)\simeq -{1\over 2}\alpha^2(\epsilon-v_Fp)
\ln {\Lambda\over T}~.
\label{76} 
\ee 
In contrast to the imaginary part of the self-energy, the real part is
determined by energy transfers $|\omega|\agt T$, which is characteristic
of the elastic virtual processes. Furthermore, ${\rm Re}\Sigma_+$ is
logarithmically divergent in the ultraviolet and is cut off by $\Lambda$: this
behavior is specific to 1d. 

It is worth stressing that in the considered case of weak interaction,
$\alpha\ll 1,$ the lowest-order contribution to the inelastic scattering rate
obtained from Eqs.~(\ref{69}),(\ref{74}),
\begin{equation}
\tau_{\rm ee}^{-1}=\pi\alpha^2 T~,
\label{77}
\end{equation}
is much smaller than temperature:
\be
T\tau_{\rm ee}\gg 1~.
\label{78}
\ee
In higher dimensionalities, this fact is commonly referred to as one of the
conditions of the existence of Fermi liquid: the inverse lifetime of a
quasiparticle is smaller than its characteristic energy. In this respect, the
weakly interacting Luttinger liquid, while being a canonical example of a
non-Fermi liquid, reveals the typical Fermi-liquid property. The
Luttinger-liquid physics is in fact encoded in the singular {\it real} part of
the perturbative self-energy, Eq.~(\ref{76}). For $\alpha^2\ln(\Lambda/T)\ll
1$, the Green's function can be written as $G^R_+(\epsilon,p)\simeq
Z/(\epsilon-v_Fp+i\pi\alpha^2T/2)$, where the single-particle weight $Z\simeq
1-(\alpha^2/2)\ln(\Lambda/T)$ is suppressed by interaction. It is in fact the
product $\alpha^2\ln(\Lambda/T)$, coming from the real part of the
self-energy, that after resummation of all orders gets exponentiated and leads
to the Luttinger-liquid power-law singularities. Those singularities have been
accounted for by the renormalization at step 1 in the present approach.

It is instructive to compare the perturbative collision rate in a clean
Luttinger liquid, Eq.~(\ref{77}), with the damping\cite{lehur02} of the exact
retarded single-particle Green's function in the coordinate-time $(x,t)$
representation,
\be
g^R_{\pm}(x,t)=2i\Theta(t)\,{\rm Im}\,\bar{g}_{\pm}(x,t)~.
\label{79}
\ee 
Considering right-movers for definiteness, we have
\begin{eqnarray} 
\bar{g}_{+}(x,t)&=&(\Lambda/2\pi u)
A^{1+\alpha_b/2}\left(x/u
-t+i/\Lambda\right)\nonumber \\
&\times& A^{\alpha_b/2}\left(x/u+t-i/\Lambda\right)~,
\label{80}
\end{eqnarray}
where 
\be
A(z)=\pi T/\Lambda\sinh (\pi Tz)
\ee
and $\alpha_b=(K+K^{-1}-2)/2$, which becomes $\alpha_b\simeq \alpha^2/2$ for
$\alpha\ll 1$.

The Green's function is peaked at the ``classical trajectory''$x=ut$. 
In the limit of large $ t \to \infty$ we have an exponential 
decay\cite{lehur02}
\be
\bar{g}_{+}(x=ut,t)\simeq  {-i\Lambda\over 2\pi u}
\left({2\pi T\over i\Lambda}\right)^{\alpha_b/2}\exp(- \pi\alpha_b T t)~,
\label{84}
\ee
which should be contrasted with a power-law $t$-dependence at zero 
temperature,
\be
\bar{g}_{+}(ut,t) ={-i \Lambda \over 2\pi u} \,{1\over (2i\Lambda
t)^{\alpha_b/2}}~. 
\label{85}
\ee
The large factor $\Lambda$ in the numerator of Eq.~(\ref{84}) is reminiscent
of the pole in the noninteracting Green's function. It is worth stressing
that, since $|A(i/\Lambda)|\to 1$, the nonanalytical singularity of the factor
$\sinh^{-\alpha_b/2}[\,\pi T(x/u-t+i/\Lambda)\,]$ in the main peak at $x=ut$
is exactly canceled by the Luttinger-liquid renormalization factor
$(T/\Lambda)^{\alpha_b/2}$ which arises due to the resummation of the real
part of the self-energy. Thus the exponential decay in Eq.~(\ref{84}) comes
solely from the factor $[A(2t-i/\Lambda)]^{\alpha_b/2}$ which describes the
tail of the ``left-moving peak'' centered at $x=-ut$. The decay $\exp
(-\pi \alpha_bTt)$ of the ``residue''
\begin{eqnarray}
[\,x-u(t-i/\Lambda)\,]g^R_+(x,t)|_{x\to ut}\propto \sinh^{-\alpha_b/2}(2\pi
Tt)\nonumber 
\label{86}
\end{eqnarray}
for $t\to\infty$ agrees with the Golden Rule expression (\ref{77}) to order
$\alpha^2$: $\pi \alpha_b T \simeq 1/2\tau_{\rm ee}.$ Therefore, the
temporal decay of the Green's function is determined by the processes of
inelastic scattering of right-movers on left-movers.

We see that the perturbative result for the scattering rate yields the
asymptotic behavior of the exact Green's function. This means that the Golden
Rule calculation of the inelastic scattering time is meaningful in Luttinger
liquid. However, the Golden Rule time $\tau_{\rm ee}$ does not cut off the WL
correction in the spinless case, as we demonstrate below in Sec.~\ref{VI}.

\section{Weak localization and dephasing: Fermionic path--integral approach}
\label{VI}
\setcounter{equation}{0}

\subsection{WL dephasing vs Golden Rule}
\label{VIa}

The notion of dephasing associated with the behavior of the single-particle
Green's function (\ref{86}) makes sense in a clean Luttinger liquid in the
ring geometry, see Sec.~\ref{VIe}. However, as we are going to show in this
section, the WL dephasing time $\tau_\phi^{\rm wl}$ that governs the cutoff of
the WL correction $\Delta\sigma_{\rm wl}$ is parametrically different from the
Golden Rule time $\tau_{\rm ee}$, Eq.~(\ref{77}).

This difference can in fact be anticipated based on the following
observation. The integral (\ref{73}) for the Golden Rule rate is determined
by the energy and momentum transfer $q,\omega=0$. On the other hand, in the
spirit of Ref.~\onlinecite{altshuler82}, soft inelastic scattering with
$qv_F,\omega\ll 1/\tau^{\rm wl}_\phi$ is not expected to contribute to the WL
dephasing. This suggests that the dephasing rate $1/\tau_\phi^{\rm wl}$
requires a self-consistent cutoff in Eq.~(\ref{63}) at $qv_F,\omega\sim
1/\tau^{\rm wl}_\phi$, similarly to Eq.~(\ref{4}) for higher-dimensional
systems, and so is parametrically different from the one in Eq.~(\ref{77}).
In particular, if one introduces, similarly to higher dimensionalities, a
sharp low-frequency cutoff in the right-hand side of Eq.~(\ref{73}), one
immediately gets zero instead of Eq.~(\ref{77}). On the other hand,
introducing disorder in $D_{+}$ and $\Pi_{-}$ in Eq.~(\ref{73}), one
effectively broadens the delta functions by $1/\tau$, and the self-consistent
equation can be solved.

Such a self-consistent equation should be treated with the utmost caution, in
view of the singular character of the ballistic propagators. In particular,
when transformed to the space-time representation, the integral in
Eq.~(\ref{73}) is determined by a single point, $x,t=0$, so that one could be
drawn to the erroneous conclusion that it is high---rather than
low---frequencies that matter. In any case, this equation is at most
qualitative. However, the message that we learn from this consideration is
correct: (i) the WL dephasing rate differs from the Golden Rule result, and
(ii) to evaluate it, we have to include disorder in the interaction
propagators. To perform a controllable quantitative analysis, we employ the
path-integral technique, which is a natural tool in the situation when the
diagrammatic calculation suffers from the infrared problems.

\subsection{Weak localization in 1d}
\label{VIb}

\begin{figure}[ht]
\centerline{
\includegraphics[width=0.85\columnwidth]{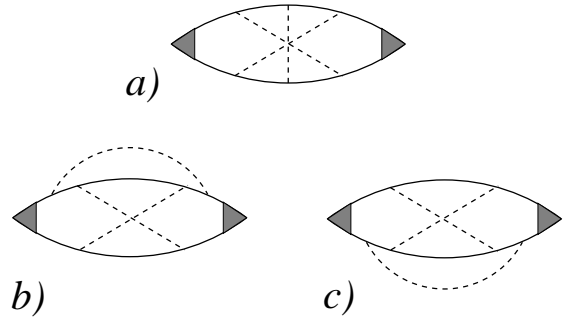}
} 
\caption{Diagrams describing the leading WL correction to the conductivity of
Luttinger liquid for $\tau_\phi^{\rm wl}\ll \tau$. The dashed lines represent
backscattering off impurities. The current vertices are dressed by impurity
ladders. The solid lines are the Green's functions with disorder-induced
self-energies. The diagrams are understood as fully dressed by e-e 
interactions. }
\label{f5} 
\end{figure}

We begin by identifying the leading contribution to the WL correction to the
conductivity in the regime of strong dephasing, $\tau_\phi^{\rm wl}/\tau\ll
1$.  This regime takes place at sufficiently high temperatures---the precise
condition is found below in Sec.~{IIIc}, where we evaluate $\tau_\phi^{\rm
wl}$. The diagrams that give the main contribution to $\Delta\sigma_{\rm wl}$
are shown in Fig.~\ref{f5}.

\begin{figure}[ht]
\centerline{
\includegraphics[width=0.85\columnwidth]{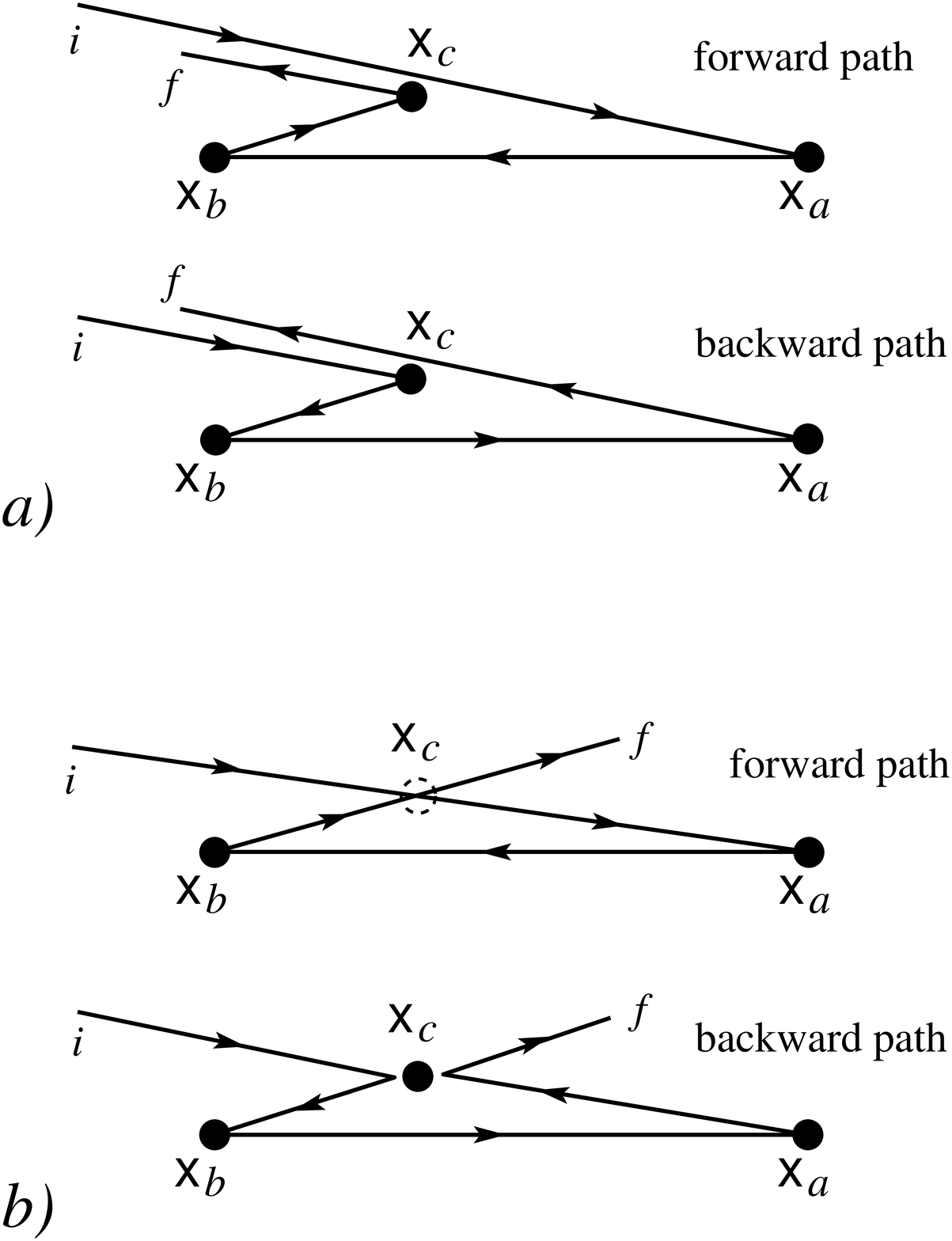}
} 
\caption{
(a) Interference paths described by the Cooperon diagram, Fig.~\ref{f5}a.
Black dots: the backscattering off impurities. The one-dimensional graphs are
stretched vertically for ease of visualization. The two waves start at point
$i$ and interfere at point $f$ after traversing the closed loop in opposite
directions. (b) Interfering paths for the diagram in Fig.~\ref{f5}c. The
dashed circle denotes the absence of backscattering off the impurity at point
$x_c$. The paths corresponding to the diagram in Fig.~\ref{f5}b are obtained
by interchanging the upper and lower plots in Fig.~\ref{f6}b. Note that
similar paths describe the WL in the ballistic regime in higher
dimensionalities.\cite{gasparyan85}
}
\label{f6} 
\end{figure}

\begin{figure}[ht]
\centerline{
\includegraphics[width=0.95\columnwidth]{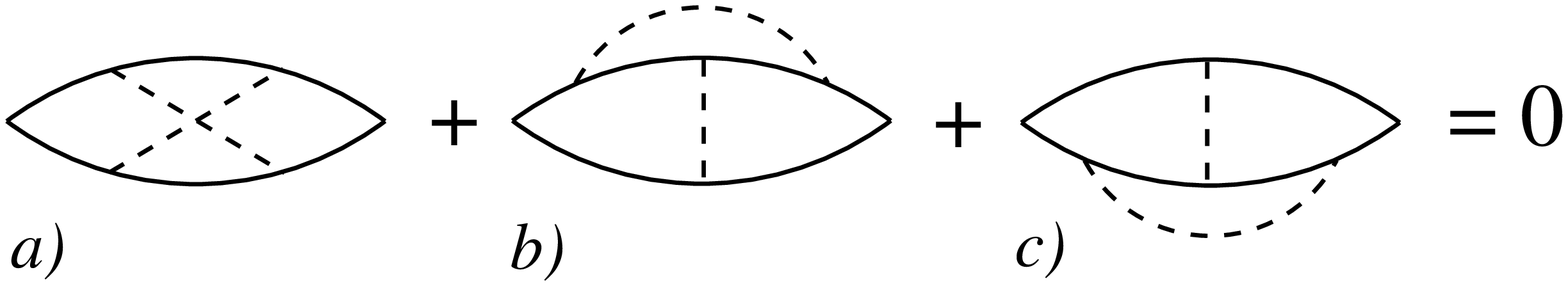}
} 
\caption{Two-impurity non-Drude diagrams. }
\label{f7} 
\end{figure}

The ``three-impurity Cooperon" in Fig.~\ref{f5}a describes the propagation of
two electron waves along the path connecting three impurities (``minimal
loop") in time-reversed directions, Fig.~\ref{f6}a. Indeed, one can
easily check that the interference paths involving only two backscattering
impurities are impossible merely due to geometrical reasons. This means that
the Cooperon with two backscattering impurity lines\cite{coopeven} yields zero
contribution to the conductivity to leading order in $(\epsilon_F\tau)^{-1}$.
If one retains the forward scattering and does not gauge it out, the
two-impurity Cooperon diagram (Fig.~\ref{f7}a) is canceled by the
contribution of two other non-Drude diagrams with two impurity lines
(Fig.~\ref{f7}b,c). The diagrams represented in Fig.~\ref{f5}b,c (the
corresponding trajectories are shown in Fig.~\ref{f6}b) are of the same
order as the Cooperon diagram.


In the absence of interaction, all higher-order diagrams (including
non-Cooperon diagrams) which describe quantum-interference processes involving
a larger number of impurities ($4,5,\ldots$) are equally important and
together with the diagrams in Fig.~\ref{f5} sum up to exactly cancel
$\sigma_{\rm D}$, which spells a complete localization.\cite{berezinskii73} In
the WL regime, they only yield subleading corrections through a systematic
expansion in powers of the small parameter $\tau_\phi^{\rm wl}/\tau$. Thus,
the WL in 1d is governed by the ``shortest possible Cooperon'', namely, the
Cooperon with three impurity lines. In essence, this is due to the fact that
the WL condition $1/\tau_{\phi}^{\rm wl}\gg \Delta_\xi$ reduces in 1d to the
condition of the ``ballistic dephasing'', $\tau_\phi^{\rm wl}/\tau\ll 1$.  In
turn, this means that the contribution to $\Delta\sigma_{\rm wl}$ of typical
configurations of three impurities separated by a distance of order $v_F\tau$
is strongly suppressed by dephasing. In the compact configurations that do
contribute the three impurities are anomalously close to each other, with a
characteristic distance between them being much smaller than the mean free
path.

\subsection{WL dephasing in 1d: Qualitative discussion of the leading-order
cancellation} 
\label{VIc}

Before turning to the path-integral evaluation of the WL dephasing rate in a
spinless 1d system, we present a simple argument demonstrating the
cancellation of the leading contribution $1/\tau_{\rm ee}$ to
$1/\tau_\phi^{\rm wl}$. For the purpose of a qualitative discussion, we can
consider inelastic scattering within the picture of a {\it classical} thermal
noise with characteristic frequencies $|\omega|\ll T$. Let $U_0^\pm(x)$ be a
given realization of the fluctuating scalar potential $U^\pm(x,t=0)$ created
by right-/left-movers at time $t=0$.  The peculiarity of the 1d geometry is
that in the clean case the initial density profiles of right-/left-movers,
whose dispersion relation is linear, remain unchanged: they move with the
plasmon velocity either to the left or to the right as a whole, and so does
the fluctuating potential:
\be
U^\pm(x,t)=U^\pm_0(x\mp ut)~.
\label{87}
\ee 

\begin{figure}[ht]
\centerline{
\includegraphics[width=7cm]{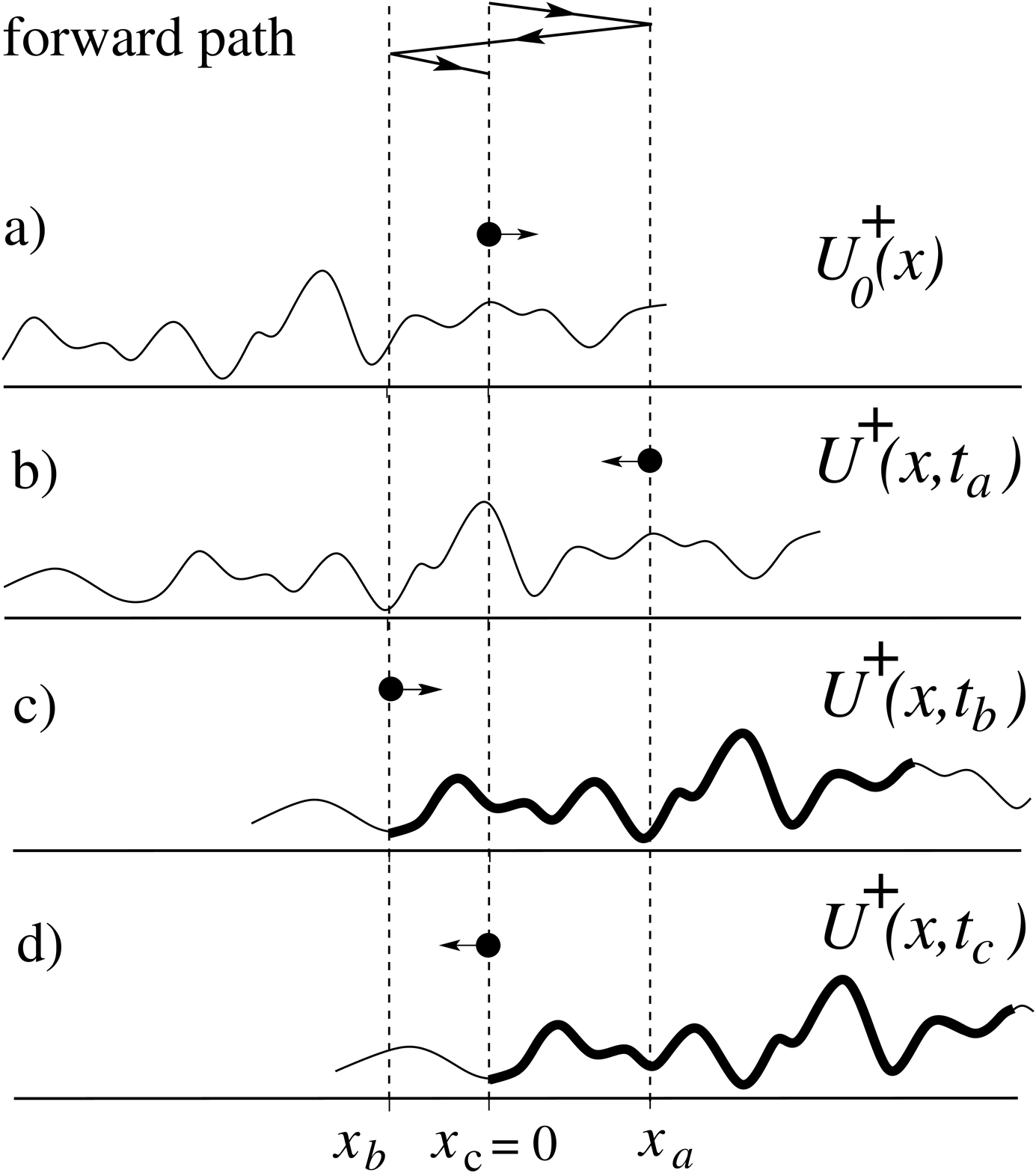}
} 
\caption{
Illustration of the quasiclassical electron dynamics in the presence of three
impurities (whose positions are denoted by the dashed lines) and a
right-moving component of the fluctuating scalar potential $U^{+}(x,t)$.  Top:
the forward path corresponds to that in Fig.~\ref{f6}a. The position (black
dots) and the direction of the velocity (arrows) of the particle, as well as
the profile of the fluctuating potential, are shown at different times: (a)
$t=0$; (b) $t=t_a+0$; (c) $t=t_b+0$; (d) $t=t_c+0$. The phase $\phi_{-}^F$ of
the electronic wave is accumulated when the particle moves to the left, that
is between the snapshots (b) and (c). The segment of the fluctuating potential
that contributes to the phase shift is marked in bold.}
\label{f8} 
\end{figure}

We now demonstrate that the phases acquired by the time-reversed waves
propagating along a closed three-impurity loop are equal for {\it arbitrary}
$U_0^\pm (x)$ and hence cancel each other in the interference contribution
even before the averaging over fluctuations of $U_0^\pm (x)$. As we show in
Appendix~\ref{aE}, the velocity of electrons moving along the ``Cooperon path"
is renormalized by the interaction and coincides with $u$. Referring to the
paths shown in Fig.~\ref{f6}a, we first follow the ``forward" trajectory
$x_c=0\to x_a \to x_b \to 0$ for a particle which starts moving to the right
at $t=0$ and is backscattered at these points at times $t_a$, $t_b$, and
$t_c$, respectively, see Fig.~\ref{f8}. It is important to recall that the
scattering amplitude $g_4$ has been eliminated through the renormalization of
the Fermi velocity. Therefore, on the right-moving parts of the forward
trajectory (segments $0 \to x_a$ and $x_b\to 0$) the phase changes by
$\phi_+^F$ only due to the scattering off $U^-(x,t)$, while on the left-moving
parts (segment $x_a\to x_b$) only the scattering off $U^+(x,t)$ matters and
yields the phase shift $\phi_-^F$. Similarly, there are phase shifts for the
backward path $\phi^B_\pm$ due to the scattering off $U^\mp(x,t)$.

\begin{figure}[ht]
\centerline{
\includegraphics[width=7cm]{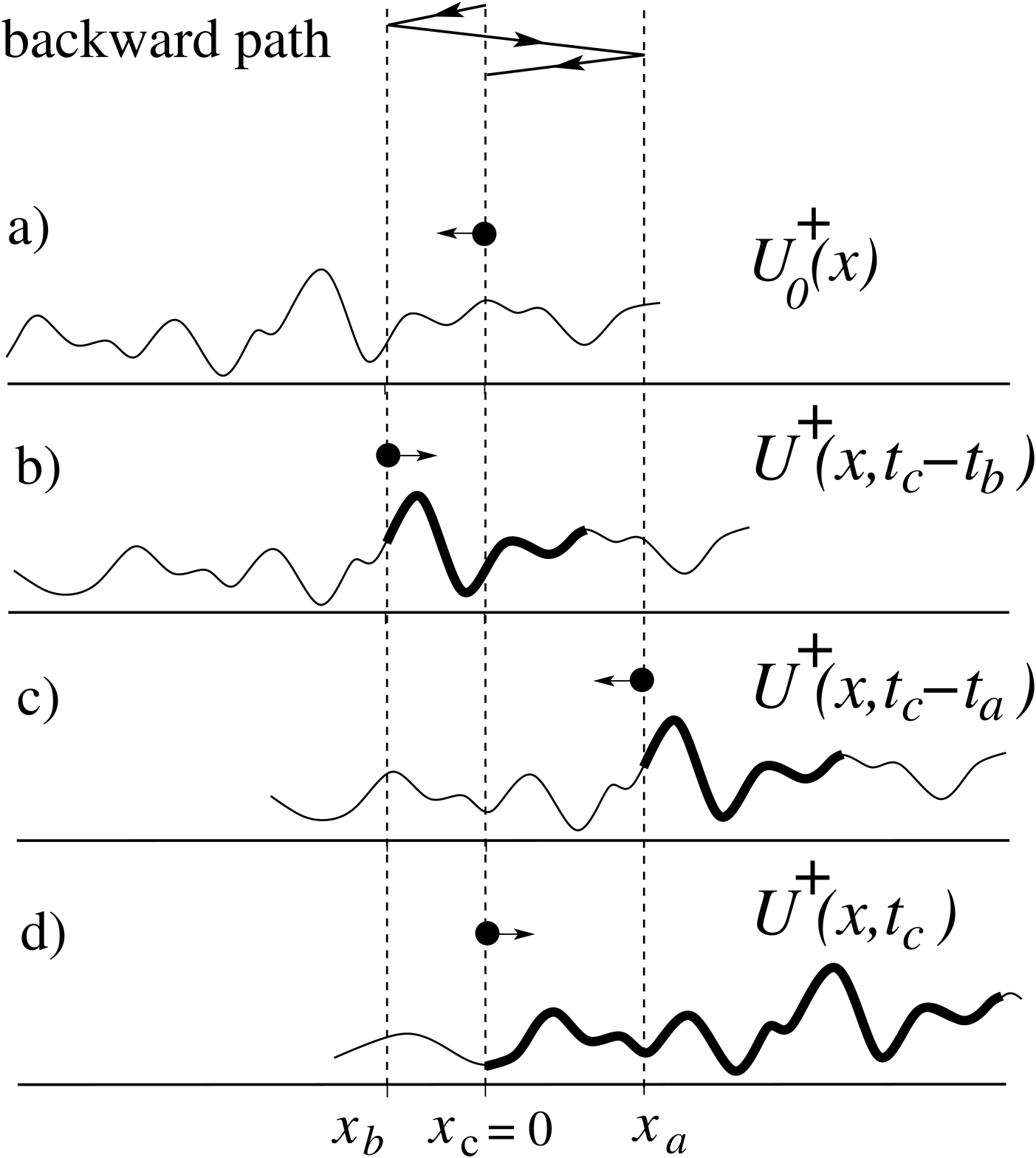}
} 
\caption{
Same as in Fig.~\ref{f8} but for the backward part,
Fig.~\ref{f6}a. (a) $t=0$; (b) $t=t_c-t_b+0$; (c) $t=t_c-t_a+0$; (d)
$t=t_c+0$.  The phase $\phi_{-}^B$ of the electronic wave is accumulated when
the particle moves between the snapshots (a) and (b) and between the snapshots
(c) and (d). }
\label{f9}
\end{figure}

The phase $\phi^F_-$ acquired on the forward path between points $x_a$ and 
$x_b$ is given by
\be
\phi^F_-=-{1\over u}
\int_{-ut_c}^0 U^+_0(x)\,dx~.
\label{88}
\ee
Here we use $t_b-t_a=t_c/2$. Electrons moving on the backward path interact 
with the same right-moving potential $U^+(x,t)$ on the segments $0\to x_b$ and
$x_a\to 0$, see Fig.~\ref{f9}. The phase $\phi^B_-$ is written as
\be
\phi^B_-=-{1\over u}\int_{-2u(t_c-t_b+t_a)}^0 U_0^+(x)\,dx=\phi^F_-~.
\label{89}
\ee
Similarly, $\phi^F_+=\phi^B_+$.  This is clearly seen in
Figs.~\ref{f8}d and \ref{f9}d, where the regions of the
potential profile that contribute to the phase shifts are identical at time
$t_c$ for the forward and backward paths. As a result, the interaction-induced
phase shifts cancel out in the total Cooperon phase
$(\phi^F_++\phi^F_-)-(\phi^B_++\phi^B_-)$ for an arbitrary profile
$U_0(x)$. This means, in particular, that in a calculation of the conductance
through a short piece of Luttinger liquid to third order in the impurity
strength, which is the order at which the interference term arises, the latter
would remain unaffected by inelastic e-e collisions. In the calculation of the
WL correction to the conductivity of an infinite quantum wire, we have,
indeed, to include disorder in the interaction propagator, despite the
ballistic character of dephasing in the WL regime and in accordance with the
argument made in Sec.~\ref{VIa}. This is done systematically within the
path-integral approach in the next subsection.

\subsection{WL dephasing in 1d: Path-integral calculation}
\label{VId}

The dephasing-induced action $S(t_c,t_a)$ acquired by the Cooperon is
accumulated on the classical (saddle-point) path, whose geometry for three
impurities if fixed by two length scales, the total length of the path $v_Ft_c$
and the distance between two rightmost impurities $v_Ft_a\leq v_Ft_c/2$
(Fig.~\ref{f10}). As shown in Appendix~\ref{aB}, the WL correction can then be
represented as
\begin{equation}
\Delta\sigma_{\rm wl}=-2\sigma_{\rm D}\!\int_0^\infty \!\!dt_c \!\int_0^\infty
\!\!dt_a P_2(t_c,t_a)\exp \left[-S(t_c,t_a)\right]~, 
\label{91} 
\end{equation} 
where 
\be
P_2(t_c,t_a)=(1/8\tau^2)\exp (-t_c/2\tau)\Theta(t_c-2t_a)
\label{92}
\ee 
is the probability density of return to point $x=0$ after two reflections at
points $x=v_Ft_a$ and $x=-v_F(t_c/2-t_a)$ (in this subsection, we ignore the
difference between $v_F$ and $u$). The contribution $S_{ij}$ to the
dephasing action associated with inelastic interaction between electrons
propagating along the paths $x_i(t)$ and $x_j(t)$ is obtained similarly to
higher dimensionalities:\cite{altshuler82,altshuler85}
\begin{eqnarray} 
&&S_{ij}=-T\int
{d\omega\over 2\pi}\int {dq\over 2\pi}\int_0^{t_c} \!dt_1\!\int_0^{t_c} \!dt_2
\,{1\over \omega}\,{\rm Im}\,V_{\mu\nu}(\omega,q) \nonumber \\ &&\times\,\,\exp
\{iq\left[\,x_i(t_1)-x_j(t_2)\,\right]-i\omega(t_1-t_2)\}~, 
\label{93}
\end{eqnarray} 
where $V_{\mu\nu}(\omega,q)$ is the dynamically screened retarded interaction.
Because of the HF cancellation of the bare interaction between electrons from
the same chiral branch ($g_4$ processes), the screened interaction
$V(\omega,q)$ acquires the indices $\mu,\nu$ denoting the direction of motion
of the interacting electrons, $\mu={\rm sgn}\,\dot{x}_i$, $\nu={\rm
sgn}\,\dot{x}_j$. More specifically, the $g_4$ processes have been
incorporated in the renormalization of the bare Fermi velocity. It is worth
stressing that if we would keep both $g_2$ and $g_4$ processes in
$V(\omega,q)$, the dephasing action in 1d could not be written in
the form of Eq.~(\ref{93}). Indeed, in that case the dephasing action would
generate the contribution of the diagrams of type (b) in Fig.~\ref{f4},
which is actually absent due to the HF cancellation (\ref{71}).

\begin{figure}[ht]
\centerline{
\includegraphics[width=7cm]{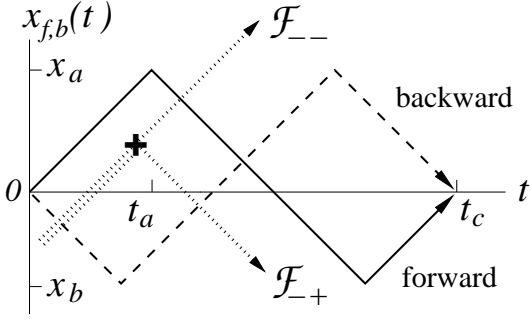}
} 
\caption{Illustration of electron dynamics governing the WL and dephasing:
Time-reversed paths $x_f(t)$ (solid) and $x_b(t)=x_f(t_c-t)$ (dashed) on
which the interaction-induced action $S$ that yields dephasing of the Cooperon
is accumulated. Dotted lines: the propagation of dynamically screened
interaction. The interaction may change the direction of propagation upon
scattering off disorder (as marked by a cross). Each interaction line gives a
contribution to $S$ proportional to $(N_f-N_b)^2$, where $N_{f,b}$ is the
number of its intersections with the forward ($f$) and backward ($b$)
paths. One sees that $N_f\neq N_b$ only due to impurity scattering in the
interaction propagator. Interaction and electron lines lying on top of each
other do not yield dephasing because of the HF cancellation.}
\label{f10} 
\end{figure}

The main steps in the derivation of Eq.~(\ref{93}) are as follows. First, the
interaction is treated within the random-phase approximation (RPA), neglecting
closed electronic bubbles with more than two interaction legs, which is
justified for the weak interaction, $\alpha\ll 1$. Second, we perform an
independent averaging of the RPA bubble and the electronic trajectories over
the disorder. This is justified if the characteristic energy transfer
$\omega\sim 1/\tau_\phi^{\rm wl}$ is much larger $1/\tau$. Further, we treat
the thermal electromagnetic fluctuations through which electrons interact with
each other as a classical field, which is a valid approximation provided that
$\omega$ is much smaller than $T$. As will be seen from the results of our
calculation, the characteristic energy transfer $\omega\sim 1/\tau_\phi^{\rm
wl}$ does satisfy both these requirements for the considered case of weak
interaction, $\alpha\ll 1$. Finally, we will neglect the influence of the
interaction on the velocity ($u$ vs $v_F$), which is again justified to the
leading order in the interaction strength.

Expanding $V_{\mu\nu}$ to second order in $\alpha$, we have
\be
{\rm Im}V_{\mu\nu}(\omega,q)=-\pi\alpha^2v_F
\omega {\cal F}_{\mu\nu}(\omega,q)~,
\label{94}
\ee
where the functions ${\cal F}_{\mu\nu}$ Fourier-transformed to $(x,\tau)$
space are given by Eqs.~(\ref{C6}),(\ref{C7}) with $\gamma=1/\tau$. The action
$S_{ij}$ can then be written in a simple form:
\be
S_{ij}=\pi\alpha^2v_FT\int_0^{t_c}\!\!dt_1\!\int_0^{t_c}\!\!dt_2\,
{\cal F}_{\mu\nu}[\,x_i(t_1)-x_j(t_2),t_1-t_2\,]~,
\nonumber
\label{95}
\ee
where, to first order in $\tau^{-1}$, ${\cal F}_{\mu\nu}(x,t)$ read
\bea
{\cal F}_{++}(x,t)&\simeq&\delta(x+v_Ft)\,(1-|t|/2\tau)~,
\label{96} \\
{\cal F}_{+-}(x,t)&\simeq&
\Theta(v_F^2t^2-x^2)/4v_F\tau~,
\label{97}
\end{eqnarray}
and ${\cal F}_{--}(x,t)={\cal F}_{++}(-x,t)$, 
${\cal F}_{-+}(x,t)={\cal F}_{+-}(x,t)$.
The total action is given by $S=2(S_{\rm ff}-S_{\rm fb})$, where $f$ and $b$
stand for ``forward" and ``backward" time-reversed paths (Fig.~\ref{f10}).

Let us first calculate $S$ for the case when the disorder in interaction
propagators (\ref{96}),(\ref{97}) is neglected, $\tau^{-1}=0$. We get
straightforwardly
\be
S_{\rm ff}=S_{\rm fb}=\pi\alpha^2 Tt_c/2~.
\label{100}
\label{sff}
\ee
One sees that $S_{\rm ff}$ reproduces the Golden Rule result,
Eqs.~(\ref{77}),(\ref{86}). As we explain in Sec.~\ref{VIe}, this rate is
relevant for the AB dephasing in the multiple-connected (ring)
geometry. However, in the present case of dephasing in a wire the self-energy
processes $(S_{\rm ff}+S_{\rm bb})$ are exactly canceled in the action by the
vertex corrections $(S_{\rm fb}+S_{\rm bf})$, yielding $S=0$. This confirms
the conclusion of Sec.~\ref{VIc}: for clean-system interaction propagators,
the WL dephasing is absent.

Therefore, the dephasing in Eq.~(\ref{91}) is only due to the dressing of the
dynamically screened interaction by impurities. A calculation of the dephasing
action $S$ to order ${\cal O}(\tau^{-1})$ is presented in
Appendix~\ref{aC}, with the result
\begin{equation}  
S(t_c,t_a)=2\pi\alpha^2 \,T\,t_a \left(t_c-2t_a\right)/\tau~.
\label{101}
\end{equation}
The dephasing vanishes for $t_a=0,\:t_c/2$, since in these cases the two paths
$f$ and $b$ are identical.

Substituting Eq.~(\ref{101}) in Eq.~(\ref{91}) we find for $\tau_\phi^{\rm
wl}\ll\tau$:
\bea
&&\hspace{-1cm}\Delta\sigma_{\rm wl} 
= -{1\over 4}\,\sigma_{\rm D} \int_0^\infty dt_c \,
{1\over \tau^2}
\exp(-t_c/2\tau)\nonumber \\ &&\hspace{-1cm}\times \int_0^{t_c/2}dt_a 
\exp[\,-4\pi \alpha^2
T t_a(t_c/2-t_a)/\tau\,]\nonumber \\ &&\hspace{-1cm}= -{1\over 4}\,\sigma_{\rm
D}\left(\tau_\phi^{\rm wl}\over \tau\right)^2\!\ln{\tau\over \tau_\phi^{\rm
wl}} \propto {1\over \alpha^2T}\,\ln(\alpha^2T)~,
\label{102}
\eea
where
\begin{equation}
{1\over \tau_\phi^{\rm wl}}=\alpha\left({\pi T\over \tau}\right)^{1/2}~,\quad
T\gg T_1={1\over \alpha^2\tau}~.
\label{103}
\end{equation}
In agreement with the above findings, $1/\tau_\phi^{\rm wl}$ vanishes in the
clean limit,\cite{curvature} in contrast to the total e-e scattering rate,
Eq.~(\ref{77}). The logarithmic factor in Eq.~(\ref{102}) is related to
the contribution of rare configurations of disorder in which two of the three
impurities are anomalously close to each other ($t_a\to 0$ or $t_a\to t_c/2$):
in these configurations the dephasing is strongly suppressed. As a result, the
characteristic $t$ in the integral (\ref{102}) are spread over the whole
range between $\tau_\phi^{\rm wl}$ and $\tau$.

\begin{figure}[ht]
\centerline{
\includegraphics[width=7cm]{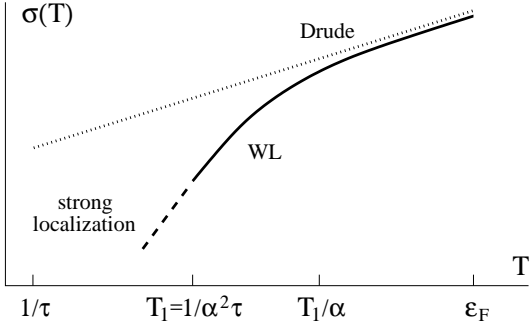}
} 
\caption{Schematic behavior of $\sigma (T)$ on the log-log scale. Dotted line:
the $T$-dependent Drude conductivity. Below
$T_1/\alpha$ the WL correction, Eq.~(\ref{102}), dominates $d\ln\sigma/d\ln
T$. Below $T_1$ the localization becomes strong. }
\label{f11} 
\end{figure}

The scale $T_1$ defined in Eq.~(\ref{103}) marks the temperature below which
the localization effects become strong. It is worth mentioning that the $T$
dependence of $\sigma(T)=\sigma_{\rm D}(T)+\Delta\sigma_{\rm wl}(T)$ starts to
be dominated, with decreasing $T$, by the WL term rather than by $\sigma_{\rm
D}(T)$ at $T\sim T_1/\alpha$, i.e., well above $T_1$ for $\alpha\ll 1$. These
results are illustrated in Fig.~\ref{f11}.

Let us briefly mention what happens below $T_1$. In the strongly localized
phase, the only peculiarity of Luttinger liquid is the $T$ dependence of the
scattering rate $\tau^{-1}(T)$ and the related $T$ dependence of the
localization length $\xi(T)$, persisting down to $T\tau(T)\sim 1$ (see
Sec.~\ref{IIIb1}); otherwise, the picture is similar to that in many-channel
quantum wires or 2d systems. If $\alpha\ll 1$, there exists a wide range of
$T$ where the localization is strong but the conductivity is a power-law
function of $T$. We call this transport regime in a weakly disordered system,
associated with interaction-induced diffusion over localized states, Power-Law
Hopping (PLH); see
Refs.~\onlinecite{gornyi05b,gogolin75,thouless77,gogolin83,basko03} for
details. An important point is that the lifetime of localized states due to
inelastic e-e scattering can be calculated in the PLH regime at the Golden
Rule level. At still lower $T$, the lifetime becomes infinite, so that the e-e
interaction can no longer support hopping via localized states and the
conductivity vanishes at a certain critical $T_c$, see
Refs.~\onlinecite{gornyi05b,basko06} for details.

\subsection{Weak localization vs Aharonov-Bohm dephasing}
\label{VIe}

\begin{figure}[ht]
\centerline{
\includegraphics[width=6.5cm]{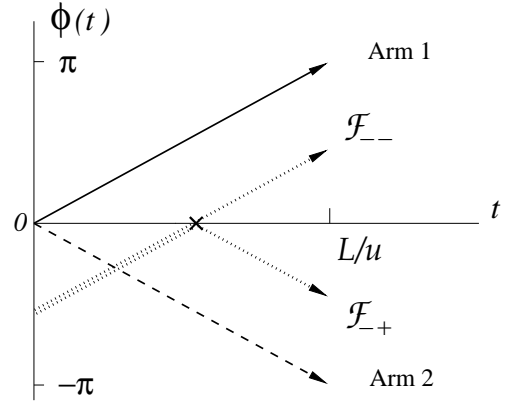}
} 
\caption{
Illustration of electron dynamics governing the dephasing of the first
harmonics $G_{\rm AB}$ of the AB oscillations in the ring geometry with two
arms of equal length $L$.  The position of a particle on the ring is
parametrized by the angle $\phi(t):$ $\phi=0$ corresponds to the point at
which the ring is connected to one lead, $\phi=\pm \pi$---to the point at
which it is connected to the other lead.  The paths $\phi_1(t)$ (solid line)
and $\phi_2(t)$ (dashed) traveling along arm 1 and arm 2, respectively,
interfere at the second contact at time $t=L/u$. Dotted lines: the propagation
of the dynamically screened interaction. The interaction may change the
direction of propagation upon scattering off the contact (as marked by a
cross). Similarly to Fig.~\ref{f11}, each interaction line gives a contribution
to the AB-dephasing action proportional to $(N_1-N_2)^2$, where $N_{1,2}$ is
the number of its intersections with the trajectory in arm 1 or arm 2. One
sees that for any dotted line the number of possible intersections in the time
domain $0<t<L/u$ is strictly one, yielding only the ``self-energy''
contribution to the action and no ``vertex part''.  Interaction and electron
lines lying on top of each other do not contribute to the dephasing because of 
the HF cancellation.
}
\label{f12} 
\end{figure}

As shown in Sec.~\ref{VId}, the dephasing rate that cuts off the WL
correction is much smaller than the Golden Rule result from Sec.~\ref{V},
which governs the damping of the single-particle Green's function,
Eq.~(\ref{86}). One may ask if the Golden Rule rate $1/\tau_{\rm ee}$ is
observable at all. The answer is yes: this type of dephasing makes sense in a
clean Luttinger liquid in the {\it ring geometry}, where it governs the decay
rate $(\tau_\phi^{\rm AB})^{-1}$ of Aharonov-Bohm (AB)
oscillations.\cite{lehur02,gornyi05a,lehur05} Indeed, the amplitude
$G_{AB}$ of the first harmonic of the AB conductance is proportional to the
product of the Green's functions in two arms of the interferometer:
\begin{eqnarray}
G_{AB}&\propto&
\left\{{1\over
\Omega_n}\int_0^{1/T}\!d\tau\,e^{i\Omega_n\tau}
\right.\nonumber \\
&\times&\left.
g_+(L_1,\tau)g_+(-L_2,-\tau)\right\}_{\Omega_n\to +0}~,
\label{104}
\end{eqnarray}
where $L_{1,2}$ are the lengths of the arms of the ring. In the case of
$L_1=L_2\equiv L$, the damping is solely due to the dephasing and is given by
\begin{equation}
G_{AB}\propto \exp (-L/u\tau_{\rm ee})\equiv \exp (-L/u\tau_\phi^{\rm
AB})~,
\label{105}
\end{equation}
i.e., $\tau_\phi^{\rm AB}=\tau_{\rm ee}$. The difference from the WL
correction is in the absence of the vertex parts of the dephasing
action. Specifically, $S_{12}=S_{21}=0$ in the AB setup [where $S_{ij}$ is
given by Eq.~(\ref{93}) with $i,j$ equal to 1 or 2 for trajectories in arms 1
or 2, correspondingly], because of the absence of interaction between the 
arms, see Fig.~\ref{f12}.
On the other hand, the
self-energy part $S_{11}=S_{22}$ (analog of $S_{\rm ff}$ and $S_{\rm bb}$ in
the WL problem) is given by Eq.~(\ref{sff}) with $t_c=L/u$. An
analogous $T^{-1}$ dependence of $\tau_\phi^{\rm AB}$ was also obtained in
Ref.~\onlinecite{seelig01} for a quantum wire coupled capacitively to metallic
gates. The preexponential factor in Eq.~(\ref{104}) depends on details of how
the AB ring is connected to the leads. A particular AB setup---two Luttinger
liquids weakly coupled to each other at two points by tunneling---was
thoroughly studied in Ref.~\onlinecite{lehur05}. To summarize, the WL and AB
dephasing rates in a Luttinger liquid are parametrically different in the
spinless case studied in this paper; specifically, $\tau_\phi^{\rm AB}\ll
\tau_\phi^{\rm wl}$. In diffusive many-channel wires, an analogous difference
was shown in Ref.~\onlinecite{ludwig04}.

\section{Functional bosonization: All-in-one approach}
\label{VII}
\setcounter{equation}{0}


Equations (\ref{102}),(\ref{103}) have been derived in two steps, emphasizing
two key ingredients of the theory: the Luttinger-liquid renormalization of
disorder and the Fermi-liquid-like dephasing of electrons scattered by the
renormalized disorder. Although both effects are generated by \mbox{e-e}
interactions, they are of essentially different origin: the renormalization of
disorder is due to virtual processes with energy transfers in the range
between $T$ and $\Lambda$, whereas the dephasing stems from real processes
with energy transfers smaller than $T$. Conceptually, an advantage of the
two-step approach is that it reveals the transparent physical picture of
transport in the disordered Luttinger liquid. Alternatively, one can develop a
unified approach where the virtual and real transitions are treated on an
equal footing, which is what we do below. A fully bosonized approach, as well
as a purely fermionic one, turns out to be not particularly advantageous for
this purpose. Instead, in this paper we employ the method of so-called
``functional bosonization'', adjusted to deal with the particulars of e-e
scattering in one spatial dimension. The method was introduced for a clean
Luttinger liquid by Fogedby\cite{fogedby76} and by Lee and Chen.\cite{lee88} A
transparent exposition and further development of the method in 1d can be
found in recent reviews by Yurkevich\cite{yurkevich02} and Lerner and
Yurkevich.\cite{lerner05} For recent applications to single-impurity problems
see Refs.~\onlinecite{fernandez99,grishin04}.

In essence, the functional bosonization is a technique based on the
conventional Hubbard-Stratonovich transformation that decouples the
density-density interaction term in the fermionic action. Bosonic fields
appear in this approach as {\it auxiliary} fields of the Hubbard-Stratonovich
transformation, whose fluctuations mediate the dynamically screened
interaction between electrons. The peculiarity of a clean Luttinger liquid is
that the RPA approximation for the dynamically screened interaction is
exact.\cite{dzyaloshinskii74} This is a drastic simplification, so that the
ease in which an arbitrary strong interaction in the clean case can be
incorporated via the fully bosonized method is also the property of the
functional-bosonization approach. In the disordered case, the
functional-bosonization method has a technical advantage, especially useful
for weak interactions, in that the conductivity is represented (in contrast to
the fully bosonized approach) as a fermionic loop, albeit dressed by
interactions in all possible ways.

\subsection{Functional bosonization: Background}
\label{VIIa}

Our purpose here is to develop the functional-bosonization scheme in the
presence of disorder and to apply it for the calculation of the
weak-localization term (\ref{102}) in the conductivity. The quantity to begin
with is the fermionic Green's function. In the absence of disorder, we write
it in the Matsubara representation for an electron moving to the right (+) or
to the left (--) as 
\begin{equation} 
g_\mu (x,\tau)=g_\mu ^{(0)}(x,\tau)\,e^{-B_{\mu\mu}(x,\tau)}~, \quad\mu=\pm~,
\label{106} 
\end{equation} 
where 
\begin{equation} 
g_\pm^{(0)}(x,\tau)=\mp {iT\over 2v_F}\,{1\over \sinh [\pi T(x/v_F\pm i\tau)]}
\label{107} 
\end{equation} 
is the free Green's function and the factor $\exp (-B_{\mu\mu})$ accounts for
interactions between electrons. The function $B_{\mu\mu}$ can be found by
using the fact that, in the absence of backscattering due to e-e interaction,
the Hubbard-Stratonovich transformation characterized by a decoupling field
$\varphi(x,\tau)$ results in an identical representation of $g_\mu (x,\tau)$
as the average over $\varphi(x,\tau)$ of a function $\tilde{g}_\mu
(x,\tau)$,\cite{lee88,yurkevich02,lerner05}
\begin{equation}
g_\mu(x,\tau)=\left<\tilde{g}_\mu (x,\tau)\right>~,
\label{108}
\end{equation}
where $\tilde{g}_\mu (x,\tau)$ describes the propagation of a particle of the
same chirality in the field $\varphi(x,\tau)$: 
\begin{equation} 
[\,\partial_\tau-i\mu v_F\partial_x-i\varphi(x,\tau)\,]\,\tilde{g}_\mu
(x,\tau)=\delta(x)\delta(\tau)~.
\label{109}
\end{equation}
The angular brackets in Eq.~(\ref{108}) and throughout Sec.~\ref{VII} below
denote the functional averaging over the field $\varphi(x,\tau)$. 
The representation (\ref{108}) is equivalent to a local gauge transformation
for the fermionic field of the right-/left-mover\cite{lerner05}
\begin{equation}
\psi_\mu(x,\tau)\to \psi_\mu(x,\tau)\exp [\,i\theta_\mu(x,\tau)\,]~,
\label{110}
\end{equation}
where the bosonic field $\theta_\mu(x,\tau)$ is related to the decoupling
field $\varphi(x,\tau)$ by
\begin{equation} 
(\partial_\tau-i\mu v_F\partial_x)\theta_\mu(x,\tau)=\varphi (x,\tau)~.
\label{111} 
\end{equation}
The Gaussian character of the Luttinger-liquid theory in the clean case means
that the averaging in Eq.~(\ref{108}) is performed with an action in which all
terms containing powers of $\varphi(x,\tau)$ higher than second vanish to
zero. The correlator $\left<\varphi(x,\tau)\varphi(0,0)\right>=V(x,\tau)$
then gives the dynamically screened RPA interaction. 

As a result, the functions $B_{\mu\mu}(x,\tau)$ are represented as
propagators of the fields $\theta_\mu(x,\tau)$:
\begin{eqnarray}
B_{\mu\nu}(x,\tau)
=\left<\,[\,\theta_\mu (0,0)-\theta_\mu (x,\tau)\,]\,
\theta_\nu (0,0)\right>~,
\label{112}
\end{eqnarray}
where for later use we also introduce the correlator of
fields $\theta_\mu$ of different chirality, $\mu,\nu=\pm$.  
It is convenient to put the bare
coupling constant $g_4=0$, shift the Fermi velocity accordingly (as discussed
in Sec.\ref{IIIa}) and deal with two different interaction propagators
$V_{++}\neq V_{+-}$ (see Appendix~\ref{aA}). Then the correlators
$B_{\mu\nu}(x,\tau)$ read
\begin{eqnarray}
&&B_{+\pm}(x,\tau)=T\sum_n\int\! 
{dq\over 2\pi}\,\left(e^{iqx-i\Omega_n\tau}-1\right)\nonumber\\
&&\times\,\,{V_{+\pm}(i\Omega_n,q)\over
  (i\Omega_n-qv_F)(i\Omega_n\mp qv_F)}~,
\label{113}
\end{eqnarray}
where $\Omega_n=2\pi nT$ is the bosonic frequency,
$B_{--}(x,\tau)=B_{++}(-x,\tau)$, and $B_{-+}(x,\tau)=B_{+-}(x,\tau)$.  The
integration with $V_{\mu\nu}$ given by Eq.~(\ref{A11}) [or, equivalently,
by Eqs.~(\ref{A13}),(\ref{A14}) with a renormalized Fermi velocity, see
discussion in Sec.~\ref{IIIa}] yields
(see Appendix~\ref{app:dephasing}):
\begin{eqnarray}
B_{\mu\mu}(x,\tau)&=&-\ln\eta_\mu(x,\tau)-{\alpha_b\over 2}{\cal L}(x,\tau)~,
\label{115}\\
B_{\mu,-\mu}(x,\tau)&=&-{\alpha_r\over 2}{\cal L}(x,\tau)~,
\label{117}
\end{eqnarray}
where 
\be
\alpha_b=(1-K)^2/2K~,\quad \alpha_r=(1-K^2)/2K~,
\label{118}
\ee
and
\begin{eqnarray}
{\cal L}(x,\tau)=\ln{(\pi T/\Lambda)^2\over \sinh [\,\pi T(x/u+i\tau)\,]
\sinh [\,\pi T(x/u-i\tau)\,]}~.\nonumber \\
\label{119}
\end{eqnarray}
The functions
\begin{equation}
\eta_\pm(x,\tau)={v_F\over u}{\sinh[\pi T(x/v_F\pm i\tau)]\over 
\sinh [\pi T(x/u\pm
i\tau)]}
\label{120}
\end{equation}
replace $v_F\to u$ in the free Green's functions
$g_\pm^{(0)}(x,\tau)$ in Eq.~(\ref{106}).

\subsection{Functional bosonization: Diagrammatic technique}
\label{VIIb}

\begin{figure}[ht] 
\centerline{
\includegraphics[width=8cm]{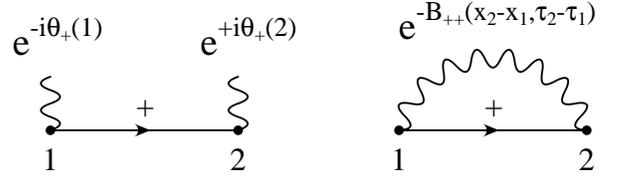}
}
\caption{
(a) The Green's function of a right-mover propagating between points
$1=(x_1,\tau_1)$ and $2=(x_2,\tau_2)$ in a given field $\varphi(x,\tau)$
acquires two phase factors $\exp [i\theta_+(2)]$ and $\exp [-i\theta_+(1)]$
(denoted as wavy lines) at the end points. Solid line: the free Green's
function $g^{(0)}_+(x_2-x_1,\tau_2-\tau_1)$. (b) The wavy line connecting the
points 1 and 2 denotes averaging of the phase factors over fluctuations of
$\theta_+(1)$ and $\theta_+(2)$. 
 } 
\label{f13} 
\end{figure}

The exactness of the RPA approximation in 1d suggests the following
diagrammatic technique. The Green's function (\ref{106}) for a right-mover
propagating from point $1=(x_1,\tau_1)$ to point $2=(x_2,\tau_2)$ can be
represented by the diagram in Fig.~\ref{f13}, where the solid line is
$g_+^{(0)}(x_2-x_1,\tau_2-\tau_1)$ and the wavy line connecting the end points
denotes the pairing of two fluctuating factors $\exp [i \theta_+(2)]$ and
$\exp [-i\theta_+(1)]$ with the Gaussian fields $\theta_+(2)$ and
$\theta_+(1)$:
\begin{eqnarray}
&&\left<\,\exp
\{\,i[\,\theta_+(2)-\theta_+(1)\,]\,\}\,\right>\nonumber \\
&&=\exp \,[\,-B_{++}(x_2-x_1,\tau_2-\tau_1)\,]~.
\label{121}
\end{eqnarray}
The fields $\theta_+(2)$ and $\theta_+(1)$ enter the exponent of
Eq.~(\ref{121}) with different signs, according to the direction of the arrow
on the solid line.

\begin{figure}[ht] 
\centerline{
\includegraphics[width=8cm]{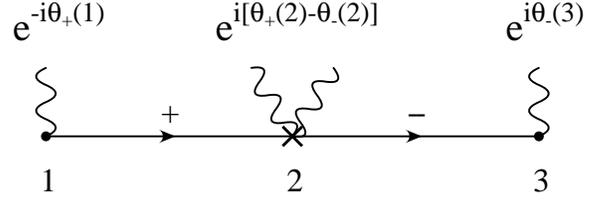}
}
\caption{
The Green's function with a backscattering vertex (marked by a cross) between 
the end points. The vertex is dressed by a local fluctuating factor $\exp
\{i[\theta_+(2)-\theta_-(2)]\}$.  
}
\label{f14} 
\end{figure}

\begin{figure}[ht] 
\centerline{
\includegraphics[width=8cm]{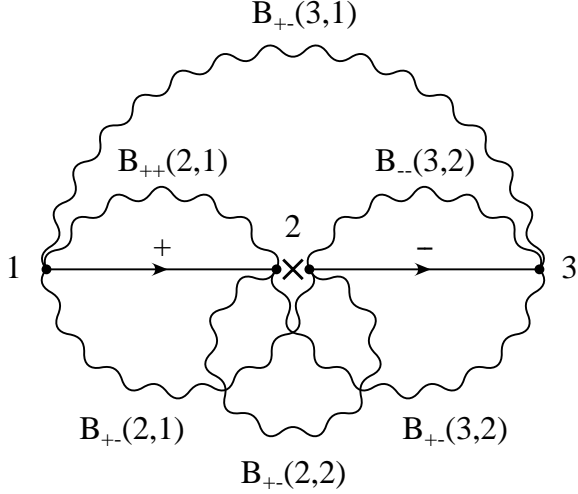}
}
\caption{
Illustration of the diagrammatic technique with the propagators $B_{\mu\nu}$:
the correction $\delta g(x_3,x_1,\tau_3-\tau_1)$ [Eq.~(\ref{122})] to the
Green's function due to the backscattering at point 2. The wavy lines denote
averaging of the fluctuating factors $\exp (\pm\theta_\mu)$ shown in
Fig.~\ref{f14}. The pairing of all the fields $\theta_\mu$ with each other
according to Eq.~(\ref{112}) yields the corresponding contributions
$B_{\mu\nu}$ to the factor (\ref{123}). For ease of visualization, the ends of
the free Green's functions are split off from the backscattering vertex at
point 2.
}
\label{f15} 
\end{figure}

Consider now a backscattering vertex which describes reflection of the
right-mover from an impurity (Figs.~\ref{f14} and \ref{f15}, before and after
averaging over the fluctuations of $\theta_\mu$, respectively). The Green's
function propagating from point 1 to point 3 upon backscattering on the static
potential $U_b(x)$ at point 2 is given by
\begin{eqnarray}
&&\delta g(x_3,x_1,\tau_3-\tau_1)
=\int_0^{1/T} \!\!\!\!\!d\tau_2\int dx_2 \,\, U_b(x_2)\nonumber \\
&\times& g_+^{(0)}(x_2-x_1,\tau_2-\tau_1)g_-^{(0)}(x_3-x_2,\tau_3-\tau_2)
\nonumber \\
&\times& 
\left<\,\exp
\{\,i[\,\theta_+(2)-\theta_+(1)+\theta_-(3)-\theta_-(2)\,]\,\}\,\right>~.
\nonumber
\\ 
\label{122}
\end{eqnarray}

A direct consequence of Eq.~(\ref{110}) is that the vertex at point 2 is
dressed by the local fluctuating factor $\exp \{i[\theta_+(2)-\theta_-(2)]\}$.
The correlations of the fields $\theta_\mu$ at different points in $x$ space
in the last factor of Eq.~(\ref{122}) account for e-e interaction. Apart from
the averages (\ref{121}) of the fields of the same chirality, there appear 
averages of the type
\begin{eqnarray}
&&\left<\,\exp
\{\,i[\,\theta_+(2)-\theta_-(1)\,]\,\}\,\right>\nonumber \\
&&=\exp \,[\,-B_{+-}(x_2-x_1,\tau_2-\tau_1)-M_0\,]~,
\label{122a}
\end{eqnarray}
where 
\begin{equation}
M_0=\left<\theta_+^2(0,0)\right>-\left<\theta_+(0,0)\theta_-(0,0)\right>~.
\label{122b}
\end{equation}
Note that the factors $\exp (\pm M_0)$ cancel out in any closed fermionic
loop, so that in the diagrammatic technique for observables one can omit them
from the very beginning.  The Gaussian character of the correlations means
that the exponent of the last factor in Eq.~(\ref{122}) is represented as a 
sum of the correlators $B_{\mu\nu}$:
\begin{eqnarray}
&&\left<\,\exp
\{\,i[\,\theta_+(2)-\theta_+(1)+\theta_-(3)
-\theta_-(2)\,]\,\}\,\right>\nonumber\\
&&=\exp \,[\,-B_{++}(x_2-x_1,\tau_2-\tau_1)\nonumber \\
&&+B_{+-}(x_2-x_1,\tau_2-\tau_1)
-B_{--}(x_3-x_2,\tau_3-\tau_2)\nonumber \\
&&+B_{+-}(x_3-x_2,\tau_3-\tau_2)
-B_{+-}(x_3-x_1,\tau_3-\tau_1)\,]~,\nonumber \\
\label{123}
\end{eqnarray}
as illustrated in Fig.~\ref{f15} [we used in Eq.~(\ref{123}) the fact that
$B_{+-}(2,2)=0$].

The appearance of the combination $\theta_+(2)-\theta_-(2)$ for backscattering
at point 2 in Eq.~(\ref{122}) and the factorization of the correlations
(\ref{123}) in the simplest case of a single backscattering vertex suggests a
way to proceed to two or more vertices. In effect, the e-e interaction can be
completely gauged out to the impurity vertices, as emphasized in
Ref.~\onlinecite{lerner05}. Specifically, one can, in principle, calculate
observables (given by closed fermionic loops) perturbatively to any order in
the strength of disorder and exactly in the strength of e-e interaction by
attaching a factor
\begin{equation}
\exp \,\{\,\pm i\,[\,\theta_+(N)-\theta_-(N)\,]\,\}
\label{124}
\end{equation}
to each backscattering vertex, where $N$ labels the vertex, and pairing all
the fields $\theta_\mu(N)$ with each other. The averaging over the
fluctuations of $\theta_\mu(N)$ is then performed with the correlators
(\ref{115})--(\ref{117}) calculated for a homogeneous system. The sign $\pm$
in Eq.~(\ref{124}) should be chosen to correspond to the chirality with which
an electron is incident on the impurity.

The WL correction to the conductivity is given by the fermionic loop with 6
backscattering vertices (Fig.~\ref{f5}), each of which should be dressed with
the factors (\ref{124}).  However, taking disorder into account at this level
would not be sufficient. As we showed in Sec.~\ref{VI}, in order to calculate
the effect of dephasing for spinless electrons, disorder should also be
included, to first order in the strength of disorder $\gamma$, in the loops
that make up the RPA series for the effective interaction.  We recall that in
the RPA approximation, substantiated in Sec.~\ref{VI}, no more than two
external interaction lines connect to each of the fermionic loops that
constitute the interaction propagator and the averaging of the interaction
propagator over disorder is performed in such a way that no impurity lines
connect different fermionic loops. As we show in Appendix~\ref{aD}, in the
language of the functional bosonization, the RPA approximation in the presence
of disorder translates into a disorder-induced renormalization of the
correlators $B_{\mu\nu}(x,\tau)$. Specifically, the renormalization consists
in adding a disorder-induced damping to the propagators $V_{++}$ and $V_{+-}$
in Eq.~(\ref{113}). The renormalized interaction propagators are
derived in Appendix~\ref{aA} [see Eqs.~(\ref{A32})--(\ref{A33})] and are given
in the Matsubara representation by
\begin{eqnarray}
V_{++}(i\Omega_n,q)&=&-{g_2^2\over 2\pi v_F}\,{q^2v_F^2-iqv_F\Omega_n
+|\Omega_n|\gamma/2\over
  q^2u^2+\Omega_n^2+|\Omega_n|(\gamma+\delta\gamma)}~,\nonumber\\ 
\label{125}\\
V_{+-}(i\Omega_n,q)&=&g_2\,{q^2v_F^2+\Omega_n^2
+|\Omega_n|(\gamma+\delta\gamma/2)\over
  q^2u^2+\Omega_n^2+|\Omega_n|(\gamma+\delta\gamma)}~,\nonumber\\
\label{126}
\end{eqnarray}
where $\gamma$ denotes the transport scattering rate, Eq.~(\ref{A20}), and
$\delta\gamma$ is given by Eq.~(\ref{A34}) (at variance with the preceding
part of the paper, in this section we do not use the notation $\tau$ for the
transport scattering time in order to avoid a possible confusion with the
Matsubara time).  

\begin{figure}[ht] 
\centerline{
\includegraphics[width=7cm]{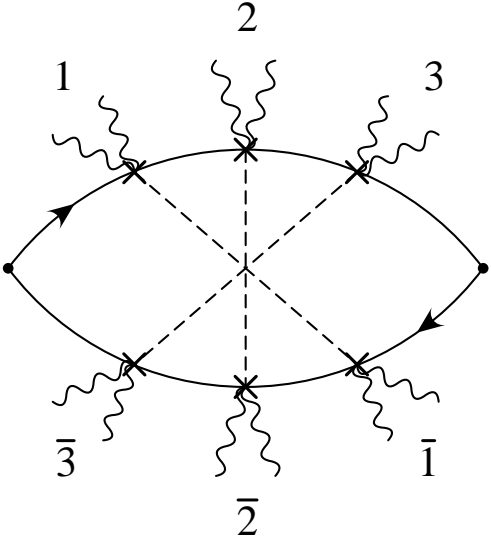}
}
\caption{
The three-impurity Cooperon diagram with interaction effects encoded in the
fluctuating factors $\exp (\pm\theta_\mu)$ (denoted by the wavy lines)
attached to the backscattering vertices. Each impurity vertex $N=(x_N,\tau_N)$
is characterized by two fluctuating fields $\theta_\pm(N)$, and similarly for
$\bar{N}=(x_N,\bar{\tau}_N)$. Pairing of all the fields with each other 
(similarly to Figs.~\ref{f14},\ref{f15}) describes both the elastic 
renormalization of the three impurities and the dephasing of the Cooperon.
}\label{f16} 
\end{figure}

Thus, our strategy in deriving Eq.~(\ref{102}) via the functional bosonization
is this: find the disorder-renormalized correlators $B_{\mu\nu}(x,\tau)$ to
order ${\cal O}(\gamma g_2^2)$ and use them to average over fluctuations of
the fields $\theta_\mu$ ``attached'' to 6 impurity vertices of the 3-impurity
Cooperon (Fig.~\ref{f16}). The diagrams (b) and (c) in Fig.~\ref{f5} are
calculated in the same way.

\subsection{Functional bosonization with a disorder-renormalized RPA}
\label{VIIc}

We now turn to the calculation of the effective (disorder-renormalized)
propagators $B_{\mu\nu}(x,\tau)$ to order ${\cal O}(\gamma g_2^2)$ by using
Eq.~(\ref{113}) and the interaction propagators $V_{++}$ and
$V_{+-}$ given by Eqs.~(\ref{125}),(\ref{126}). At second order in
interaction, $B_{++}(x,\tau)$ reads
\begin{eqnarray}
&&B_{++}(x,\tau)\simeq -{g_2^2\over 2\pi v_F}\,T\,
\sum_n\int \!{dq\over 2\pi}\,\left(e^{iqx-i\Omega_n\tau}-1\right)\nonumber
\\
&&\times\,\,{q^2v_F^2-iqv_F\Omega_n+|\Omega_n|\gamma/2\over
  (-i\Omega_n+qv_F)^2\,(q^2v_F^2+\Omega_n^2+|\Omega_n|\gamma)}~.
\label{127}
\end{eqnarray}
Let us represent $B_{++}$ as a sum $B_{++}={\cal B}_0+{\cal B}_>+{\cal B}_<$,
where ${\cal B}_0(x)$ is the contribution of $\Omega_n=0$, ${\cal
B}_>(x,\tau)$ and ${\cal B}_<(x,\tau)={\cal B}_>(-x,-\tau)$ are the
contributions of $\Omega_n>0$ and $\Omega_n<0$, respectively. In turn, ${\cal
B}_>(x,\tau)$ is a sum of two terms:
\begin{equation}
{\cal B}_>={\cal B}_>^{dp}+{\cal B}_>^{sp}~,
\label{128}
\end{equation}
where ${\cal B}_>^{dp}$ is the contribution of the double pole at
$qv_F=i\Omega_n$ and ${\cal B}_>^{sp}$ is the contribution of the single
poles at $qv_F=\pm i(\Omega_n^2+\Omega_n\gamma)^{1/2}$.

The term coming from zero frequency does not depend on $\gamma$:
\bea
{\cal B}_0(x)=-{g_2^2\over 2\pi v_F}T\int \!{dq\over
2\pi}\, {e^{iqx}-1\over q^2v_F^2}=\alpha^2{\pi T|x|\over v_F}~,
\label{129} 
\eea
where $\alpha=g_2/2\pi v_F$, as before. The contribution of the double pole
does not depend on $\gamma$ either,
\begin{eqnarray}
{\cal B}_>^{dp}(x,\tau)&=&\alpha^2
{\pi Tx\over v_F}\,\Theta(x)
\nonumber \\
&\times&{1\over \exp\,
    [\,2\pi T(x/v_F +i\tau)\,]-1}~,
\label{130}
\end{eqnarray}
and is an expansion of $\ln\eta_+$ (\ref{120}) in powers of $\alpha$.
Expanding the single-pole contribution to first order in $\gamma$, we have
\begin{eqnarray}
{\cal B}_>^{sp}(x,\tau)&\simeq& -
{\alpha^2\over 4}\left\{\ln{2\pi T/\Lambda\over 1-\exp [-2\pi
      T(|x|/v_F+i\tau)]}\right. \nonumber \\
&-& 
{\gamma |x|\over 2v_F}\ln {1\over 1-\exp [-2\pi
      T(|x|/v_F+i\tau)]}
\nonumber \\
&-&\left. 
{\gamma\over 4\pi T} F\,\left[\,2\pi T\left({|x|\over v_F}+i\tau\right)\,
\right]
\right\}~,
\label{131}
\end{eqnarray}
where 
\begin{equation}
F(z)=\sum_{n=1}^\infty {\exp (-nz) -1\over n^2}~.
\label{132}
\end{equation}
The sum (\ref{132}) is defined for ${\rm Re}\,z>0$. At small $|z|\ll 1$, it
behaves as $F(z)\simeq z\ln |z|$, which cancels the second term in
Eq.~(\ref{131}) at the ultraviolet cutoff $\tau\to 0,\,|x|\to
1/v_F\Lambda$. Below, we will need the analytical continuation of
$B_{\mu\nu}(x,\tau)$ in the complex plane of $\tau$. In particular, an
important contribution to the dephasing will come from the analytical
continuation of $F(z)$ to large negative ${\rm Re}\,z<0$. The latter can be
done, e.g., by means of the integral representation
\begin{equation}
F(z)=\int_0^z\!dz'\, \ln [\,1-\exp (-z')\,]~.
\label{133}
\end{equation}

Piecing all the terms together, $B_{++}(x,\tau)$ with a disorder-induced
correction of order ${\cal O}(\gamma g_2^2)$ is written as
\begin{equation}
B_{++}=B_{++}|_{\gamma=0}\,+b~,
\label{134}
\end{equation}
where $B_{++}|_{\gamma=0}$ is given by Eq.~(\ref{115}), $b=b_1+b_2$, and
\begin{eqnarray}
&&b_1(x,\tau)\simeq\alpha^2{\gamma\over 16\pi T}\left\{ \left({2\pi T x\over
  v_F}\right)^2 +{2\pi T|x|\over v_F}\right.\nonumber \\
&\times&\left.\ln{1\over 4\sinh [\,\pi T(x/v_F+i\tau)\,]
\sinh [\,\pi T(x/v_F-i\tau)\,]}\right\}~,\nonumber \\ 
\label{135}
&&b_2(x,\tau)\simeq \alpha^2{\gamma\over 16\pi T}
\left\{
F\left[2\pi T\left({|x|\over v_F}+i\tau\right)\right]\right.\nonumber \\&&
\left.\hspace{1.2cm}+\,
F\left[2\pi T\left({|x|\over v_F}-i\tau\right)\right]\right\}~. 
\label{136}
\end{eqnarray}
Note that the second term in Eq.~(\ref{135}) can be viewed as an $x$
dependent renormalization of $\alpha$ induced by disorder,
$\delta\alpha=-\alpha\gamma |x|/4v_F$, as follows from a comparison with
Eq.~(\ref{115}). This is consistent with the RG flow of the interaction
constant (\ref{60}). However, Eqs.~(\ref{135}),(\ref{136}) contain more
information and allow us to study the dephasing missed in the RG equations.

Consider now the disorder-renormalized propagator $B_{+-}(x,\tau)$. To
second order in $g_2$,
\begin{eqnarray}
&&B_{+-}(x,\tau)\simeq -T\, \sum_n\int \!{dq\over 2\pi}\,
{e^{iqx-i\Omega_n\tau}-1\over \Omega_n^2+q^2v_F^2}
\nonumber \\
&&\times\,\,\left(\, g_2 + {g_2^2\over 4\pi
v_F}{|\Omega_n|\gamma\over
q^2v_F^2+\Omega_n^2+|\Omega_n|\gamma}\right)~.
\label{137}
\end{eqnarray}
Expanding Eq.~(\ref{137}) to linear order in $\gamma$, we readily obtain a
correction to $B_{+-}(x,\tau)$ [Eq.~(\ref{117})] which is again given by
Eqs.~(\ref{135}),(\ref{136}) but with the opposite sign, so that finally
\begin{equation}
B_{+\pm}=B_{+\pm}|_{\gamma=0}\,\pm b~.
\label{138}
\end{equation}

\subsection{Cooperon in a dynamic environment}
\label{VIId}

\begin{figure}[ht] 
\centerline{\includegraphics[width=7cm]{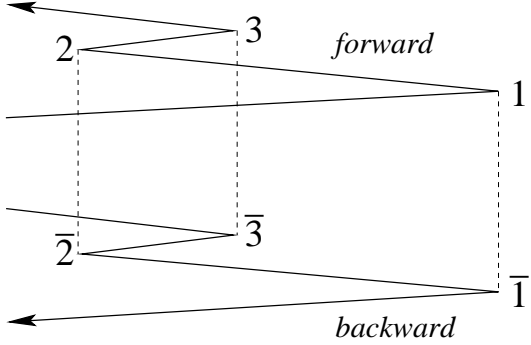}}
\caption{
Two time-reversed paths whose interference constitutes the three-impurity
Cooperon. There are three backscatterings off impurities on each of the paths
(the positions of the impurities are marked by the dashed lines). The
one-dimensional paths are stretched vertically for ease of visualization,
with the ``time axis" directed upwards for the forward path and downwards for
the backward path. Note that this representation of the paths is identical to
the graphs in the Berezinskii technique.\cite{berezinskii73} }
\label{f17} 
\end{figure}

Now that we have found the propagators $B_{\mu\nu}(x,\tau)$, let us turn to
the calculation of the WL correction to the conductivity. The interaction
induces the factor
\begin{equation}
\exp (-S_C)=\left<\exp [\,i(\theta_f-\theta_b)\,]\right> 
\label{139}
\end{equation}
to the Cooperon loop, where $\theta_f$ and $\theta_b$ are the phases
accumulated by an electron propagating along the ``forward'' and ``backward''
paths (Fig.~\ref{f17} represents the paths from Fig.~\ref{f6}a
in a way which is more convenient here) and the averaging is performed with
the disorder-renormalized correlators (\ref{138}). Again, $S_C$ accounts for
both the dephasing and the elastic renormalization of impurities. The factor
(\ref{139}) is the same for all 3 diagrams for $\Delta\sigma_{\rm wl}$ in
Fig.~\ref{f5}.

The phases $\theta_{f,b}$ are given by
\begin{eqnarray}
\theta_f&=&\sum_{N=1}^3(-1)^{N+1}\,[\,\theta_+(N)-\theta_-(N)\,]~,
\label{140}\\
\theta_b&=&\sum_{\bar{N}=1}^3(-1)^{\bar{N}+1}\,[\,\theta_+(\bar{N})-\theta_
-(\bar{N})\,]~,
\label{141}
\end{eqnarray}
where $N,\bar{N}$ are the impurity-vertex numbers, for the forward and
backward paths, respectively; as shown in Fig.~\ref{f6}. The exponent in
Eq.~(\ref{139}) is represented as
\begin{equation}
S_C=-2(M_{ff}-M_{fb})~,
\label{142}
\end{equation}
where 
\begin{eqnarray}
M_{ff}&=&-{1\over 2}\left<\theta_f^2\right>\nonumber \\
&=&{1\over
2}\sum_{NN'}(-1)^{N+N'}M(N,N') - M_0~, 
\label{143}\\
M_{fb}&=&-{1\over 2}\left<\theta_f\theta_b\right>\nonumber \\
&=&{1\over
2}\sum_{N\bar{N}}(-1)^{N+\bar{N}}M(N,\bar{N})-M_0~, 
\label{144}
\end{eqnarray}
and all the pairings between the phases associated with impurities $N$ and
$N'$ are given by the combination
\begin{eqnarray}
M(N,N')&=&B_{++}(x_N-x_{N'},\tau_N-\tau_{N'})\nonumber \\
&+&B_{--}(x_N-x_{N'},\tau_N-\tau_{N'})\nonumber \\
&-&2B_{+-}(x_N-x_{N'},\tau_N-\tau_{N'})~,
\label{145}
\end{eqnarray}
and similarly for the pairings between $N$ and $\bar N$. The constant $M_0$
given by Eq.~(\ref{122b}) cancels out in $S_C$. Note that on the closed
contour $\left<\theta_f^2\right>=\left<\theta_b^2\right>$, which is accounted
for by the factor of 2 in Eq.~(\ref{142}).

\subsection{Diffuson ladder and renormalization of disorder}
\label{VIIe}

Before calculating the three-impurity Cooperon, it is instructive to look at
the diffuson diagrams. More specifically, let us consider first the ladder
diagrams without including self-energy corrections.

\subsubsection{One impurity line}
\label{VIIe1}

\begin{figure}[ht] 
\centerline{
\includegraphics[width=7cm]{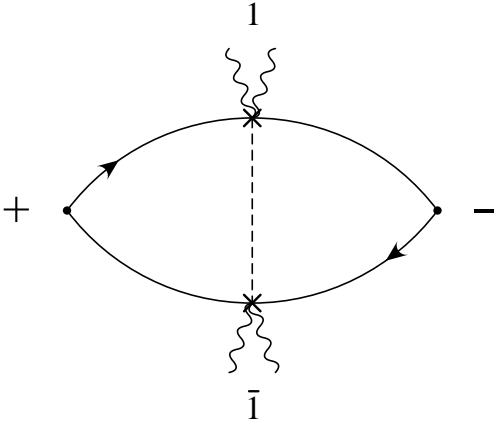}
}
\caption{
Density-density correlator of first order in the impurity strength. The
backscattering vertices are marked by the crosses. Pairing of the fluctuating
factors $\exp(\pm \theta_\mu)$ (denoted by the wavy lines) with each other
yields the factor $\exp [-M(1,\bar{1})]$ in Eq.~(\ref{146}), which
describes the renormalization of the impurity strength by the interaction.
}
\label{f18} 
\end{figure}

The simplest diagram for the density-density correlator of order ${\cal
O}(\gamma)$ is shown in Fig.~\ref{f18} and written as
\begin{eqnarray}
d_1&=&{v_F\gamma_0\over 2}
\int_0^{1/T}\!\!\!d\tau_1\int_0^{1/T}\!\!\!d\bar{\tau}_1\,\,
\nonumber \\
&\times&
g_+^{(0)}(x_1,\tau_1)g_-^{(0)}(x-x_1,\tau-\tau_1)\,\exp
[-M(1,\bar{1})]\nonumber\\
&\times&g_+^{(0)}(-x_1,-\bar{\tau}_1)g_-^{(0)}(x_1-x,\bar{\tau}_1-\tau)~.
\label{146}
\end{eqnarray}
The dashed line gives the bare strength of backscattering $v_F\gamma_0/2$. In
Eq.~(\ref{146}) we only integrate over internal time variables, keeping other
coordinates in $(x,\tau)$ space fixed. The factor $\exp [-M(1,\bar{1})]$,
which is taken at the same spatial point, depends on $\tau_1-\bar{\tau}_1$
only:
\begin{eqnarray}
&&\exp \,[-M(1,\bar{1})\,]=\left\{\left({\Lambda\over \pi T}\right)^2\sin
\left[\pi T \left(\tau_1-\bar{\tau}_1+{i\over
\Lambda}\right)\right]\right.\nonumber \\ &&\left.\hspace{1cm}  
\times\,\,\sin \left[\pi T \left(\tau_1-\bar{\tau}_1-{i\over
\Lambda}\right)\right]\right\}^{\alpha_e}~,
\label{147}
\end{eqnarray}
where we used the ultraviolet cutoff from Eq.~(\ref{D14a}) and
\begin{equation}
\alpha_e=\alpha_r-\alpha_b=1-K~.
\label{147a}
\end{equation} 
Note that the interaction-induced phase factors cancel out at the external
vertices which do not change chirality. In particular, this means that in
Eq.~(\ref{146}) there is no renormalization of the velocity by the factors
(\ref{120}). The latter is a general property of fermionic loops with a larger
number of impurities: the Green's functions connecting to the external
vertices with a small momentum transfer are characterized by the bare Fermi
velocity $v_F$.

\begin{figure}[ht] 
\centerline{\includegraphics[width=8cm]{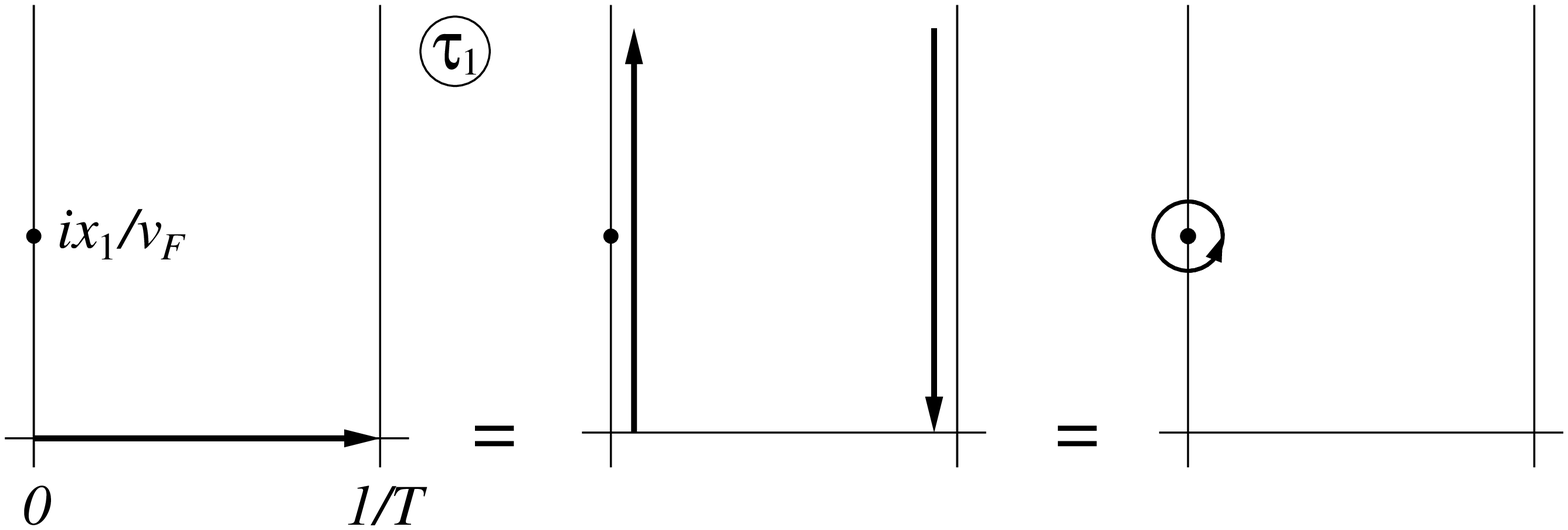}}
\caption{
Analytical structure of the integrand in Eq.~(\ref{149}), with a pole at
$\tau_1=ix_1/v_F$, for $x_1>0$ and $x<x_1$. Closing the contour of integration
upwards, the integral along the real axis of the Matsubara time is represented
as a sum of two integrals along the imaginary axis at $\tau_1=+0$ and
$\tau_1=1/T-0$. Using the periodicity with respect to the shift
$\tau_1\to\tau_1\pm 1/T$, the sum gives a residue at the pole.}
\label{f19} 
\end{figure}

In the absence of e-e interaction, $d_1$ is equal to 
\begin{equation}
d_1^{(0)}={v_F\gamma_0\over 2}I^2~,
\label{148}
\end{equation}
where
\begin{eqnarray}
&&\hspace{-1cm}I=\int_0^{1/T}\!\!\!d\tau_1\,
g_+^{(0)}(x_1,\tau_1)g_-^{(0)}(x-x_1,\tau-\tau_1)\nonumber \\ 
&&\hspace{-1cm}
=\,-{T\over 2v_F^2}\,{1\over \sinh\{\pi T [\,(2x_1-x)/v_F +i\tau\,]\}}
\nonumber \\ 
&&\hspace{-1cm}
\times\,\,[\,\Theta(x_1)\Theta(x_1-x)-\Theta(-x_1)\Theta(x-x_1)\,]~.
\label{149}
\end{eqnarray}
What is important to us here is that $I$ is given by the pole on the {\it
classical} trajectory $t_1=x_1/v_F$ of a right-moving electron (for $x_1>0$)
or a left-moving hole (for $x_1<0$), where $t_1=-i\tau_1$ (Fig.~\ref{f19}). If
we proceed to the $\Omega$ representation, the diagram for nonzero $x$ and
$x_1$ is written as
\begin{eqnarray}
&&\hspace{-3mm}\int_0^{1/T}\!\!\!d\tau \,e^{i\Omega_n\tau} d_1^{(0)}
={\gamma_0\over 4\pi v_F^3}|\Omega_n|e^{-|\Omega_n(2x_1-x)|}\nonumber\\
&&\hspace{-3mm}\times\,[\,\Theta(x_1)\Theta(x_1-x)\Theta(\Omega_n)+
\Theta(-x_1)\Theta(x-x_1)\Theta(-\Omega_n)\,]\nonumber 
\end{eqnarray}
and again is given by the (double) pole on the {\it classical} trajectory
$t=(2x_1-x)/v_F$, where $t=-i\tau$ and $|2x_1-x|$ is the total length of the
trajectory from point $x=0$ to point $x$ with backscattering at point $x_1$.
This simple consideration suggests that the two-particle correlators in the
presence of a weak e-e interaction should still be determined by a close
vicinity of the classical trajectory.

\begin{figure}[ht] 
\centerline{\includegraphics[width=8cm]{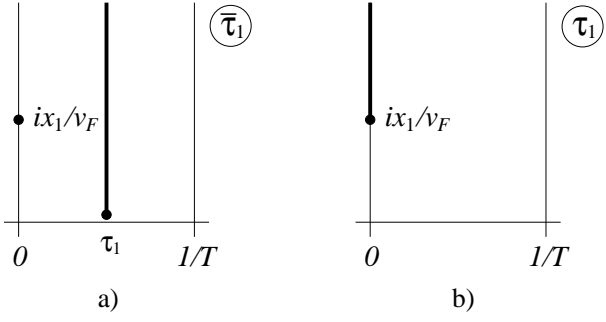}}
\caption{
Analytical structure of the integrand in Eq.~(\ref{146}). (a) The contour of
integration over $\bar{\tau}_1$ is closed upwards where the singularities of
the integrand for $x_1>0$ and $x<x_1$ are a pole at $\bar{\tau}_1=ix_1/v_F$ and
a branch cut which starts at $\bar{\tau}_1=\tau_1+i/\Lambda$ and runs
upwards. (b) The integral over $\tau_1$ of the contribution of the pole in
Fig.~\ref{f20}a is given exactly by the branch cut starting at
$\tau_1=ix_1/v_F$ and running upwards along the imaginary axis of $\tau_1$.
}
\label{f20} 
\end{figure}

To evaluate the integrals (\ref{146}) at $\alpha_e\neq 0$, we assume for
definiteness that $x_1>0$ and $x<x_1$. The singularities of the integrand of
Eq.~(\ref{146}) in the upper half-plane of $\bar{\tau}_1$ for $-0<{\rm
Re}\,\bar{\tau}_1<1/T-0$ are then a pole at $\bar{\tau}_1=ix_1/v_F$ and a
branch cut which comes from the factor (\ref{147}). The branch cut starts at
$\bar{\tau}_1=\tau_1+i/\Lambda$ and is sent upwards, as shown in
Fig.~\ref{f20}a. The contour of integration over $\bar{\tau}_1$ can be closed
upwards, so that the integral over $\bar{\tau}_1$ along the real axis from
$\bar{\tau}_1=0$ to $\bar{\tau}_1=1/T$ is given exactly by the contributions
of the pole and the branch cut. The contribution of the latter is proportional
to $\alpha_e$ and can be omitted in the limit of weak interaction, after which
the remaining integral over $\tau_1$ reads
\begin{eqnarray}
d_1&\simeq &{\gamma_0\over 2}\left({T\over 2v_F}\right)^3{(\Lambda/\pi
  T)^{2\alpha_e}\over \sinh\{\pi T[\,(2x_1-x)/v_F+i\tau\,]\}}\nonumber \\
&\times&\int_C\!d\tau_1 \,{\sin^{-1+2\alpha_e}[\,\pi T(\tau_1-ix_1/v_F)\,]\over
\sin\{\pi T [\,\tau_1-\tau +i(x_1-x)/v_F\,]\}}~.\nonumber \\
\label{151}
\end{eqnarray}
Here the contour of integration $C$ in the complex plane of $\tau_1$ runs
anticlockwise around the branch cut along the imaginary axis of $\tau_1$,
Fig.~\ref{f20}b. For $|\alpha_e|\ll 1$ the singularity in Eq.~(\ref{151}) is
almost a pole on the classical trajectory $t_1=x_1/v_F$ with $t_1=-i\tau_1$,
so that, indeed, the system dynamics is determined by a close vicinity of the
classical trajectory. We observe, however, that taking the
interaction-dependent factor in the integrand straight at $t_1\to x_1/v_F$
yields an ultraviolet singularity that exactly cancels the factor
$(\Lambda/\pi T)^{2\alpha_e}$ in front of the integral in Eq.~(\ref{151}),
which returns us to the noninteracting result. It is thus the integration in a
close vicinity $t_1-x_1/v_F\alt 1/T$ of the classical trajectory along the
branch cut that gives the renormalization of disorder by interaction.  If
$|\alpha_e|\ll 1$, Eq.~(\ref{151}) reduces to
\begin{equation}
d_1\simeq d_1^{(0)}\left(\Lambda/
\pi T\right)^{2\alpha_e}~.
\label{152}
\end{equation}
The difference between Eq.~(\ref{152}) and the noninteracting result consists
solely in the last factor which describes the renormalization of the
disorder strength $\gamma_0\to \gamma$, where 
\begin{equation}
\gamma=\gamma_0\left(\Lambda/\pi T\right)^{2\alpha_e}~.
\label{153}
\end{equation}

\subsubsection{``Two-impurity diffuson"}
\label{VIIe2}

\begin{figure}[ht] 
\centerline{
\includegraphics[width=7cm]{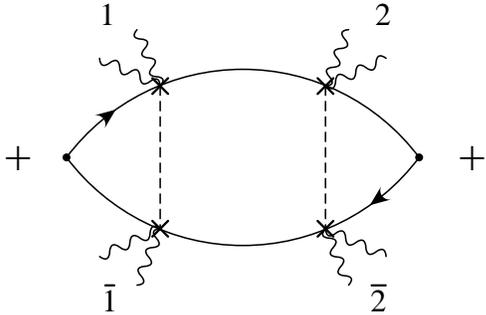}
}
\caption{``Two-impurity" diffuson with the fluctuating factors 
$\exp [\pm\theta_\mu (N)\,]$ (denoted by the wavy lines) attached to the
backscattering vertices (marked by the crosses).}
\label{f21} 
\end{figure}

Before generalizing to the Cooperon diagram, it is useful to consider a
diagram of order ${\cal O}(\gamma^2)$ with two parallel impurity lines,
Fig.~\ref{f21}. What is new in this ``two-impurity diffuson'' as compared to
Eq.~(\ref{146}) is that now there are factors $\exp (\pm M)$ which connect
different impurities. Specifically, the interaction-induced factor is
\begin{eqnarray}
&&\hspace{-1cm}\exp\,[\,-M(1,\bar{1})-M(2,\bar{2})\nonumber \\
&&\hspace{-1cm}-M(1,2)-M(\bar{1},\bar{2})+M(1,\bar{2})+M(\bar{1},2)\,]~.
\label{154}
\end{eqnarray}
Each of the cross-terms (the last 4 terms above), when taken on the classical
trajectory, may give a large contribution to the exponent of
Eq.~(\ref{154}). An important observation, however, is that on the classical
trajectory they cancel each other exactly. This can be seen immediately since
on the classical trajectory $\tau_1=\bar{\tau}_1$ and at this point
$M(1,2)=M(\bar{1},2)$ and $M(\bar{1},\bar{2})=M(1,\bar{2})$. The cancellation
can be verified for diffuson diagrams of higher order in $\gamma$ and is a
manifestation of the mere fact that for the density-density correlator there
is no dephasing in the diffuson channel. No dephasing is generated both for
the bare correlators $B_{\mu\nu}(x,\tau)$ and for those renormalized by
disorder at the level of RPA.

\begin{figure}[ht] 
\centerline{\includegraphics[width=8.5cm]{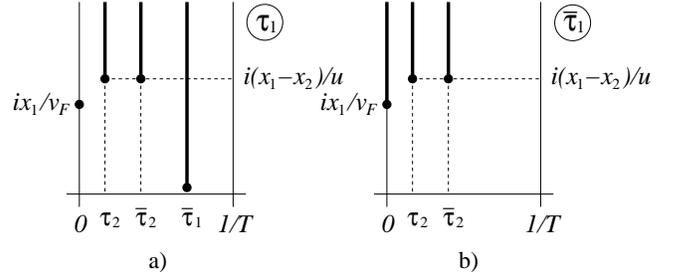}}
\caption{Integration over the internal times in
the two-impurity diffuson shown in Fig.~\ref{f21}. (a) The analytical
structure of the integral in the complex plane of $\tau_1$: the contour of
integration runs around a pole and three
vertical branch cuts. (b) The contour of integration in the complex plane of
$\bar{\tau}_1$ runs around three vertical branch cuts.  }
\label{f22} 
\end{figure}

As far as the integration around the classical trajectory of the two-impurity
diffuson is concerned, it suffices for our purposes here to analyze the case
of weak interaction. The analytical structure of the integral over four time
variables, $\tau_1$, $\tau_2$, $\bar{\tau}_1$, and $\bar{\tau}_2$ can be
simplified by selecting the singularities that are strongest in the limit
$|\alpha_e|\ll 1$. Let us assume for definiteness that $x_1>0$ and
$x_2<x_1$. In the plane of $\tau_1$ we have for ${\rm Im}\,\tau_1\geq 0$: a
pole at $\tau_1=ix_1/v_F$ and three branch cuts starting at
$\tau_1=\bar{\tau}_1+i/\Lambda$, $\tau_2+i(x_1-x_2)/u$, and
$\bar{\tau}_2+i(x_1-x_2)/u$, as shown in Fig.~\ref{f22}. For weak
interaction, these branch cuts are ``weak'' in the sense that they are
characterized by a small exponent $\alpha_e$. The approximation we make
consists in neglecting all weak cuts and keeping only ``strong'' cuts
characterized by exponents close to 1. Taking into account the weak cuts
yields corrections small in $\alpha_e\ll 1$.

Picking up $\tau_1$ at the pole $\tau_1=ix_1/v_F$ yields the following
singularities in the plane of $\bar{\tau}_1$ for ${\rm Im}\,\bar{\tau}_1\geq
0$: a strong branch cut at $\bar{\tau}_1=ix_1/v_F$ and two weak branch cuts at
$\bar{\tau}_1=\tau_2+i(x_1-x_2)/u$ and $\bar{\tau}_2+i(x_1-x_2)/u$.
Neglecting the weak cuts and integrating along the strong one similarly to
Eq.~(\ref{151}) gives the renormalization of the impurity at point
$x_1$. During the integration along the strong cut, we pick up other factors
in the integrand at the classical trajectory, i.e., at the starting point of
the strong cut, which yields a cancellation of the cross-terms connecting two
impurities. The remaining double integral over $\tau_2$ and $\bar{\tau}_2$
with the factor $\exp [-M(2,\bar{2})]$ has the same structure as in the case
of a single impurity and gives the renormalization of the impurity at point
$x_2$. The net result is that the two-impurity diffuson $d_2$ for
$|\alpha_e|\ll 1$ simply acquires two identical factors $(\Lambda/\pi
T)^{2\alpha_e}$ which renormalize the impurities:
\begin{equation}
d_2\simeq d_2^{(0)}\left(\Lambda/ \pi T\right)^{4\alpha_e}~,
\label{155}
\end{equation}
where
\begin{eqnarray}
d_2^{(0)}&=& \left({\gamma_0\over 2v_F}\right)^2{(T/2v_F)^2\over \sinh^2\{\pi
    T\,[\,(x+2x_1-2x_2)/v_F+i\tau\,]\}}\nonumber \\
    &\times&[\Theta(x_1)\Theta(x_1-x_2)\Theta(x-x_2)\nonumber \\
    &+&\Theta(-x_1)\Theta(x_2-x_1)\Theta(x_2-x)]~.
\label{156}
\end{eqnarray}
Self-energy corrections are incorporated in the same way leading to an
additional factor $d_2\to d_2\exp (-\gamma|x+2x_1-2x_2|/v_F)$, where $\gamma$
is given by Eq.~(\ref{153}). This procedure is straightforwardly extended to
the diffuson ladder with three or more impurity legs, so that the difference
of the full diffuson ladder in the presence of e-e interaction from that in
the absence of interaction is only in the change $\gamma_0\to\gamma$ 
[Eq.~(\ref{153})].

\subsection{Dephasing action and weak-localization correction}
\label{VIIf}

Let us now return to the three-impurity Cooperon whose interaction-induced
factor is given by Eqs.~(\ref{139}),(\ref{142})--(\ref{145}). As the analysis
of the two-impurity diffuson demonstrated, the analytical structure of the
diagrams with increasing number of factors $\exp (\pm M)$ rapidly gets quite
complicated. The same example suggested, however, a systematic way to deal
with the limit of weak interaction by ignoring in the first approximation all
weak branch cuts. The integration around strong cuts then leads to the
renormalization of impurities, while picking up nonsingular factors at the
classical trajectory (the starting points of the strong cuts) generates the
dephasing action. At this level, the calculation of the Cooperon is similar to
that of the diffuson, with one essential difference: in contrast to the
diffuson, the Cooperon factor $\exp (-S_C)$, Eq.~(\ref{139}), does {\it not}
vanish on the classical trajectory and hence leads to dephasing. 

Another important issue is related to the definition of the classical
trajectory for the Cooperon; more specifically, to the question of what is the
velocity that characterizes this trajectory. To answer this question, one has
to carefully treat the factors $\eta_\pm$ (\ref{120}) in the integration over
the internal times of the Cooperon. As shown in Appendix~\ref{aE}, the
classical Cooperon trajectory is characterized by a single velocity and this
is $u$. Having established the fact that there is the unique velocity on the
whole classical trajectory of the Cooperon, we can, in the leading
approximation, ignore below the difference between $u$ and $v_F$ in the
functions $M(N,N')$.

Neglecting first the disorder-induced corrections $b(x,\tau)$ in
Eq.~(\ref{138}) and summing up according to Eqs.~(\ref{143}),(\ref{144}) all
12 terms each of which is given by Eq.~(\ref{145}), we observe that the
difference $M_{ff}-M_{fb}$ in this approximation ``miraculously'' cancels out
to zero on the classical trajectory. This is in agreement with
Eq.~(\ref{103}) saying that the dephasing rate for the Cooperon vanishes in
the limit $\gamma\to 0$. 

Combining Eqs.~(\ref{138}) and (\ref{145}), the disorder-renormalized
correlators $M$ are represented as 
\begin{equation} 
M=M\vert_{\gamma=0}+4b~,
\label{157} 
\end{equation} 
where $b=b_1+b_2$ is given by Eqs.~(\ref{135}),(\ref{136}). In order to
evaluate the dephasing action on the classical trajectory, we analytically
continue $b_{1,2}(x,\tau)$ in the complex plane of $\tau$. The analytical
continuation of $b_2(x,\tau)$ is effected via Eq.~(\ref{133}). On the
imaginary axis of $\tau$, the functions $b_{1,2}$ at order ${\cal O}(\gamma
\alpha^2)$ read 
\begin{eqnarray} 
b_1(x,it)&\simeq& {\pi\over 4}{\alpha^2 T\over v_F^2}\gamma|x|\,
\left[\,(|x|-v_Ft)\,\Theta(-|x|+v_Ft)\right.\nonumber \\
&+&\left. (|x|+v_Ft)\,\Theta(-|x|-v_Ft)\,\right]~,
\label{158}\\ 
b_2(x,it)&\simeq&
-{\pi\over 8}{\alpha^2 T\over v_F^2}\gamma\,
\left[\,(|x|-v_Ft)^2\,\Theta(-|x|+v_Ft)\right.\nonumber\\
&+&\left. (|x|+v_Ft)^2\,\Theta(-|x|-v_Ft)\,\right] 
\label{159} 
\end{eqnarray} 
for both $|x|/v_F$ and $|t|\gg 1/T$. Taking this limit is justified for the
calculation of the dephasing rate since upon averaging over the positions of
impurities in Eq.~(\ref{91}) the characteristic distance between impurities
is $v_F\tau_\phi^{\rm wl}$ or larger [in fact it is spread over the range
between $v_F\tau_\phi^{\rm wl}$ and $v_F\tau$ because of the logarithmic
factor in Eq.{(\ref{102})], which in turn is much larger than $v_F/T$
according to Eq.~(\ref{103}). The sum of Eqs.~(\ref{158}),(\ref{159}) gives
\begin{eqnarray} 
&&b(x,it)\simeq {\pi\over 8}{\alpha^2 T\over v_F^2} \gamma\,
(x^2-v_F^2t^2)\nonumber \\ &&\times\,\,
[\,\Theta(-|x|+v_Ft)+\Theta(-|x|-v_Ft)\,]~.
\label{160}
\end{eqnarray} 
Introducing a short-hand notation 
\begin{equation} 
\beta (N,N')={8\over \pi \alpha^2T\gamma }\,b[\,x_N-x_{N'},i(t_N-t_{N'})\,]~,
\label{161} 
\end{equation} 
where $b$ is given by the asymptote (\ref{160}), we see that $\beta(N,N')=0$
on the classical trajectory of the Cooperon for all $N,N'$ except for
\begin{equation} 
\beta(1,\bar{2})=\beta(2,\bar{3})=-2t_a(t_c-2t_a)~.  
\label{162}
\end{equation} 
Summing up the terms (\ref{162}) with the signs prescribed by
Eqs.~(\ref{144}),(\ref{157}) we obtain the dephasing action: 
\begin{equation} 
S_C^{\rm deph}\simeq 2\pi\alpha^2T\gamma t_a(t_c-2t_a)~.  
\label{163}
\end{equation} 
This result coincides with Eq.~(\ref{101}) for the dephasing action derived
via the path-integral calculation in Sec.~\ref{VId}. The integration along
the cuts around the classical trajectory yields the renormalization of
disorder on short spacial scales of order $v_F/T$ around impurities, similarly
to the renormalization of the diffuson ladder in
Eqs.~(\ref{152}),(\ref{155}). The three-impurity Cooperon with both the
renormalization of disorder and the dephasing included is thus given by
\begin{equation} 
c_3\simeq c_3^{(0)}\left(\Lambda/\pi T\right)^{6\alpha} 
\exp[-2\pi\alpha^2T\gamma t_a(t_c-2t_a)]~, 
\label{164} 
\end{equation}
where 
\begin{eqnarray} 
c_3^{(0)}&=&\left({\gamma_0\over 2v_F}\right)^3{(T/2v_F)^2\over \sinh^2[\,\pi
T(x_\Sigma/v_F+i\tau)\,]}\nonumber \\
&\times&\left[\,\Theta(x_1)\Theta(x_1-x_2)\Theta(x_3-x_2)\right.  \nonumber \\
&\times&\:\:\Theta(x_3-x)\Theta(x_3)\Theta(x_1-x)\nonumber \\
&+&\:\:\Theta(-x_1)\Theta(x_2-x_1)\Theta(x_2-x_3)\nonumber \\
&\times&\left.\:\,\Theta(x-x_3)\Theta(-x_3)\Theta(x-x_1)\,\right]
\label{165} 
\end{eqnarray} 
is the three-impurity Cooperon bubble without interaction, 
\begin{equation}
x_\Sigma=2(x_1+x_3-x_2)-x~, 
\label{166}
\end{equation} $|x_\Sigma|$ is the total length of
the classical trajectory along which the density fluctuations propagate from
point $x=0$ to point $x$. The length of the Cooperon loop $v_Ft_c$ and the
distance between two right-most impurities $v_Ft_a$ are related to the
positions of impurities by $v_Ft_c/2=\max\{x_3,x_1\}-x_2$ and
$v_Ft_a=|x_3-x_1|$. Equation (\ref{164}) describes the Cooperon bubble with
scalar vertices---the current vertices are straightforwardly incorporated
below.

To calculate the conductivity, one has to add also self-energy corrections to
the Green's functions constituting the Cooperon bubble. In the presence of e-e
interaction, this is done in essentially the same way as the renormalization
of impurities in Sec.~\ref{VIIe} and results in the additional factor $\exp
(-\gamma|x_\Sigma|/v_F)$, where $\gamma$ is the renormalized rate of
backscattering, which the Cooperon acquires on the length $|x_\Sigma|$. The
WL correction $\Delta\sigma_{\rm wl}$ is then expressed in 
terms of $c_3$ from Eq.~(\ref{164}) as
\begin{eqnarray}
&&\Delta\sigma_{\rm wl}=-2\times {\left(1\over 2\right)^2}\times 
\left\{{e^2v_F^2\over
\Omega_n}\int_0^{1/T}\!\!\!d\tau\int \!dx \,\,e^{i\Omega_n\tau}\right.
\nonumber \\ 
&&\times\,\left.\int\!dx_1\!\int\!dx_2\!\int\!dx_3\,\,c_3\,e^{-\gamma
|x_\Sigma|/v_F}\right\}_{\Omega_n\to +0},
\label{167}
\end{eqnarray}
where the factor of 2 accounts for the contribution of the diagrams (b) and
(c) in Fig.~\ref{f5}, which give exactly the same contribution as the
Cooperon, and the factor of $(1/2)^2$ comes from the dressing of the
current-vertices by the diffuson ladders, similarly to Eq.~(\ref{B4}). The
current vertices in Eq.~(\ref{167}) are simply given by $\pm v_F$, since we
formulate our model in terms of a linear electron spectrum from the very
beginning.\cite{velocity} Integrating (\ref{167}) we reproduce Eq.~(\ref{102})
which was obtained in Sec.~\ref{VId} in two distinct steps by treating the
renormalization of disorder and dephasing separately. Here, we have
demonstrated how these two effects arise ``side by side" from the technical
point of view in the framework of the functional bosonization. Specifically,
the analytical structure of the density-density correlator in the time domain
is such that the dephasing comes from the classical trajectory, whereas the
renormalization of impurities is associated with the integration along strong
branch cuts around the classical trajectory.

\section{Summary}
\label{VIII}

In this paper, we have studied the transport properties of interacting
spinless electrons in a disordered quantum wire within the framework of the
Luttinger-liquid model. Our main result is the weak-localization correction
(\ref{102}), governed by the dephasing rate (\ref{103}).

We have developed two alternative approaches for a systematic analytic
treatment of the problem. One is a two-step procedure which combines the
bosonic-RG treatment of high-energy renormalization processes at the first
step with the subsequent analysis of low-energy real processes in the
fermionic language. The other approach is based on the method of ``functional
bosonization'' which makes it possible to treat both types of effects
simultaneously.

We have demonstrated that the notion of weak localization is applicable to the
strongly correlated one-dimensional electron system. This finding is of
conceptual importance, showing that the famous non-Fermi-liquid character of
the Luttinger liquid does not prevent this system from exhibiting the features
characteristic of conventional mesoscopic electron structures. Our approach
thus provides a framework for systematically studying the mesoscopic phenomena
in strongly correlated electron systems. Further directions of research
include mesoscopic fluctuations and noise in a disordered Luttinger liquid,
spin-related effects (in particular, the magnetoresistance), several-channel
quantum wires, transport in a disordered system in the limit of strong
interaction, transport in a ``granular Luttinger liquid" (strong impurities).

\section{Acknowledgments}
\label{grateful}

We thank A.~Altland, D.~Aristov, T.~Giamarchi, L.~Glazman, M.~Grayson,
K.~Le~Hur, D.~Maslov, A.~Nersesyan, A.~Tsvelik, A.~Yashenkin, O.~Yevtushenko,
and I.~Yurkevich for interesting discussions. The work was supported by the
Schwerpunktprogramm ``Quanten-Hall-Systeme" and the Center for Functional
Nanostructures of the Deutsche Forschungsgemeinschaft, by RFBR, by the Program
``Leading Russian Scientific Schools" (Grant No.\ 2192.2003.2), and by the
Program of RAS ``Condensed Matter Physics". The authors acknowledge the
hospitality of the Condensed Matter and Statistical Physics Section of the
Abdus Salam ICTP (Trieste), where part of this work was performed. The work of
IVG, conducted as part of the project ``Quantum Transport in Nanostructures"
made under the EUROHORCS/ESF EURYI Awards scheme, was supported by funds from
the Participating Organisations of EURYI and the EC Sixth Framework Programme.
ADM acknowledges the hospitality of the Kavli Institute for Theoretical
Physics at Santa Barbara during the completion of the manuscript and partial
support by NSF under Grant No.\ PHY99-07949.

\appendix

\section{Polarization operator and screening in disordered wires}
\label{aA}
\renewcommand{\theequation}{A.\arabic{equation}}
\setcounter{equation}{0}

In this Appendix, we derive the polarization operator and the dynamically
screened RPA-interaction for a disordered 1d electron system. As everywhere in
the paper, we consider spinless electrons. The Luttinger-liquid power-law
singularities only show up in the renormalized strength of disorder $\gamma$
and do not make any other difference in the derivation below.

We begin by considering the clean case.  According to Dzyaloshinskii-Larkin
theorem,\cite{dzyaloshinskii74} the RPA equations for screened interaction are
then exact. The propagators of interaction between right-movers ($V_{++}$)
and between right- and left-movers ($V_{+-}$) obey
\bea
V_{++}&=&g_4-g_4\ \Pi_{+}\ V_{++}-g_2\ \Pi_{-}\ V_{+-}~,
\label{A1}
\\
V_{+-}&=&g_2-g_2\ \Pi_{+}\ V_{++}-g_4\ \Pi_{-}\ V_{+-}~,
\label{A2}
\eea
where 
\[
\Pi_{\pm}(i\Omega_n,q)={1\over 2\pi}\frac{q}{qv_F\mp i\Omega_n}
\]
are the chiral polarization operators for right(+)/left(-) movers,
$\Omega_n=2\pi n T$ is the bosonic Matsubara frequency. The total
polarization operator is a sum of the chiral terms:
\be
\Pi(i\Omega_n,q)= \rho \frac{q^2v_F^2}{q^2v_F^2+\Omega_n^2}~,
\label{A4}
\ee
where $\rho=1/\pi v_F$ is the thermodynamic density of states, so that
$\Pi(0,q)=\rho$. For later use we also introduce the two-particle
propagator
\be
D_{\pm}(i\Omega_n,q)=\frac{i\Omega_n}{|\Omega_n|}\,{\pi 
\rho \over i\Omega_n\mp qv_F}~,
\label{A5}
\ee
which satisfies 
\bea
&&\int \!{dp\over 2\pi}\,
G^{(0)}_\pm(i\epsilon_m+i\Omega_n,p+q)G^{(0)}_\pm(i\epsilon_m,p)
=D_{\pm}(i\Omega_n,q)
\nonumber \\
&&\times\left[\,\Theta(-\epsilon_m)\Theta(\epsilon_m+\Omega_n)+
\Theta(\epsilon_m)\Theta(-\epsilon_m-\Omega_n)\,\right]~,
\nonumber
\eea
where $G^{(0)}_\pm(i\epsilon_m,p)=(i\epsilon_m\mp v_Fp)^{-1}$ are the free
Green's functions for a given chirality and the ultraviolet momentum cutoff is
sent to infinity. The polarization operator and the two-particle propagator
are related to each other via
\[
\Pi_{\pm}(i\Omega_n,q)={\rho\over 2}-
{|\Omega_n|\over 2\pi} \,D_{\pm}(i\Omega_n,q)~.
\]

The solution to Eqs.~(\ref{A1}) and (\ref{A2}) for arbitrary
$g_2$ and $g_4$ reads\cite{solyom79}
\begin{eqnarray*}
V_{++}&=&\left[\,g_4+(g_4^2-g_2^2)\,
{1\over 2\pi v_F}\,\frac{qv_F}{i\Omega_n+qv_F}\,\right]
\nonumber
\\
&\times& (\Omega_n^2+q^2v_F^2)/(\Omega_n^2+ q^2u^2)~,
\label{A8}
\\
V_{+-}&=&g_2\,(\Omega_n^2+q^2v_F^2)/(\Omega_n^2+q^2u^2)~,
\label{A9}
\end{eqnarray*}
where
\[
u=v_F\left[\left(1+g_4/2\pi v_F\right)^2-\left(g_2/2\pi
      v_F\right)^2\right]^{1/2}~. 
\label{A10}
\]
For $g_4=g_2$ we have
\bea
V_{++}&=&V_{+-}=g_2\,\frac{\Omega_n^2+q^2v_F^2}{\Omega_n^2+u^2 q^2}~,
\label{A11}
\\
u&=&v_F\sqrt{1+{g_2\over \pi v_F}}~,
\label{A12}
\eea
while for $g_4=0$ the propagators $V_{++}$ and $V_{+-}$ are different from
each other:
\bea
V_{++}&=&{g_2^2\over 2\pi v_F}\,
\frac{qv_F(i\Omega_n-qv_F)}{\Omega_n^2+u^2 q^2}~,
\label{A13}
\\
V_{+-}&=&g_2\,\frac{\Omega_n^2+q^2v_F^2}{\Omega_n^2+u^2 q^2}~,
\label{A14}
\\
u&=&v_F\left[1-\left(g_2/2\pi v_F\right)^2\right]^{1/2}~.
\label{A15}
\eea
The interaction between left-movers is given by $V_{++}$ with $q\to -q$.

Let us now turn to the disordered case. We only take into account the backward
scattering induced by disorder, since the forward scattering can be completely
gauged out in the calculation of the conductivity. We average the RPA bubbles
over disorder separately by using the ladder approximation, see
Sec.~\ref{VId}. The backscattering off disorder makes it necessary to
introduce two indices $\mu,\nu=\pm$ in the polarization bubble $\Pi_{\mu\nu}$,
corresponding to the chirality of the vertices of the bubble. The RPA
equations for the interaction propagators in the presence of disorder read
\bea
V_{++}&=&g_4-(g_4\Pi_{++}+g_2\Pi_{-+})V_{++}\nonumber \\
&-&(g_2\Pi_{--}+g_4\Pi_{+-})V_{+-}~,
\label{A16}
\\[0.2cm]
V_{+-}&=&g_2-(g_2\Pi_{++}+g_4\Pi_{-+})V_{+-}\nonumber \\
&-&(g_4\Pi_{--}+g_2\Pi_{+-})V_{++}~,
\label{A17}
\eea
where
\[
\Pi_{\mu\nu}(i\Omega_n,q)={\rho \over 2}\delta_{\mu\nu}-{|\Omega_n|\over
  2\pi}D_{\mu\nu}(i\Omega_n,q). 
\label{A18}
\]

We now calculate the two-particle propagators $D_{\mu\nu}$ in the RPA.
Performing the analytical continuation to real frequencies $i\Omega_n\to
\omega+i0$, we introduce the retarded two-particle propagator
$[D^{R}_{\mu\nu}(\omega,q)]^{(0)}$ with only self-energy impurity lines
included,
\be
[D^{R}_{++}(\omega,q)]^{(0)}
= \frac{i \pi \rho}{\omega - qv_F+i\gamma/2}~,
\label{A19}
\ee
where 
\be
\gamma=1/\tau
\label{A20}
\ee
and $\tau$ is the transport scattering time. Note the factor of $1/2$ in front
of $\gamma$ in Eq.~(\ref{A19}), which reflects the fact that only
backscattering off impurities is considered, i.e., $\tau_q=2\tau$, where
$\tau_q$ is the quantum scattering time, in contrast to the case of isotropic
scattering, where $\tau_q=\tau$. The propagator $[D^{R}_{--}(\omega,q)]^{(0)}$
is given by Eq.~(\ref{A19}) with the change $q\to -q$, whereas the nondiagonal
propagators $[D^{R}_{\mu,-\mu}(\omega,q)]^{(0)}=0$.

The equations for dressed two-particle retarded propagators read
\bea
D^{R}_{--}(\omega,q)&=&[D^{R}_{--}(\omega,q)]^{(0)}\left[\,1+{\gamma\over
    2 \pi \rho}D^{R}_{+-}(\omega,q)\,\right]~, 
\nonumber \\
\label{A21}
\\
D^{R}_{+-}(\omega,q)&=&[D^{R}_{++}(\omega,q)]^{(0)}{\gamma\over  2 \pi
  \rho}D^{R}_{--}(\omega,q)~,
\label{A22}
\eea
and
\bea
D^{R}_{++}(\omega,q)&=&D^{R}_{--}(\omega,-q)~, 
\label{A23} \\
D^{R}_{+-}(\omega,q)&=&D^{R}_{-+}(\omega,q)~.
\label{A24}
\eea
Solving Eqs.~(\ref{A21})--(\ref{A24}), we get
\bea
D^{R}_{--}(\omega,q)&=&i\pi \rho\,
\frac{\omega-q+i\gamma/2}{\omega^2-q^2v_F^2+i\omega\gamma}~, 
\label{A25}
\\
D^{R}_{+-}(\omega,q)&=&-\pi \rho \,{\gamma\over 2}\,
\frac{1}{\omega^2-q^2v_F^2+i\omega\gamma}~, 
\label{A26}
\eea
and
\bea
\Pi_{--}(\omega,q)&=&\Pi_{++}(\omega,-q)
\nonumber \\
&=& {\rho\over 2} \ 
\frac{q^2v_F^2-qv_F\omega-i\omega\gamma/2}{q^2v_F^2-\omega^2-i\omega\gamma}~,
\label{A27} 
\\
\Pi_{+-}(\omega,q)&=&\Pi_{-+}(\omega,q)\nonumber \\
&=& {\rho\over 2}\ \frac{i\omega\gamma/2}{q^2v_F^2-\omega^2-i\omega\gamma}~,
\label{A28}
\\
\Pi(\omega,q)&=&\sum_{\mu\nu}\Pi_{\mu\nu} =\rho\,
\frac{q^2v_F^2}{q^2v_F^2-\omega^2-i\omega\gamma}~. 
\nonumber \\
\label{A29}
\eea
Equation (\ref{A29}) reduces to Eq.~(\ref{A4}) in the clean limit $\gamma\to
0$ and acquires the conventional diffusive form
\[
\Pi^{\rm diff}(\omega,q)=\rho Dq^2/(Dq^2-i\omega)
\label{A30}
\]
in the limit $\omega\ll \gamma,qv_F$\,, where the 1d diffusion constant
$D=v_F^2/\gamma$.


From Eqs.~(\ref{A16}), (\ref{A17}), (\ref{A27}), and (\ref{A28}), we obtain
the retarded interaction propagators for $g_2=g_4$:
\bea
V_{++}^R&=&V_{+-}^R=\frac{g_2}{1+g_2\Pi}\nonumber \\ &=& g_2
\,\frac{q^2v_F^2-\omega^2-i\omega\gamma}{q^2u^2-\omega^2 -i\omega\gamma}
\label{A31}
\eea
with $u$ given by Eq.~(\ref{A12}). For $g_4=0$ we have 
\bea
V_{++}^R&=&-g_2^2\  \frac{ \Pi_{--} }{ (1+g_2\Pi_{+-})^2
-g_2^2\Pi_{--}\Pi_{++} }
\nonumber 
\\ 
\nonumber
\\
&=&-{g_2^2\over 2\pi v_F} \ \frac{q^2v_F^2-qv_F\omega-i\omega\gamma/2}
{q^2u^2-\omega^2-i\omega(\gamma+\delta\gamma)}~,
\label{A32}
\\[0.3cm]
V_{+-}^R&=& g_2\ \frac{ 1+g_2\Pi_{+-} }{ (1+g_2\Pi_{+-})^2
-g_2^2\Pi_{--}\Pi_{++}
  }
\nonumber \\
\nonumber \\
&=&g_2 \ 
\frac{q^2v_F^2-\omega^2-i\omega(\gamma+\delta\gamma/2)}
{q^2u^2-\omega^2-i\omega(\gamma+\delta\gamma)}~,
\label{A33}
\eea
where $u$ is given by Eq.~(\ref{A15}) and
\be
\delta\gamma=-\gamma {g_2\over 2\pi v_F}~.
\label{A34}
\ee

\section{Weak-localization correction to the conductivity in 1d}
\label{aB}
\renewcommand{\theequation}{B.\arabic{equation}}
\setcounter{equation}{0}

In this Appendix, we calculate the WL correction to the conductivity in a
noninteracting 1d system, assuming that the phase coherence is broken by an
external source of dephasing and the dephasing time is small; specifically,
$\tau_\phi\ll \tau$.  Under this condition, the conductivity can be expanded
in powers of $\tau_\phi/\tau$. The lowest-order diagrams are depicted in
Fig.~\ref{f5}. The sum of the diagrams with two impurity lines, shown in
Fig.~\ref{f7}, yields a zero contribution to order
$(\epsilon_F\tau)^{-1}$.

We formally consider only backward scattering, i.e., the impurity line
corresponds to the correlation function of the backscattering random
potential
\[
L^{-1}\langle |{\cal U}_b(q)|^2\rangle=1/2\pi \rho \tau
\label{B1}
\]
($L$ is the system size) and changes chirality at every impurity vertices. For
the anisotropic scattering, we have $\tau=\tau_q/2$, where $\tau$ is the
transport time and $\tau_q$ is the quantum lifetime. The retarded
single-particle Green's function expressed in terms of $\tau$ reads
\[ 
G_{\pm}^R(\epsilon,p)=(\epsilon\mp v_Fp+i/2\tau_q)^{-1}
=(\epsilon\mp v_Fp+i/4\tau)^{-1}~.  
\label{B2}
\] 
The anisotropic scattering also introduces the current vertex renormalization:
each current vertex is multiplied by a factor $\tau/\tau_q=1/2$.

The sum of the diagrams (b) and (c) in Fig.~\ref{f5} with an impurity line
covering the two-impurity Cooperon (two crossed impurity lines) is equal to
the contribution $\sigma_{C3}$ of the diagram with the three-impurity
Cooperon. The total three-impurity WL correction is thus given by
\[
\Delta\sigma_{\rm wl}=2\,\sigma_{C3}~.
\label{B3}
\]
The expression for $\sigma_{C3}$ reads
\be
\sigma_{C3}=-2 \left({\tau\over \tau_q}\right)^2 e^2 v_F^2 \int{d\epsilon\over
2\pi}\left(-{\partial f_\epsilon\over \partial\epsilon}\right) \int {dQ\over
2\pi} J(Q)~,
\label{B4}
\ee
where $f_\epsilon$ is the Fermi distribution function, the overall minus sign
reflects the fact that the product of the current vertices is negative in the
Cooperon with three backscattering lines, and the factor of 2 is related to
two possibilities of setting chiralities at the vertices: $(+-)$ and
$(-+)$. The function $J(Q)$ describes the fermionic loop:
\bea
&&J(Q)={1\over(2\pi \rho \tau)^3}\int{dp\over 2\pi}\int{dp_1\over
  2\pi}\int{dp_2\over 2\pi} 
\nonumber \\
&&\times\,\, G^R_{+}(p)G^R_{-}(p_1)G^R_{+}(p_2)G^R_{-}(-p+Q)
\nonumber \\
&&\times\,\,
G^A_{+}(p)G^A_{-}(-p_2+Q)G^A_{+}(-p_1+Q)G^A_{-}(-p+Q)~.\nonumber\\
\label{B5}
\eea
Using the equality
\[
G^R_{\pm}(p)G^A_{\pm}(p)= 2i\tau\left[\,G^R_{\pm}(p)-G^A_{\pm}(p)\,\right],
\label{B6}
\]
we simplify Eq.~(\ref{B5}):
\[
J(Q)=4\tau^2 |P(Q)|^2[P(Q)+P^*(Q)]~,
\label{B7}
\]
where 
\[
P(Q)={1\over 2\pi \rho \tau}\int{dp\over 2\pi}G^R_{+}(p)G^A_{-}(-p+Q)~.
\label{B8}
\]
Now we introduce a phenomenological dephasing time $\tau_\phi$ through an
additional decay of the Green's function:
\[
G_{\pm}^R(\epsilon,p)\to (\epsilon\mp v_Fq+i/4\tau+i/2\tau_\phi)^{-1}~,
\label{B9}
\]
which yields
\bea
P(Q)&=&{1\over 2\tau}\ \frac{1}{i v_F Q+1/2\tau+1/\tau_\phi}~,
\label{B10}\\
J(Q)&=&{1\over \tau}\left({1\over 2\tau}+{1\over \tau_\phi}\right)
\frac{1}{[\,(v_F Q)^2+(1/2\tau+1/\tau_\phi)^2\,]^2}~.
\nonumber \\
\label{B11}
\eea
Substituting Eq.~(\ref{B11}) in Eq.~(\ref{B4}), we find for 
$\tau_\phi\ll \tau$: 
\be
\Delta\sigma_{\rm wl}=-{1\over 8} \sigma_{\rm D} 
\left({\tau_\phi\over \tau}\right)^2,
\label{B13}
\ee
where  $\sigma_{\rm D}=e^2v_F\tau/\pi$ is the Drude conductivity.

Equation (\ref{B13}) for the weak-localization correction can be cast in the
form of an integral over the time needed for a particle to return to the
starting point after two backscatterings:
\be
\Delta\sigma_{\rm wl}= 
- 2 \sigma_{\rm D} \int_0^\infty \!dt_c \,P_2(t_c) \exp(-t_c/\tau_\phi)~.
\label{B14}
\ee
Here 
\[
P_2(t)\simeq t/16\tau^2~,\qquad t\ll\tau
\label{B15}
\]
is the return probability after two reflections (hence the subscript 2) for a
particle which starts moving either to the right or to the left and the
overall factor of 2 in Eq.~(\ref{B14}) comes from the summation over the sign
of the velocity at the starting point.

Let us now turn from the phenomenological factor $\exp(-t_c/\tau_\phi)$ to
the actual dephasing factor. The latter depends not only on the total time
$t_c$ but also on the geometry of the Cooperon path. We thus have to perform
averaging over the geometry of the path at given $t_c$. For this purpose we
need the probability density of return $P_2(t_c,t_a)$ for a particle which
starts moving to the right at point $x=0$ under the condition that the total
length is $v_Ft_c$ and the first reflection occurs at point $x=v_Ft_a$. This
function is written as
\bea
&&P_2(t_c,t_a)=v_F\int_0^\infty \!dt_b\, P_1(t_a) P_1(t_b-t_a) 
\,\Theta(t_b-t_a)
\nonumber \\
&&\times\,\, \exp[\,-(t_c-t_b)/2\tau\,]\,\Theta(t_c-t_b)\, \delta[x(t_c)]~.
\label{B16}
\eea
Here 
\[
P_1(t)={d\over dt}\left(1-e^{-t/2\tau}\right)
\label{B17}
\]
is the probability density of being reflected at point $v_F t$ (again, note
that $\tau$ is the transport scattering time and only the backscattering is
taken into account). The factor $\exp\,[-(t_c-t_b)/2\tau\,]$ in
Eq.~(\ref{B16}) describes the probability of avoiding the backscattering on
the last segment of the path and
\[
x(t_c)=2v_F(t_a-t_b)+v_F t_c
\label{B18}
\]
for $t_c>t_b$. We obtain
\bea
&&P_2(t_c,t_a)={1\over 4\tau^2}\,e^{-t_c/2\tau}\nonumber \\
&&\times\,\,
\int_0^\infty \!\!dt_b\, \Theta(t_b-t_a)\Theta(t_c-t_b)\,
\delta[\,2(t_a-t_b)+t_c\,]
\nonumber \\
&&=\, {1\over 8\tau^2}\,e^{-t_c/2\tau}\,\Theta(t_c-2t_a)~.\nonumber
\label{B19}
\eea
For the dephasing action which depends on $t$ and $t_a$ we have
\[
\Delta\sigma_{\rm wl} = - 2 \sigma_{\rm D} 
\int_0^\infty \!dt_c \int_0^\infty \!dt_a
\,P_2(t_c,t_a) \exp[-S(t_c,t_a)]~, 
\label{B20}
\]
which reduces to Eq.~(\ref{B14}) for the phenomenological action
$S(t_c,t_a)=t_c/\tau_\phi$. 

\section{Dephasing action in the path-integral approach}
\label{aC}
\renewcommand{\theequation}{C.\arabic{equation}}
\setcounter{equation}{0}

In this Appendix, we present details of the calculation of the dephasing
action defined by Eq.~(\ref{93}) and given in the final form by
Eq.~(\ref{101}). For the imaginary parts of the retarded interaction
propagators (\ref{A32}),(\ref{A33}) which enter Eq.~(\ref{93}), we have,
retaining only the terms of order ${\cal O}(g_2^2)$:
\bea
{\rm Im}V_{+\pm}^R(\omega,q)\simeq-{g_2^2\over 4\pi v_F}\omega\gamma
\frac{(qv_F-\omega)(qv_F\mp\omega)}
{(q^2v_F^2-\omega^2)^2+\omega^2\gamma^2}~.\nonumber
\eea
Next, we introduce functions ${\cal F}_{+\pm}(\omega,q)$ which obey 
\be
{\rm Im}V_{+\pm}^R(\omega,q)=-{g_2^2\over 4\pi v_F}\,\omega
{\cal F}_{+\pm}(\omega,q)~. 
\ee
Putting $g_2=0$ in ${\cal F}_{+\pm}$ and doing the Fourier transform to 
$(x,t)$ space, we get
\bea
{\cal F}_{++}(x,t)&=&e^{-\gamma|t|/2}\left[\,\delta(x_+)\right.
\nonumber\\
&+&\left.\frac{\gamma^2|x_-|}{16v_F^2}f_{++}\left(-
  \frac{\gamma^2}{16v_F^2}x_{+}x_-\right)\,\right]~,\label{C6}
\\
{\cal F}_{+-}(x,t)&=&{\gamma\over 4v_F}e^{-\gamma|t|/2}f_{+-}\left(-
  \frac{\gamma^2}{16v_F^2}x_{+}x_-\right)~.\nonumber\\
\label{C7}
\eea
Here
$x_\pm=x\pm v_Ft$,
\bea
f_{++}(z)&=&{1\over \sqrt{z}} I_1(2\sqrt{z})\,\Theta(z)~,
\nonumber\\
f_{+-}(z)&=&I_0(2z)\,\Theta(z)~,
\nonumber
\eea
$I_0(z)$ and $I_1(z)$ are the modified Bessel functions. Since the dephasing
action becomes of order unity at times $t\sim\tau_\phi^{\rm wl} \ll \tau,$ we
expand ${\cal F}_{++}(x,t)$ and ${\cal F}_{+-}(x,t)$ to the first order in the
disorder strength $\gamma$, which yields Eqs.~(\ref{96}),(\ref{97}).

Let us now turn to the calculation of the dephasing action $S(t,t_a)$ which
can be represented as
\bea
S(t,t_a)=S_{\rm ff}+S_{\rm bb}-S_{\rm fb}+S_{\rm bf}=2(S_{\rm ff}-S_{\rm
fb})~,
\nonumber
\eea
where $S_{ij}$ are given by Eq.~(\ref{93}). We split the forward path into
three segments corresponding to the right- (R) and left- (L) moving parts,
where the coordinate $x_f(t)$ behaves as 
\bea
&&{\rm I_f},\quad R: \quad 0<t<t_a;  \quad  x_f(t)=v_Ft \nonumber \\
&&{\rm II_f},\quad L: \quad t_a<t<t_a+{t_c\over 2};  \quad
x_f(t)=v_F(2t_a-t)\nonumber \\ 
&&{\rm III_f},\quad R: \quad t_a+{t\over 2}<t<t_c;  \quad
x_f(t)=v_F(t-t_c)~.\nonumber 
\eea
Similarly, for the backward (time-reversed) path we have 
$x_b(t)=x_f(t_c-t)$:
\bea
&&{\rm I_b},\quad L: \quad 0<t<{t_c\over 2}-t_a;  \quad  x_b(t)=-v_Ft
\nonumber \\ 
&&{\rm II_b},\quad R: \quad {t_c\over 2}-t_a<t<t_c-t_a;  \quad\nonumber \\
&&\hspace{4cm}
x_b(t)=v_F(2t_a-t_c+t)\nonumber \\ 
&&{\rm III_b},\quad L: \quad t_c-t_a<t<t_c;  
\quad  x_b(t)=v_F(t_c-t)~.\nonumber 
\eea

The ``self-energy" part $S_{\rm ff}$ of the action is written as
\be
S_{\rm ff}={g_2^2\over 4\pi v_F^2} T J_{\rm ff}~,\quad
J_{\rm ff}=\sum_{n=1}^3J_n+2\sum_{n=4}^6J_n~, 
\ee
where each of the terms $J_n$ corresponds to the integration over a certain
pair of the forward-path segments (for brevity, we set $v_F=1$ in the
arguments of ${\cal F}_{\mu\nu}$ below):
\bea
{\rm I_f}-{\rm I_f}:\qquad  J_1&=&v_F\int_0^{t_a}\!dt_1\int_0^{t_a}\!dt_2
\nonumber \\
&\times&
{\cal F}_{++}(t_1-t_2,t_1-t_2)~,
\nonumber\\
{\rm II_f}-{\rm II_f}:\qquad
J_2&=&v_F\int_{t_a}^{t_a+t_c/2}\!dt_1\int_{t_a}^{t_a+t_c/2}\!dt_2\nonumber \\ 
&\times&{\cal F}_{--}(-t_1+t_2,t_1-t_2)~,
\nonumber\\
{\rm III_f}-{\rm III_f}:\qquad J_3&=&v_F\int_{t_a+t_c/2}^{t_c}
\!dt_1\int_{t_a+t_c/2}^{t_c} \!dt_2\nonumber \\ 
&\times&{\cal F}_{++}(t_1-t_2,t_1-t_2)~,
\nonumber\\
{\rm I_f}-{\rm II_f}:\qquad  2J_4&=&v_F\int_{0}^{t_a} \!dt_1
\int_{t_a}^{t_a+t_c/2} \!dt_2  
\nonumber \\
&\times&\left[\,{\cal F}_{+-}(t_1+t_2-2t_a,t_1-t_2)\right. 
\nonumber \\
&+& \left. {\cal F}_{-+}(-t_1-t_2+2t_a,t_1-t_2) \,\right]~,\nonumber \\
{\rm I_f}-{\rm III_f}:\qquad 2J_5&=&v_F\int_{0}^{t_a} \!dt_1 
\int_{t_a+t_c/2}^{t_c} 
\!dt_2 
\nonumber \\
&\times&\left[\,{\cal F}_{++}(t_1-t_2+t_c,t_1-t_2)\right. \nonumber \\
&+& \left.{\cal F}_{++}(-t_1+t_2-t_c,t_1-t_2) \,\right]~,\nonumber \\
{\rm II_f}-{\rm III_f}:\qquad 2J_6&=&v_F\int_{t_a}^{t_a+t_c/2} \!dt_1
\int_{t_a+t_c/2}^{t_c} \!dt_2  
\nonumber \\
&\times&\left[\,{\cal F}_{-+}(-t_1-t_2+2t_a+t_c,t_1-t_2)\right. \nonumber \\
&+& \left. {\cal F}_{+-}(t_1+t_2-2t_a-t_c,t_1-t_2) \,\right]~,\nonumber 
\eea

The ``vertex" part $S_{\rm fb}$ of the action is given by
\be
S_{\rm fb}={g_2^2\over 4\pi v_F^2} T J_{\rm fb}~,
\quad J_{\rm fb}=\sum_{n=7}^{15}J_n~,
\label{C22}
\ee
where 
\bea
{\rm I_f}-{\rm I_b}:\qquad J_7&=&v_F\int_0^{t_a}\!dt_1\int_0^{t_c/2-t_a}\!dt_2
\nonumber \\
&\times&
{\cal F}_{+-}(t_1+t_2,t_1-t_2)~,\nonumber\\
{\rm I_f}-{\rm II_b}:\qquad 
J_8&=&v_F\int_{0}^{t_a}\!dt_1\int_{t_c/2-t_a}^{t_c-t_a}\!dt_2\nonumber \\ 
&\times&{\cal F}_{++}(t_1-t_2-2t_a+t_c,t_1-t_2)~,\nonumber \\
{\rm I_f}-{\rm III_b}:\qquad J_9&=&v_F\int_{0}^{t_a} \!dt_1\int_{t_c-t_a}^{t_c}
\!dt_2\nonumber \\ 
&\times&{\cal F}_{+-}(t_1+t_2-t_c,t_1-t_2)~,\nonumber \\
{\rm II_f}-{\rm I_b}:\qquad J_{10}&=&v_F\int_{t_a}^{t_a+t_c/2} \!dt_1
\int_{0}^{t_c/2-t_a} \!dt_2  
\nonumber \\
&\times&{\cal F}_{--}(-t_1+t_2+2t_a,t_1-t_2)~,\nonumber \\
{\rm II_f}-{\rm II_b}:\qquad J_{11}&=&v_F\int_{t_a}^{t_a+t_c/2} \!dt_1
\int_{t_c/2-t_a}^{t_c-t_a} \!dt_2  
\nonumber \\
&\times&{\cal F}_{-+}(-t_1-t_2+t_c,t_1-t_2)~,\nonumber \\
{\rm II_f}-{\rm III_b}:\qquad J_{12}&=&v_F\int_{t_a}^{t_a+t_c/2} \!dt_1
\int_{0}^{t_c/2-t_a} \!dt_2  
\nonumber \\
&\times&{\cal F}_{--}(-t_1+t_2+2t_a,t_1-t_2)~,\nonumber \\
{\rm III_f}-{\rm I_b}:\qquad J_{13}&=&v_F\int_{t_c/2+t_a}^{t_c} \!dt_1
\int_{0}^{t_c/2-t_a} \!dt_2  
\nonumber \\
&\times&{\cal F}_{+-}(t_1+t_2-t_c,t_1-t_2)~,\nonumber \\
{\rm III_f}-{\rm II_b}:\qquad J_{14}&=&v_F\int_{t_c/2+t_a}^{t_c} \!dt_1
\int_{t_c/2-t_a}^{t_c-t_a} \!dt_2  
\nonumber \\
&\times&{\cal F}_{++}(t_1-t_2-2t_a,t_1-t_2)~,\nonumber
\\
{\rm III_f}-{\rm III_b}:\qquad J_{15}&=&v_F\int_{t_a+t_c/2}^{t_c} \!dt_1
\int_{t_c-t_a}^{t_c/2-t_a} \!dt_2  
\nonumber \\
&\times&{\cal F}_{+-}(t_1+t_2-2t_c,t_1-t_2)~.\nonumber
\eea
Carrying out the integration yields:
\bea
J_1&=&{t_a\over 2}~,\nonumber \\
J_2&=&{t_c\over 4}~,\nonumber \\
J_3&=&{1\over 2}\left({t_c\over 2}-t_a\right)~,\nonumber \\
J_4&=&\gamma {t_a t_c\over 8}~,\nonumber \\
J_6&=&\gamma {t_c\over 8} \left({t_c\over 2}-t_a\right)~,\nonumber \\
J_8&=&{t_a\over 2}-\gamma {t_a\over 4} \left({t_c\over 2}-t_a\right)~,
\nonumber
\\
J_9&=&\gamma {t_a^2\over 4},\nonumber \\
J_{10}&=&{1\over 2}\left({t_c\over 2}-t_a\right)-\gamma{t_a\over 4}
\left({t_c\over 2}-t_a\right)~, \nonumber \\ 
J_{11}&=&{\gamma \over 4} \left[\,\left({t_c\over 2}-t_a\right)^2 +
  t_a^2\,\right]~, \nonumber \\ 
J_{12}&=&{t_a\over 2} -\gamma{t_a\over 4} \left({t_c\over 2}-t_a\right)~, 
\nonumber \\
J_{13}&=&{\gamma \over 4}\left({t_c\over 2}-t_a\right)^2~,  
\nonumber \\
J_{14}&=& {1\over 2}\left({t_c\over 2}-t_a\right)-\gamma{t_a\over 4}  
\left({t_c\over 2}-t_a\right)~,  \nonumber \\  
J_5&=&J_7=J_{15}=0~, \nonumber
\eea
and
\bea
J_{\rm ff}&=&{t_c\over 2}+\gamma{t_c^2\over 8}~,\\
J_{\rm fb}&=&{t_c\over 2}+{\gamma \over 4} \left[\,{t_c^2\over
    2}-8t_a\left({t_c\over 2}-t_a\right)\,\right],\\ 
J_{\rm ff}-J_{\rm fb}&=&2\gamma t_a\left({t_c\over 2}-t_a\right).
\eea
We see that the difference $J_{\rm ff}-J_{\rm fb}$ vanishes at $\gamma\to 0$,
i.e., in the absence of disorder in the interaction propagators.  Finally, we
arrive at the expression for the total action
\bea
S(t,t_a)&=&2\times {g_2^2\over 4\pi v_F^2} T (J_{\rm ff}-J_{\rm fb})
\nonumber\\
&=&{g_2^2\over\pi v_F^2} T\gamma
t_a\left({t_c\over 2}-t_a\right)~. 
\eea
Using $\alpha=g_2/2\pi v_F$, we obtain Eq.~(\ref{101}).

\section{Functional bosonization in the clean case}
\label{app:dephasing}
\renewcommand{\theequation}{D.\arabic{equation}}
\setcounter{equation}{0}

In this Appendix, we calculate the correlators $B_{\mu\nu}(x,\tau)$ in a clean
Luttinger liquid for an arbitrary strength of interaction. Following the
discussion in Sec.~\ref{IIIa}, we consider only the $g_2$ interaction, while
the $g_4$ interaction is accounted for in the shift of the Fermi velocity,
Eq.~(\ref{21}). Substituting the interaction propagator $V_{++}(i\Omega_n,q)$
given by Eq.~(\ref{A13}) [with $v_F$ understood here and below as the 
renormalized velocity
(\ref{21})] in Eq.~(\ref{113}), we have 
\begin{eqnarray}
&&B_{++}(x,\tau)= -T\sum_n\, \int {dq\over 2\pi}\, \frac{g_2^2}{2\pi v_F}\,
\frac{qv_F}{\Omega_n^2+q^2u^2}\nonumber \\ 
&&\times\,\,
{\displaystyle{e^{iqx-i\Omega_n\tau}-1\over -i\Omega_n+qv_F}} 
\label{D1}
={\cal B}_0+{\cal B}_>+{\cal B}_<~, 
\end{eqnarray} 
where ${\cal B}_0(x)$, ${\cal B}_>(x,\tau),$ and ${\cal B}_<(x,\tau)$ are the
contributions of terms with $\Omega_n=0$, $\Omega_n>0,$ and $\Omega_n<0$ in
the Matsubara sum over the bosonic frequency $\Omega_n=2\pi n T$,
respectively. The sums over $\Omega_n>0$ and $\Omega_n<0$ are related to each
other as
\be
{\cal B}_<(x,\tau)={\cal
B}_>(-x,-\tau)~.
\label{D3}
\ee

To accurately treat the ultraviolet cutoff for an arbitrary ratio $u/v_F$, we
introduce a finite range of interaction $d$, so that $g_2(q)$ depends on the
transferred momentum $q$ and vanishes for $qd\gg 1$. The RPA approximation,
leading to Eq.~(\ref{D1}), is exact in the Luttinger-liquid model for
arbitrary $g_2(q)$. In the integrand of Eq.~(\ref{D1}), $g_2(q)$, $v_F(q)$,
and $u(q)$ are related to each other at given $q$ in precisely the same way,
Eq.~(\ref{33}), as in the case of $g_2(q)={\rm const}$. Everywhere below in
this Appendix, except for Eq.~(\ref{D12}), when writing $g_2$, $v_F$, or $u$,
we understand them as taken at $q=0$.

The term in Eq.~(\ref{D1}) with $\Omega_n=0$ at $|x|\gg d$ reads
\begin{equation}
{\cal B}_0(x)
=\frac{v_F^2-u^2}{2u^2}{2\pi T\over v_F}|x|~.
\label{D5}
\end{equation}
In the integration over $q$ at $\Omega_n>0$, there are poles at
$q=i\Omega_n/v_F$ and $q=\pm i\Omega_n/u$. The first only contributes to
the integral if $\Omega_n\alt v_F/d$, whereas the latter---only if
$\Omega_n\alt u/d$, which sets the limits of summation over $n$:
\begin{eqnarray}
&&{\cal B}_>(x,\tau)
= 2 \pi T \,\Theta(x)\sum_{n=1}^\infty \frac{e^{-\Omega_n/\Lambda_v}
 [e\,^{-\Omega_n(x/v_F +i\tau)}-1\,]}{\Omega_n}\nonumber \\
&&- \pi T \left({v_F\over u}+{\rm sgn}\,x\right)\sum_{n=1}^\infty
 \frac{e^{-\Omega_n/\Lambda} [\,e^{-\Omega_n (|x|/u +i\tau)}-1\,]}{\Omega_n}~,
\nonumber 
\\
\label{D6}
\end{eqnarray}
where there appear two ultraviolet scales $\Lambda_v=v_F/d$ and $\Lambda=u/d$
and we approximate the cutoffs by the exponential factors $\exp
(-\Omega_n/\Lambda_v)$ and $\exp (-\Omega_n/\Lambda)$ (the exact shape of
the cutoffs plays no role in the infrared physics we deal with in this paper).

Using for $N\gg 1$ and ${\rm Re}\,z\gg 1/N$ the formula
\[
\sum_{n=1}^{N} \frac{e^{-n z}-1}{n}\simeq 
-\ln(1-e^{-z})-\ln N~, 
\label{D7}
\]
we write ${\cal B}_>(x,\tau)$ for $|x|\gg d$ as
\begin{eqnarray}
{\cal B}_>(x,\tau)&=& 
\Theta(x)\ln\left\{{\Lambda\over \Lambda_v}{1
-\exp \,[-2\pi T(x/u+i\tau)]\over 1-\exp \,[-2\pi T(x/v_F+i\tau)]}\right\}
\nonumber \\
&-&\frac{v_F-u}{2u}
\ln{2\pi T/\Lambda\over 1-\exp \,[-2\pi T(|x|/u+i\tau)]}~.
\nonumber \\
\label{D8}
\end{eqnarray}
Combining Eqs.~(\ref{D5}), (\ref{D8}), and (\ref{D3}), we obtain
$B_{++}(x,\tau)$ given by Eq.~(\ref{115}). Note that the factor $v_F/u$ in
$\eta_+(x,\tau)$ [Eq.~(\ref{120})] comes from the ratio $\Lambda/\Lambda_v$ in
Eq.~(\ref{D8}). The remaining cutoff in ${\cal L}(x,\tau)$ is given by
$\Lambda$.

The Green's function [Eq.~(\ref{106})] of a right-mover then reads for 
$|x|\gg d$:
\bea
&&g_{+}(x,\tau)=g_{+}^{(0)}(x,\tau)\exp\,[-B_{++}(x,\tau)\,]
\nonumber \\
&&=\,-\frac{i T}{2u}\,\frac{1}{\sinh\,[\,\pi T(x/u+i\tau)\,]}  
\nonumber 
\\ 
&&\times\,\,
\frac{(\pi T/\Lambda)^{\alpha_b}}{\{\sinh \,[\,\pi T(x/u+i\tau)\,]
\sinh \,[\,\pi T(x/u-i\tau)\,]\}^{\alpha_b/2}}~.
\nonumber 
\label{D11}
\eea
Note that the Fermi velocity $v_F$ drops out completely in the product
$g_{+}^{(0)}(x,\tau)\eta_{+}(x,\tau)$. The Green's function in the real-time
domain, Eq.~(\ref{80}), is obtained by means of the Wick rotation $\tau\to i
t+{\rm sgn}\,t/\Lambda$.

The calculation of $B_{+-}(x,\tau)$ is performed in a similar way:
substituting Eq.~(\ref{A14}) in Eq.~(\ref{113}), we get
\begin{eqnarray}
B_{+-}(x,\tau)=
-T\sum_n\, \int {dq\over 2\pi}\, g_2\,
 {\displaystyle{e^{iqx-i\Omega_n\tau}-1\over
  q^2u^2+\Omega_n^2}}~.
\label{D12}
\end{eqnarray}
Note that the denominator $\Omega_n^2+q^2v_F^2$ in Eq.~(\ref{113}) is canceled
by the numerator in Eq.~(\ref{A14}), so that $v_F$ disappears from the problem
right away and the integration over $q$ in Eq.~(\ref{D12}) is determined at
$\Omega_n\neq 0$ solely by the plasmon poles $q=\pm i\Omega_n/u$. For $|x|\gg
d$, we arrive at Eq.~(\ref{117}).

In Sec.~\ref{VII} and in Appendix~\ref{aD}, we do contour integrals in the
complex plane of the Matsubara time $\tau$ and should be careful about the
behavior of $B_{\mu\nu}(x,\tau)$ at $x\to 0$ for $\tau\to 0$. When integrating
in the complex plane of $\tau$, it is convenient to have the ultraviolet
cutoffs in $B_{\mu\nu}(x,\tau)$ that are independent of $\tau$. Our choice is
to use Eq.~(\ref{D6}), which yields a cutoff which depends on $x$:
\begin{equation} x\to x+{u\over \Lambda}\,{\rm sgn}\,\, x=x+d\,\,{\rm
sgn}\,\,x~, \label{D14a} \end{equation} i.e., $|x|\to|x|+d$ in the propagators
$B_{\mu\nu}(x,\tau)$ in Eqs.~(\ref{115})--(\ref{117}), and similarly in the
disorder-induced corrections to them in Eqs.~(\ref{135}),(\ref{136}).

\section{``Dirty RPA" in the functional bosonization}
\label{aD}
\renewcommand{\theequation}{E.\arabic{equation}}
\setcounter{equation}{0}

In this Appendix, we demonstrate that the correlators (\ref{112}) in the
presence of disorder are modified according to Eq.~(\ref{113})
with the effective RPA interaction $V_{\mu\nu}$ (\ref{125}),(\ref{126})
whose dynamical properties reflect the disorder-induced backscattering
(``dirty RPA"). For this purpose, we {\it directly} calculate the
disorder-induced correction to the averages of the type $\left<\exp
\{i[\theta_\mu (x,\tau)-\theta_\nu (0,0)]\}\right>$ in the framework of the
diagrammatic technique formulated in Sec.~\ref{VIIb}. In the expansion in
powers of $\gamma$, the rules of the technique prescribe that impurity
vertices are connected with each other via the correlators
$B_{\mu\nu}(x,\tau)\vert_{\gamma=0}$ in a spatially homogeneous system. To
obey Wick's theorem, multiple fermionic loops can only be present in this
technique if each of them contains backscattering vertices, since all
fermionic loops without impurities have been accounted for by the ``clean RPA"
propagators. We show below that the result for $B_{\mu\nu}$ obtained this way
coincides for $\alpha\ll 1$ with the result of the ``dirty RPA",
Eq.~(\ref{138}).

\begin{figure}[ht] 
\centerline{
\includegraphics[width=8cm]{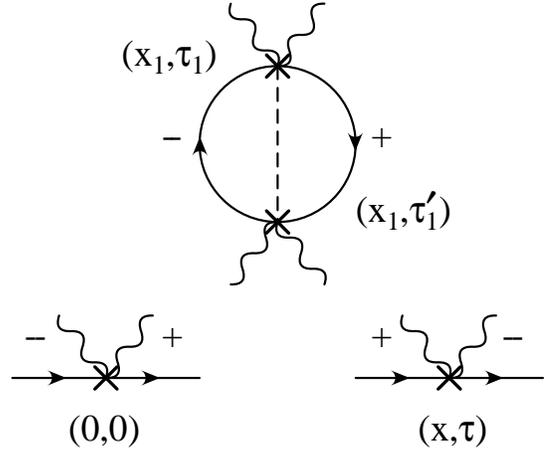}
}
\caption{
Diagram in the functional-bosonization technique, which describes the addition
of disorder in the RPA interaction propagator. Two backscattering vertices
$(0,0)$ and $(x,\tau)$ are connected to each other via the fermionic bubble 
with a single impurity at point $x_1$.  
}
\label{f23} 
\end{figure}

Let us consider the average of first order in $\gamma$,
\begin{eqnarray}
&&\left<e^{i[\theta_+(x,\tau)-\theta_+(0,0)]}\right>-\left<e^
{i[\theta_+(x,\tau)-\theta_+(0,0)]}\right>_{\gamma=0}\nonumber\\
&&\simeq\int \!dx_1 I(x,\tau,x_1)~,
\label{E1}
\end{eqnarray}
given by the diagram in Fig.~\ref{f23}. We are interested in the
pairings between the fields $\theta_+(0,0)$ and $\theta_+(x,\tau)$ that
connect up the points $(0,0)$ and $(x,\tau)$ via the fermionic bubble that
contains two backscatterings at point $x_1$:
\begin{eqnarray}
&&I(x,\tau,x_1)={v_F\gamma_0\over 2}e^{-B_{++}(x,\tau)}\int_0^{1/T}\! 
d\tau_1\int_0^{1/T}\!
d\tau'_1 \nonumber \\
&&\times \,\,g_+^{(0)}(0,\tau'_1-\tau_1)g_-^{(0)}(0,\tau_1-\tau'_1)\nonumber \\
&&\times \,\,e^{-B_{++}(0,\tau_1-\tau'_1)-B_{--}(0,\tau_1-\tau'_1)
+2B_{+-}(0,\tau_1-\tau'_1)}\nonumber
\\&&
\times \left[\,e^{-B_{++}(x_1,\tau'_1)+B_{+-}(x_1,\tau'_1)
+B_{++}(x_1,\tau_1)-B_{+-}(x_1,\tau_1)}-1\,\right]\nonumber\\
&&\times \left[\,e^{B_{++}(x-x_1,\tau-\tau'_1)
-B_{+-}(x-x_1,\tau-\tau'_1)}\right.\nonumber\\
&&\left.\times\,\, e^{-B_{++}(x-x_1,\tau-\tau_1)
+B_{+-}(x-x_1,\tau-\tau_1)}-1\,\right]~.
\label{E2}
\end{eqnarray}

\begin{figure}[ht] 
\centerline{\includegraphics[width=8cm]{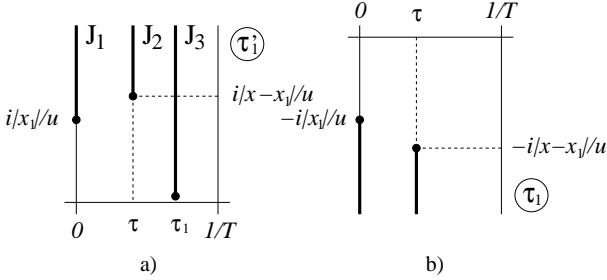}}
\caption{
Analytical structure of the integrand in Eq.~(\ref{E2}) in the complex plane
of $\tau'_1$ and $\tau_1$. (a) The contour of integration over $\tau'_1$
is closed upwards and runs around three branch cuts [$J_{1,2,3}$ are the
corresponding terms in Eq.~(\ref{E7})]. (b) The contour of
integration over $\tau_1$ for the first two terms in Eq.~(\ref{E7}) is closed
downwards and runs around two branch cuts.}
\label{f24} 
\end{figure}

\noindent All functions $B_{\mu\nu}$ in Eq.~(\ref{E2}) are understood as taken
at zero disorder. The analytical structure of the integrand in the complex
plane of $\tau_1$ and $\tau'_1$ is similar to that in the integrals analyzed
in Secs.~\ref{VIIe},\ref{VIIf}. Specifically, in the plane of $\tau'_1$ we
have six vertical branch cuts starting at $\tau'_1=\pm i|x_1|/u$, $\tau'_1=\pm
i|x-x_1|/u$, and $\tau'_1=\tau_1\pm i/\Lambda$, and going to ${\rm
Im}\,\tau'_1=\pm i\infty$, Fig.~\ref{f24}a. On top of that, there are
singularities associated with the functions $\eta_\mu$ (\ref{120}). Note that
the strong singularity (a double pole if $\alpha\to 0$) at $\tau'_1=\tau_1$
(if $\Lambda$ is sent to infinity)
that comes from the product of two free Green's functions
$g^{(0)}_\pm[\,0,\pm(\tau'_1-\tau_1)\,]$ is largely canceled by the vanishing
of the rest of the integrand at this same point. Evaluation of the integrals
(\ref{E2}) at arbitrary $\alpha$ leads to rather cumbersome expressions which
are not particularly interesting for our purposes. Let us extract the factor
that renormalizes the impurity strength,
\begin{eqnarray}
I(x,\tau,x_1)=&-&{v_F\gamma_0\over 2u^2}\left({\Lambda\over \pi
T}\right)^{2\alpha_e}e^{-B_{++}(x,\tau)\vert_{\gamma=0}}
\nonumber \\&\times&J(x,\tau,x_1)~,\nonumber
\label{E3}
\end{eqnarray}
and analyze the behavior of the factor $J(x,\tau,x_1)$ at small
$\alpha$. By introducing a short-hand notation $s(z)=\sinh(\pi T z)$,
$J(x,\tau,x_1)$ is written as
\begin{eqnarray}
&&J(x,\tau,x_1)=-\left({T\over 2}\right)^2
\int_0^{1/T}\! d\tau_1\int_0^{1/T}\!d\tau'_1\nonumber\\
&&\times\,{1\over s^{1-\alpha_e}[i(\tau'_1-\tau_1)]
s^{1-\alpha_e}[-i(\tau'_1-\tau_1)]}\nonumber\\ \nonumber \\
&&\times\,\left\{\left[{s(|x_1|/u+i\tau'_1)s(|x_1|/u-i\tau'_1)\over
s(|x_1|/u+i\tau_1)s(|x_1|/u-i\tau_1)}\right]^{\alpha_e/
2}\chi_1-1\right\}\nonumber \\
&&\times\,\left\{\left[{s(|x-x_1|/u+i\tau-i\tau_1)\over
s(|x-x_1|/u+i\tau-i\tau'_1)}\right]^{\alpha_e/2}\right.\nonumber \\
&&
\times\,\left.\left[{s(|x-x_1|/u-i\tau+i\tau_1)\over 
s(|x-x_1|/u-i\tau+i\tau'_1)}\right]^{\alpha_e/
2}\chi_2-1\right\}~,
\label{E4}
\end{eqnarray}
where 
\begin{eqnarray}
\chi_1&=&{s(x_1/v_F+i\tau'_1)s(x_1/u+i\tau_1)\over
s(x_1/u+i\tau'_1)s(x_1/v_F+i\tau_1)}~,\nonumber
\label{E5}\\
\chi_2&=&{s[(x-x_1)/v_F+i(\tau-\tau_1)]\over
s((x-x_1)/u+i(\tau-\tau_1)]}\nonumber \\ \nonumber \\
&\times&{s[(x-x_1)/u+i(\tau-\tau'_1)]\over s[(x-x_1)/v_F+i(\tau-\tau'_1)]}~.
\label{E6}\nonumber
\end{eqnarray}

In Sec.~\ref{VIIc}, while calculating the disorder-induced correction to
$B_{\mu\nu}$ to order ${\cal O}(\gamma\alpha^2)$, we neglected the difference
between $v_F$ and $u$ in the interaction propagators. To compare with the
result of Sec.~\ref{VIIc}, we neglect the difference between the two velocities
here as well; in particular, by putting $\chi_1=\chi_2=1$. We discuss the role
of such factors in the calculation of the dephasing rate in Appendix~\ref{aE}.

The integral over $\tau'_1$ in Eq.~(\ref{E4}) can be represented as a sum of
three terms, 
\begin{equation}
J=-{T\over 2}\int_0^{1/T}\!d\tau_1\, (J_1+J_2+J_3)~, 
\label{E7}
\end{equation}
each of which is the integral around a branch cut in the complex plane of
$\tau'_1$, Fig.~\ref{f24}a. Inspection of Eq.~(\ref{E4}) shows that $J_3$,
the term related to the branch cut starting at
$\tau'_1=\tau_1+i/\Lambda$, can be omitted for $\alpha_e\ll 1$, since $J_3\sim
{\cal O}(\alpha_e^3)$, which gives a subleading contribution to $J$ [one power
of $\alpha_e$ comes from the difference of the integrand on two sides of the
cut and two more come from the expansion of the expressions in the curly
brackets in Eq.~(\ref{E4}) in $\tau'_1-\tau_1$]. The
integrals $J_{1,2}$ are written for $\alpha_e\ll 1$ as
\begin{widetext}
\begin{eqnarray*}
&&J_1\simeq \alpha_e\, {\pi T\over 2} \int_0^\infty\!\!
dt'_1 {1\over s^{2(1-\alpha_e)}\left(t'_1+{|x_1|\over v_F}
+i\tau_1\right)}
\left[\,{
s(t'_1)
s\left(t'_1+2{|x_1|\over v_F}\right)
\over
s\left({|x_1|\over v_F}+i\tau_1\right)
s\left({|x_1|\over v_F}-i\tau_1\right)
}\,\right]^{\alpha_e/2}\nonumber \\
&&\times\,\left\{\left[\,{
s\left({|x-x_1|\over v_F}+i\tau-i\tau_1\right)
s\left({|x-x_1|\over v_F}-i\tau+i\tau_1\right)
\over
s\left(t'_1+{|x-x_1|\over v_F}+{|x_1|\over v_F}+i\tau\right)
s\left(t'_1-{|x-x_1|\over v_F}+{|x_1|\over v_F}+i\tau\right)
}\,\right]^{\alpha_e/2}\!\!\!e^{i\pi\alpha_e/2}-1\right\}~,
\nonumber \\ \label{E8} \\
&&J_2\simeq -\alpha_e\, {\pi T\over 2} \int_0^\infty\!\!
dt'_1 {1\over s^{2(1-\alpha_e)}\left(t'_1+{|x-x_1|\over v_F}+i\tau_1
-i\tau\right)}\,
\left[\,{
s\left({|x-x_1|\over v_F}+i\tau-i\tau_1\right)
s\left({|x-x_1|\over v_F}-i\tau+i\tau_1\right)
\over
s(t'_1)
s\left(t'_1+2{|x-x_1|\over v_F}\right)
}\,\right]^{\alpha_e/2}\nonumber \\
&&\times\,\left\{\left[\,{
s\left(t'_1+{|x-x_1|\over v_F}-{|x_1|\over v_F}-i\tau\right)
s\left(t'_1+{|x-x_1|\over v_F}+{|x_1|\over v_F}-i\tau\right)
\over
s\left({|x_1|\over v_F}+i\tau_1\right) 
s\left({|x_1|\over v_F}-i\tau_1\right)
}\,\right]^{\alpha_e/2}\!\!\!e^{-i\pi\alpha_e/2}-1\right\}~,
\nonumber \\ \label{E9}
\end{eqnarray*}
In turn, each of the integrals of $J_1$ and $J_2$ over $\tau_1$ can be split
into a sum of two integrals along branch cuts in the complex plane of $\tau_1$
(Fig.~\ref{f24}b), so that in total we have $J\simeq
(J_{1a}+J_{1b})+(J_{2a}+J_{2b})$, where for $\alpha_e\ll 1$:
\begin{eqnarray} 
&&
J_{1a}\simeq 
-\alpha_e^2 \left({\pi T\over 2}\right)^2\int_0^\infty
\!dt_1\int_0^\infty\!dt'_1\,
{1\over s^{2(1-\alpha_e)}\left(t_1+t'_1+2{|x_1|\over v_F}\right)}
\left[\,{
s(t'_1)
s\left(t'_1+2{|x_1|\over v_F}\right)
\over
s\left(t_1\right)
s\left(t_1+2{|x_1|\over v_F}\right)
}\,
\right]^{\alpha_e/2}\nonumber \\
&&
\times\,\left\{\left[\,{
s\left(t_1+{|x-x_1|\over v_F}+{|x_1|\over v_F}-i\tau\right)
s\left(t_1-{|x-x_1|\over v_F}+{|x_1|\over v_F}-i\tau\right)
\over
s\left(t'_1+{|x-x_1|\over v_F}+{|x_1|\over v_F}+i\tau\right)
s\left(t'_1-{|x-x_1|\over v_F}+{|x_1|\over v_F}+i\tau\right)
}\,\right]^{\alpha_e/2}
\!\!\!-1\right\}~, \label{E10} \\
&&
J_{1b}\simeq 
\alpha_e^2 \left({\pi T\over 2}\right)^2 \int_0^\infty
\!dt_1\int_0^\infty \!dt'_1\,
{1\over s^{2(1-\alpha_e)}
\left(t_1+t'_1+{|x_1|\over v_F}+{|x-x_1|\over v_F}+i\tau\right)}
\left[\,
{s(t'_1)
\over
s\left(t_1+{|x_1|\over v_F}+{|x-x_1|\over
v_F}+i\tau\right)}\,\right]^{\alpha_e/2}
\nonumber \\ 
&&\times\,\left[\,
{s\left(t'_1+2{|x_1|\over v_F}\right)
s\left(t_1\right)
s\left(t_1+2{|x-x_1|\over v_F}\right)
\over 
s\left(t_1-{|x_1|\over v_F}+{|x-x_1|\over v_F}+i\tau\right)
s\left(t'_1+{|x-x_1|\over v_F}+{|x_1|\over v_F}+i\tau\right)
s\left(t'_1-{|x-x_1|\over v_F}+{|x_1|\over
v_F}+i\tau\right)}\,\right]^{\alpha_e/2}~,
\label{E11} \\
&&J_{2a}\simeq 
\alpha_e^2 \left({\pi T\over 2}\right)^2 \int_0^\infty
\!dt_1\int_0^\infty \!dt'_1\,
{1\over s^{2(1-\alpha_e)}\left(t_1+t'_1+{|x_1|\over v_F}+{|x-x_1|\over v_F}
-i\tau\right)} \left[\,{s\left(t'_1+{|x-x_1|\over v_F}-{|x_1|\over
v_F}-i\tau\right)
\over
s\left(t_1\right)}\,\right]^{\alpha_e/2}
\nonumber \\ 
&&\times\left[\,
{
s\left(t'_1+{|x-x_1|\over v_F}+{|x_1|\over v_F}-i\tau\right)
s\left(t_1+{|x_1|\over v_F}+{|x-x_1|\over v_F}-i\tau\right)
s\left(t_1+{|x_1|\over v_F}-{|x-x_1|\over v_F}-i\tau\right)
\over
s\left(t_1+2{|x_1|\over v_F}\right)
s\left(t'_1\right)
s\left(t'_1+2{|x-x_1|\over v_F}\right)
}
\,\right]^{\alpha_e/2}~,
\label{E12} 
\end{eqnarray}
\end{widetext}
and $J_{2b}$ is obtained from $J_{1a}$ by interchanging $x_1\leftrightarrow
x-x_1$. The terms $J_{1a}$ and $J_{2b}$ on one hand and $J_{1b}$ and $J_{2a}$
on the other give essentially different contributions to $J(x,\tau,x_1)$. The
integrals $J_{1a}$ and $J_{2b}$, as a function of the position of the impurity
$x_1$, are not exponentially small only in a close vicinity of width $\sim
u/T$ around $x_1=0$ and $x_1=x$, respectively. After averaging over $x_1$,
these ``local" terms yield a contribution of order ${\cal O}(\alpha_e^2)$
which does not depend, in the leading approximation, on $x$ and $\tau$.  On
the contrary, $J_{1b}$ and $J_{2a}$ for small $\alpha_e$ do not depend on
$x_1$ for $x_1$ between points $x_1=0$ and $x_1=x$. The integration over $x_1$
for $|x|/v_F\gg 1/T$ yields then the leading contribution to
Eq.~(\ref{E1}). Only the latter terms are thus of importance to us. In the
limit of small $\alpha_e$, their contribution to $J$ is represented as
\begin{eqnarray}
&&J(x,\tau,x_1)\simeq \alpha_e^2 \left({\pi T\over 2}\right)^2 \int_0^\infty
\!dt_1\int_0^\infty \!dt'_1\nonumber \\
&&\times\,\left\{\sinh^{-2}\,\left[\,\pi
T\left(t_1+t'_1+{|x_1|\over v_F} 
+{|x-x_1|\over v_F}+i\tau\right)\,\right]\right.\nonumber \\
&&\left. \hspace{2.5cm}
+\,(\tau\to -\tau)\,\right\}~. 
\label{E14}
\end{eqnarray}
Analytically continuing $J(x,\tau,x_1)$ onto the imaginary axis of $\tau$, we
observe that $J(x,it,x_1)$ is exponentially small for large positive values of
$(T/v_F)(|x_1|+|x-x_1|\pm v_Ft)$ and is a linear function of this sum when it
is large and negative. Since we are interested in the limit of $|x|/v_F,|t|\gg
1/T$, we can approximate $J(x,it,x_1)$ as
\begin{eqnarray}
&&J(x,it,x_1)\simeq -{\pi\over 2}\alpha_e^2T\left(t-{|x_1|\over
v_F}-{|x-x_1|\over v_F}\right) \nonumber \\
&&\times\,\Theta\left(-|x_1|
-|x-x_1|+v_Ft\right)+\,(\,t\to -t\,)~.
\label{E15}
\end{eqnarray}
Integrating over $x_1$ we obtain
\begin{eqnarray}
\int\!dx_1I(x,it,x_1)=-e^{-B_{++}(x,it)\vert_{\gamma=0}}\,b(x,it)~,
\label{E16}
\end{eqnarray}
where $b(x,it)$ is given by Eq.~(\ref{160}). We thus provided a direct
verification of the RPA form of Eq.~(\ref{113}) by deriving the first term in
the expansion of $\exp (-B_{++})$ in powers of $\gamma$ without using the
disorder-renormalized effective interaction (\ref{125}). This procedure can
be extended to higher powers of $\gamma$ and to the correlator $B_{+-}$.

An important point concerns the factors $\sinh^{\pm\alpha_e/2}(\pi Tz)$, where
$z$ denotes the arguments of the functions $s(z)$ in
Eqs.~(\ref{E11}),(\ref{E12}). These were omitted in the limit of small
$\alpha_e$ in Eq.~(\ref{E14}). However, even for $\alpha_e\ll 1$, this step
cannot be justified for arbitrary $x,x_1$, and $\tau$, e.g., if $\alpha_e
|x|T/v_F\gg 1$. What is important to us is that all these factors actually
cancel each other after the analytical continuation onto the imaginary axis of
$\tau\to it$ for large positive values of $-|x_1| -|x-x_1|\pm ut$. As
Eq.~(\ref{E15}) shows, only this range of $t$ is relevant to $J(x,it,x_1)$,
which is the quantity used in the calculation of the dephasing rate. The RPA
for pairing of all four fields at points $(0,0)$ and $(x,\tau)$, shown in
Fig.~\ref{f24}, namely of the combinations
$\theta_+(x,\tau)-\theta_-(x,\tau)$ and $\theta_+(0,0)-\theta_-(0,0)$ with
each other, via the fermionic bubble with two backscatterings at $x=x_1$ can
be reproduced in a similar way.

\section{Classical trajectory for the Cooperon}
\label{aE}
\renewcommand{\theequation}{F.\arabic{equation}}
\setcounter{equation}{0}

In this Appendix, we substantiate an important step in our derivation of the
interaction-induced action $S_C$ of the Cooperon within the
functional-bosonization scheme. Specifically, we analyze the role of the
factors $\eta_{\pm}(x,\tau)$, Eq.~(\ref{120}), which contain both the Fermi
velocity $v_F$ and the plasmon velocity $u$. We demonstrate that in the
spinless case, considered in this paper, the velocity of electrons on the
``Cooperon path" which contributes to $\Delta\sigma_{\rm wl}$ is given by $u$
and not by $v_F$. Put another way, the velocity of the interfering
quasiclassical trajectories is renormalized by the interaction and coincides
with that of the RPA propagators. As a result, one can omit the potentially
dangerous factors $\eta_{\pm}(x,\tau)$ in the calculation of the WL dephasing.

The function $c_3$ in Eq.~(\ref{167}), which gives the three-impurity Cooperon
loop, is explicitely written as
\begin{eqnarray}
&&c_3(x,\tau_f-\tau_i,x_1,x_2,x_3)
=\left({v_F\gamma_0\over 2}\right)^3 \nonumber
\\
&&\times \,
\int_0^{1/T}\!\!\!\!d\tau_1\int_0^{1/T}\!\!\!\!d\bar{\tau}_1
\int_0^{1/T}\!\!\!\!d\tau_2\int_0^{1/T}\!\!\!\!d\bar{\tau}_2
\int_0^{1/T}\!\!\!\!d\tau_3\int_0^{1/T}\!\!\!\!d\bar{\tau}_3
\nonumber
\\
&&\times\,\,\left\langle
g_+^{(0)}(x_1,\tau_1-\tau_i)\,e^{i[\,\theta_+(x_1,\tau_1)
-\theta_-(x_1,\tau_1)\,]}\right.
\nonumber
\\
&&\times \quad g_-^{(0)} (x_2-x_1,\tau_2-\tau_1)\,
e^{-i[\,\theta_+(x_2,\tau_2)-\theta_-(x_2,\tau_2)\,]}\nonumber
\\
&&\times\quad g_+^{(0)}(x_3-x_2,\tau_3-\tau_2)
\,e^{i[\,\theta_+(x_3,\tau_3)-\theta_-(x_3,\tau_3)\,]}\, \nonumber
\\
&&\times \quad g_-^{(0)} (x-x_3,\tau_f-\tau_3)\, \nonumber
\\
&&\times\quad g_-^{(0)}(x_1-x,\bar{\tau}_1-\tau_f)\,
e^{-i[\,\theta_+(x_1,\bar{\tau}_1)-\theta_-(x_1,\bar{\tau}_1)\,]}\,
\nonumber
\\
&&\times \quad  g_+^{(0)} (x_2-x_1,\bar{\tau}_2-\bar{\tau}_1)
e^{i[\,\theta_+(x_2,\bar{\tau}_2)-\theta_-(x_2,\bar{\tau}_2)\,]}
\nonumber
\\
&&\times\quad g_-^{(0)} (x_3-x_2,\bar{\tau}_3-\bar{\tau}_2)\,
e^{-i[\,\theta_+(x_3,\bar{\tau}_3)-\theta_-(x_3,\bar{\tau}_3)\,]}
\nonumber
\\
&&\times \quad  g_+^{(0)} (-x_3,\tau_i-\bar{\tau}_3) \left.\right\rangle~,
\label{F1}
\end{eqnarray}
where we write the external times of the bubble as $\tau_i$ and $\tau_f$
[in Eq.~(\ref{167}), $\tau_f-\tau_i=\tau$].  
To examine the role of the factors $\eta_\pm$, it is convenient to introduce 
the function
\[
C_3=\int_0^{1/T}\!\!\!d\tau_i \int_0^{1/T}\!\!\! d\tau_f\,
 c_3 \,\,e^{i\Omega_n(\tau_f-\tau_i)}
\label{F2}
\]
by integrating first over the external times. 

The averaging $\langle\ldots\rangle$ over the fluctuating fields 
$\theta_\pm (x,\tau) $ yields the factor
\[
\exp \,[-2(M_{ff}-M_{fb})\,]~,
\label{F3}
\]
defined in Eqs.~(\ref{143})--(\ref{145}), in the integrand of
Eq.~(\ref{F1}). In the absence of interaction ($M_{ff}=M_{fb}=0$), the
integrals over the internal times are determined by the poles of the free
Green's functions [specified by Eq.~(\ref{F10}) below with $u\to v_F$].
For given coordinates $x_1$, $x_2$, $x_3$, and $x$, we refer to this
set of times as the ``classical trajectory" of the Cooperon with velocity
$v_F$.

Each interaction-induced term $M(N,N')$ in Eqs.~(\ref{143}),(\ref{144})
contains the combination $B_{++}+B_{--}$, and hence depends on both $u$ and
$v_F$. If one neglects disorder in the RPA propagators (``clean RPA"), the
factor $\exp[-M(N,N')]$ is represented as
\begin{eqnarray}
e^{-M(N,N')}&=&e^{- \alpha_e {\cal
L}(x_N-x_{N'},\tau_N-\tau_{N'})}
\nonumber\\
&\times& \zeta(x_N-x_{N'},\tau_N-\tau_{N'})~,
\label{F5}
\end{eqnarray}
where $\zeta (x,\tau)=
\eta_+(x,\tau)\,\eta_-(x,\tau)$,
$\alpha_e=1-K$, and the functions ${\cal L}(x,\tau)$ and $\eta_\pm
(x,\tau)$ are given by Eqs.~(\ref{119}),(\ref{120}). The dependence on $v_F$
comes from the factors $\eta_\pm$. The ``dirty RPA" does not affect the
latter. We thus see that each pair of impurities $N$ and $N'$ in
Eq.~(\ref{F1}) generates two poles associated with the functions $\eta_\pm$:
\begin{eqnarray*}
&&\zeta (x,\tau)
=\left( {v_F\over u} \right)^2 \nonumber \\
&&\times \,\,{\sinh \,[\,\pi T (x/v_F+ i\tau)\,] \,
\sinh \,[\,\pi T (x/v_F-i\tau)\,] 
\over \sinh \,[\,\pi T (x/u + i\tau)\,] 
\,\sinh \,[\,\pi T (x/u - i\tau)\,] }~.
\nonumber\\
\label{F7}
\end{eqnarray*}
The singularities in Eq.~(\ref{F1}) coming from the functions $\zeta(x,\tau)$
are analogous to those in the bare Green's functions
$g_{\pm}^{(0)}(x,\tau)$. However, they can appear either as ``$u$-poles" at
$i\tau=\pm x/u$ or as ``$v_F$-poles" at $i\tau=\pm x/v_F$, depending on the
sign of the factor $(-1)^{N+N'}$ or $(-1)^{N+\bar{N}}$ in 
Eqs.~(\ref{143}),(\ref{144}).

We can readily integrate over $\tau_i$ and $\tau_f$ the free Green's functions
that connect to the external vertices. Since there are no interaction-induced
factors at the external vertices, the integration yields again the free Green's
functions. The function $C_3$ then becomes
\begin{eqnarray}
&&C_3=\left({v_F\gamma_0\over 2}\right)^3\,{1\over v_F^2}\nonumber
\\
\nonumber \\
 &&\times \,
\int_0^{1/T}\!\!\!\!d\tau_1\int_0^{1/T}\!\!\!\!d\bar{\tau}_1
\int_0^{1/T}\!\!\!\!d\tau_2\int_0^{1/T}\!\!\!\!d\bar{\tau}_2
\int_0^{1/T}\!\!\!\!d\tau_3\int_0^{1/T}\!\!\!\!d\bar{\tau}_3
\nonumber
\\
&&\times\quad g_+^{(0)}(x_1-x_3,\tau_1-\bar{\tau}_3)
\,\zeta(x_1-x_3,\tau_1-\bar{\tau}_3)\nonumber \\
&&\times\quad g_-^{(0)}(x_2-x_1,\tau_2-\tau_1)
\,\zeta(x_2-x_1,\tau_2-\tau_1)\nonumber \\
&&\times\quad g_+^{(0)}(x_3-x_2,\tau_3-\tau_2)
\,\zeta(x_3-x_2,\tau_3-\tau_2)\nonumber \\
&&\times\quad g_-^{(0)}(x_1-x_3,\bar{\tau}_1-\tau_3)
\,\zeta(x_1-x_3,\bar{\tau}_1-\tau_3)\nonumber \\
&&\times\quad g_+^{(0)}(x_2-x_1,\bar{\tau}_2-\bar{\tau}_1)
\,\zeta(x_2-x_1,\bar{\tau}_2-\bar{\tau}_1)\nonumber \\
&&\times\quad g_-^{(0)}(x_3-x_2,\bar{\tau}_3-\bar{\tau}_2)
\,\zeta(x_3-x_2,\bar{\tau}_3-\bar{\tau}_2)\nonumber \\
&&\times\quad
\zeta^{-1}(x_1-x_3,\tau_1-\tau_3)\,
\zeta^{-1}(x_1-x_3,\bar{\tau}_1-\bar{\tau}_3)
\nonumber
\\
&&\times\quad
\zeta^{-1}(x_1-x_2,\tau_1-\bar{\tau}_2)\,
\zeta^{-1}(x_3-x_2,\tau_3-\bar{\tau}_2)
\nonumber
\\
&&\times\quad
\zeta^{-1}(x_2-x_1,\tau_2-\bar{\tau}_1)\,
\zeta^{-1}(x_2-x_3,\tau_2-\bar{\tau}_3)
\nonumber
\\
&&\times\quad\zeta (0,\tau_1-\bar{\tau}_1)\,
\zeta (0,\tau_2-\bar{\tau}_2)\,
\zeta (0,\tau_3-\bar{\tau}_3)\nonumber \\
&&\times\quad R(x_1,x_2,x_3,\tau_1,\tau_2,\tau_3,
\bar{\tau}_1,\bar{\tau}_2,\bar{\tau}_3)\nonumber \\
&&\times\quad W(x_1,\tau_1,x_3,\bar{\tau}_3,\Omega_n)\nonumber \\
&&\times\quad W(x-x_3,\tau_3,x-x_1,\bar{\tau}_1,-\Omega_n)~,
\label{F8}
\end{eqnarray}
where $\zeta(0,\tau)=(v_F/u)^2$, the function $R$ combines all the factors
coming from the first factor in Eq.~(\ref{F5}), and
\begin{eqnarray*}
&&W(x_1,\tau_1,x_3,\bar{\tau}_3,\Omega_n)\nonumber \\
&=& e^{-|\Omega_n x_1|-i\Omega_n\tau_1}
[\,\Theta(\Omega_n)\Theta(x_1)-\Theta(-\Omega_n)\Theta(-x_1)\,]\nonumber \\
&+& e^{-|\Omega_n x_3|-i\Omega_n\bar{\tau}_1}
[\,\Theta(\Omega_n)\Theta(x_3)-\Theta(-\Omega_n)\Theta(-x_3)\,]~.\nonumber \\
\label{F9}
\end{eqnarray*}
In the single-particle Green's function, the Fermi velocity $v_F$ drops out
completely in the product $g_\pm^{(0)}(x,\tau)\eta_\pm(x,\tau)$.  One can see
that each free Green's function in Eq.~(\ref{F8}) is multiplied by a
corresponding factor $\eta_{\pm}$, so that all the $v_F$-poles associated with
the free Green's functions in Eq.~(\ref{F8}) are replaced by $u$-poles. One
might think that $v_F$ disappears completely from the problem, and hence the
integration over the internal times in Eq.~(\ref{F8}) straightforwardly yields
a new classical trajectory defined by
\bea
i\tau_1&=&-x_1/u~, \nonumber \\
i\tau_2&=&(x_2-2x_1)/u~, \nonumber \\
i\tau_3&=&(2x_2-2x_1-x_3)/u~, \nonumber \\
i\bar{\tau}_3&=&-x_3/u~, \nonumber \\
i\bar{\tau}_2&=&(x_2-2x_3)/u~, \nonumber \\
i\bar{\tau}_1&=&(2x_2-2x_3-x_1)/u~.\label{F10}
\eea
However, the impurity vertices are all interconnected by the interaction
propagators. Among the latter, there are propagators that connect vertices
which are not connected by the single-particle Green's functions [e.g., points
$(x_1,\tau_1)$ and $(x_2,\bar{\tau_2})$]. As a result, the $v_F$-poles in the
corresponding functions $\eta_{\pm}$ remain uncompensated.

In the noninteracting case, the sequence of points in real space
\be
1 \to 2 \to 3 \to 1 \to 2 \to 3 \to 1
\label{F11}
\ee
forms a closed loop with segments given by the Green's functions
$g_{\pm}^{(0)}$. Now that we have the interaction-induced functions
$\eta_{\pm}$ that connect every pair of impurities, other trajectories can
contribute to $C_3$, whose segments are given either by the Green's function
or by the function $\eta_{\pm}$. In view of this, the notion of a single
classical trajectory contributing to $\Delta\sigma_{\rm wl}$ requires a more
accurate justification. As we show below, the interaction itself ``chooses" a
{\it unique} classical trajectory which is defined in Eq.~(\ref{F10}), i.e.,
the one that is determined by a combination of the Green's functions with the
renormalized velocity $v_F\to u.$ The contributions of all other trajectories
are exponentially suppressed. Note that within the fermionic path-integral
approach used in Sec.~\ref{VI}, the ``correct" velocity is chosen
automatically on the saddle-point trajectory of the total action.

When analyzing the role of the factors $\eta_{\pm}$, we can, in the first
approximation, neglect other interaction-induced factors in Eq.~({\ref{F1})
by setting the exponent $\alpha_e=0$ in Eq.~(\ref{F5}), i.e., by putting
$R=1$ in Eq.~(\ref{F8}) while keeping the difference between $v_F$ and $u$ in
the functions $\zeta$. This simplifies the analytical structure of the
integrand in Eq.~(\ref{F8}) by reducing the integration along branch cuts to
the integration around simple poles only. The integration around all the poles
then yields $C_3$ represented as a sum over a whole set of trajectories, not
only the one given by Eq.~(\ref{F11}).

The integrand of Eq.~(\ref{F8}) contains products of the type $g^{(0)}_\pm
(x,it)\zeta (x,it)$ which, as a function of $t$, have poles on the classical
trajectory (\ref{F10}) at $t=\pm x/u$ with a residue equal to
\bea
&&[\,ut\mp (x+u \,\,{\rm sgn}\,x/\Lambda)\,]\,
g^{(0)}_\pm(x,it)\,\zeta (x,it)\vert_{t\to \pm x/u}\nonumber \\
\nonumber \\
&&\simeq -{1\over 2\pi i}\,{v_F\over u}\,
\exp \left({u-v_F\over u v_F}\,\pi T|x|\,\right)
\label{F12}
\eea
for $|x|/u\gg 1/T$. Writing the remaining factors $\zeta$ which are not
singular at this pole as
\begin{eqnarray}
\zeta (x,it)&\simeq& \left({v_F\over u}\right)^2\exp \left[\,\pi
T\left(|t_+^v|+|t_-^v|-|t_+^u|-|t_-^u|\right)\,\right]\nonumber \\
&\times& {\rm sgn}\,t_+^v\,\, {\rm sgn}\, t_-^v\,\, {\rm sgn}\, t_+^u\,\, 
{\rm sgn}\, t_-^u
\label{F13}
\end{eqnarray}
for $|t_\pm^{v,u}|\gg 1/T$, where $t_\pm^v=t\pm x/v$ and $t_\pm^u=t\pm x/u$,
we observe that all the exponential factors coming from
Eqs.~(\ref{F12}),(\ref{F13}) cancel out on the classical trajectory
(\ref{F10}). 

On the contrary, any other sequence of poles has at least some of the segments
of the trajectory passed with velocity $v_F$ and no such compensation occurs,
which leads to an exponential decay of the contribution to $C_3$. The exponent
depends on a particular sequence of poles but for any sequence is at least of
the order of $(|u-v_F|/v_F)Tt_c$. One sees that the contribution of any of the
``nonclassical" trajectories is exponentially suppressed either by the
``Golden-Rule dephasing" or simply by the thermal smearing. The role of the
factors $\eta_\pm$ is thus two-fold: they renormalize the velocity on the
``classical" Cooperon trajectory $(v_F\to u)$ and suppress the contribution of
all nonclassical trajectories.  Taking into account the factor $R$ in
Eq.~(\ref{F8}) within the ``clean RPA" does not give any decay on this
classical trajectory either. Including disorder within the ``dirty RPA"
(Sec.~\ref{VIIc} and Appendix~\ref{aD}) in the factor $R$ yields an
exponential decay on the classical trajectory as well, as discussed in
Sec.~\ref{VIIf}, but this decay is much weaker than on the nonclassical
trajectories, so that the latter can indeed be neglected in the calculation of
$\Delta\sigma_{\rm wl}$.


\begin{thebibliography}{1}
\bibitem{curvature} Strictly speaking, this is true for a linear dispersion
only. In reality, a finite curvature $m^{-1}$ will lead to an additional
contribution to the dephasing rate $\sim \alpha^2T^2/mv_F^2$. One sees,
however, that Eq.~(\ref{103}) gives the leading contribution in a wide
temperature range $T\ll (m^2v_F^4/\alpha^2\tau)^{1/3}$.
\bibitem{velocity} Note that if we were to start with a quadratic dispersion
and then linearize, an additional term in the current vertices would appear
proportional to $\pm \partial_\tau \theta_\pm$, as pointed out in
Ref.~\onlinecite{lerner05}. This, however, only yields an ultraviolet
renormalization of the bare Fermi velocity $v_F$ by e-e interactions with
high-energy transfers. In this sense, in our model $v_F$ already includes all
``Fermi-liquid-like" renormalizations of the electron spectrum.
\end{thebibliography}
\end{document}